\documentclass[longauth]{aa}  
\usepackage[flushleft]{threeparttable}
\usepackage{txfonts}
\usepackage{xcolor}
\usepackage{lscape}
\usepackage{xcolor}
\usepackage{natbib}
\usepackage{enumitem}   
\usepackage{soul}

\usepackage{graphicx}
\usepackage{amssymb}
\usepackage{txfonts}
\usepackage{multicol}
\usepackage[colorlinks=true,allcolors=blue]{hyperref}

\newcommand{\oiii}{\hbox{[O\,{\sc iii}]}}

\newcommand{\hb}{\hbox{H$\beta$}}

\newcommand{\kms}{\hbox{km~s$^{-1}$}\xspace}

\begin{document}

   \title{GA-NIFS: Understanding the ionization nature of EGSY8p7/CEERS-1019. Evidence for a star formation-driven outflow at $z$ = 8.6.
   }

   \titlerunning{EGSY8p7}
    

   \authorrunning{Zamora et al.}

\author{Sandra Zamora \inst{\ref{inst:SNS}}\fnmsep\thanks{E-mail: \href{mailto:sandra.zamoraarenal@sns.it}{sandra.zamoraarenal@sns.it}} 
    \and  
    Stefano Carniani  \inst{\ref{inst:SNS}}
    \and 
    Elena Bertola\inst{\ref{inst:arcetri}}
    \and
    Eleonora Parlanti \inst{\ref{inst:SNS}}
    \and
    Pablo G. Pérez-González \inst{\ref{inst:cab}}
    \and
    Santiago Arribas \inst{\ref{inst:cab}}
    \and
    Torsten Böker \inst{\ref{inst:esa_baltimore}}
    \and
    Andrew J. Bunker \inst{\ref{inst:oxford}}
    \and
    Francesco D'Eugenio \inst{\ref{inst:cambridge1},\ref{inst:cambridge2}}
    \and
    Roberto Maiolino \inst{\ref{inst:cambridge1},\ref{inst:cambridge2}}
    \and
    Michele Perna \inst{\ref{inst:cab}}
    \and 
    Bruno Rodr\'iguez Del Pino \inst{\ref{inst:cab}}
    \and
    Hannah \"Ubler \inst{\ref{inst:max_planck}}
    \and 
    Giovanni Cresci \inst{\ref{inst:arcetri}}
    \and
    Gareth C. Jones \inst{\ref{inst:cambridge1},\ref{inst:cambridge2}}
    \and
    Isabella Lamperti \inst{\ref{inst:firence},\ref{inst:arcetri}}
    \and
    Jan Scholtz	\inst{\ref{inst:cambridge1},\ref{inst:cambridge2}}
    \and
    Bartolomeo Trefoloni \inst{\ref{inst:SNS}}
    \and
    Giacomo Venturi \inst{\ref{inst:SNS}}
    }
    
   \institute{Scuola Normale Superiore, Piazza dei Cavalieri 7, I-56126 Pisa, Italy\label{inst:SNS}
        \and 
        INAF - Osservatorio Astrofisco di Arcetri, largo E. Fermi 5, 50127 Firenze, Italy\label{inst:arcetri}
        \and
        Centro de Astrobiolog\'{\i}a (CAB), CSIC-INTA, Ctra. de Ajalvir km 4, Torrej\'on de Ardoz, E-28850, Madrid, Spain\label{inst:cab}
        \and
        European Space Agency, c/o STScI, 3700 San Martin Drive, Baltimore, MD 21218, USA \label{inst:esa_baltimore}
        \and
        University of Oxford, Department of Physics, Denys Wilkinson Building, Keble Road, Oxford OX13RH, United Kingdom \label{inst:oxford}
        \and
        Kavli Institute for Cosmology, University of Cambridge, Madingley Road, Cambridge, CB3 0HA, UK \label{inst:cambridge1}
        \and
        Cavendish Laboratory - Astrophysics Group, University of Cambridge, 19 JJ Thomson Avenue, Cambridge, CB3 0HE, UK 
        \label{inst:cambridge2}
        \and
        $^{}$Max-Planck-Institut f\"ur extraterrestrische Physik (MPE), Gie{\ss}enbachstra{\ss}e 1, 85748 Garching, Germany \label{inst:max_planck}
        \and
        Dipartimento di Fisica e Astronomia, Università di Firenze, Via G. Sansone 1, 50019, Sesto F.no (Firenze), Italy 
        \label{inst:firence}
}

   \date{ }

 
  \abstract
{Understanding the physical conditions and feedback mechanisms in early massive galaxies is essential to uncover how they formed and evolved during the first billion years of the Universe. In this context, the galaxy EGSY8p7 (also known as CEERS-1019) at $z = 8.6$ provides an excellent benchmark, given its stellar mass of 10$^{9.3}$ M$_\odot$ and elevated N/O abundance despite its sub-solar metallicity, a characteristic common among the high-z galaxy population. Previous JWST/NIRspec slit observations have suggested the presence of an active galactic nucleus (AGN) in this galaxy, making EGSY8p7 a potential precursor of the massive quasars observed at $z > 6$.
In this study, we present new JWST/NIRSpec observations, obtained in integral field spectroscopy (IFS) mode and part of the program GA-NIFS. These data offer the first spatially resolved spectroscopy of this galaxy, with higher sensitivity and spectral resolution than previous studies. We identify broad (${\rm FWHM} \sim 650~{\rm km~s^{-1}}$) H$\beta$ and [O III] emission components whose emission is located between the two rest-frame UV clumps of the galaxy and extended over a distance of $\sim 1$ kpc. The morphology and kinematics of these components indicate that the broad emission arises from outflowing gas rather than from an AGN broad-line region.
The kinetic energy injection rate from stellar feedback is an order of magnitude higher than that of the outflow, while the radiation pressure rate is comparable to the outflow momentum rate. These results suggest that stellar feedback alone can drive the outflow, with radiation pressure potentially providing the required momentum transfer. 
We derive a low mass-loading factor ($\eta = 0.16$) and ionizing photon escape fraction ($f_{\mathrm{esc}} = 0.021 \pm 0.014$). Together with the high electron density measured ($n_{\mathrm{e}} \sim 2200~{\rm cm^{-3}}$), these results support the interpretation that most of the gas remains confined within the galaxy.
Comparisons of diagnostic emission-line ratios with photoionization and shock models support a star-formation–driven ionization scenario, ruling out any excitation by AGN radiation. Finally, the absence of detectable Wolf–Rayet features suggests that alternative mechanisms must be considered to explain the high N/O ratio in this galaxy. 
}


   \maketitle
%
\section{Introduction}

Understanding the physical conditions and feedback mechanisms in early galaxies is essential for unveiling the processes that shaped galaxy formation and evolution in the first billion years of cosmic history. Deep spectroscopic observations with the \textit{James Webb} Space Telescope (JWST) have recently opened a new window into this early epoch, providing critical insights into the properties of galaxies in this epoch ($z>6$). 

One particularly notable target in this context is CEERS-1019, also called EGSY8p7, a galaxy initially identified through photometric dropout by \citet[][z$_{phot}$ = 8.57$^{+0.22}_{-0.43}$]{RobertsBorsani2016}. Its redshift was spectroscopically confirmed through strong Ly$\alpha$ emission \citep[z$_{spec}$ = 8.683$^{+0.001}_{-0.004}$;][]{Zitrin2015}, making it one of the most distant Ly$\alpha$ emitters currently known. This exceptional Ly$\alpha$ emission has been interpreted as evidence for an ionized bubble or an overdense environment \citep{Larson2022,Leonova2022,Whitler2024}. Additional indications of a hard ionizing spectrum come from a tentative detection of NV$\lambda$1240 \AA\ at 4.6 $\sigma$ \citep{Mainali2018}.  

The discovery of this peculiar emission prompted an investigation into its nature, revealing a high N/O abundance for its sub-solar metallicity \citep[see][]{Marques-Chaves2024}. The origin of such an exceptional nitrogen enric  hment has been extensively debated for other high redshift sources, with proposed “classical” scenarios such as enrichment from massive star winds (WR stars) or AGB stars, as well as more “exotic” mechanisms including pollution from Pop~III star formation, tidal disruption of stars by black holes, or mixing the ejecta from super massive stars formed through collisions in dense clusters \citep[see][]{cameron2023,Watanabe2024,Nagele2023}. 

More recent JWST observations have provided a far more detailed view of this galaxy. In particular, multi-band NIRCam and MIRI imaging have been analyzed by several groups \citep{Tang2023,Nakajima2023,Larson2023}. These studies report three distinct stellar clumps within $\sim$ 1.1 kpc that are remarkably compact with effective radii between 100 and 145 pc. Stellar population modeling suggests ages of 4, 6, and 15 Myr for the three clumps, with high star formation rate surface densities of log($\Sigma_{SFR}$) $\sim$ 2.8 -- 3.3 M$_\odot$/yr/kpc$^2$. Their stellar masses range between (4.6 -- 8.7) $\times$ 10$^8$ M$_\odot$ \citep{Marques-Chaves2024} 
that, combined with their remarkably small sizes, produce stellar mass surface densities log($\Sigma_{\rm M}$/[M$_\odot$/pc$^2$]) = 3.6~--~4.1. Then, they are comparable to the densest stellar systems observed
at high redshift \citep{Claeyssens2023,Capak2015,Harikane2025,Prieto-Jimenez2025} and several orders of magnitude above the levels found in local young clusters \citep{Brown2021}. 
This extreme compactness and high stellar density 
are comparable with those found in the nuclear star clusters at the center of most local galaxies \citep{Neumayer2020}
and may reflect the conditions expected in very young, massive clusters—potential proto-globular clusters \citep[e.g.][]{Senchyna2024,Charbonnel2023}.

Interestingly, NIRSpec multi-object spectroscopy (MOS) medium-resolution observations by \citet[][R $\sim$ 1000]{Larson2023} report a tentative (2.5$\sigma$) detection of a broad H$\beta$ component \citep[see also][]{Trefoloni2025}, which they interpret as originating from a broad-line region (BLR) around an active galactic nucleus (AGN). Based on this, they classify the source as an accreting supermassive black hole (SMBH), with an estimated mass of $\log(M/M_\odot)=6.95 \pm 0.37$ and an accretion rate of $1.2\pm0.5$ times the Eddington limit. 
However, subsequent studies have highlighted the need for deeper observations to confirm this result.

In this work, we present the first spatially resolved spectroscopic analysis of EGSY8p7, made possible by the high spatial and spectral resolution (R $\sim$ 2700) of JWST/NIRSpec integral field spectroscopy (IFS).
Our analysis includes a detailed characterization of the physical properties of the source, which have been the subject of extensive debate in the literature. We explore possible signatures of AGN-driven feedback, offering new insights into the nature of this high-redshift system. Furthermore, we revisit and refine measurements reported in earlier studies, placing them in the context of a multi-clump galaxy morphology to build a more complete and physically coherent picture of EGSY8p7.

The observations used in this work are presented in Section \ref{sec:obs}, with an overview of the data reduction procedures. In Section \ref{sec:broad}, we describe previous detections of broad components in Balmer emission lines in this galaxy and compare them with our new high spectral resolution data. This section also explains both the integrated and spatially resolved analyses performed in this study. Sections \ref{sec:ism} to \ref{sec:star} focus on the results obtained for the galaxy, including an analysis of the interstellar medium, the physical conditions of the ionized gas, and the properties of the stellar populations. The implications of these results are discussed in Section \ref{sec:disc}, and the final conclusions are presented in Section \ref{sec:conc}.

Throughout this work, we assume a concordance cosmology with $\Omega_{m} = 0.274$, $\Omega_{\Lambda} = 0.726$, and $H_0 = 70~\mathrm{km~s^{-1}~Mpc^{-1}}$, giving a cosmology corrected scale of 4.57 kpc/arcsec at z = 8.683.

\section{JWST observations and data processing}\label{sec:obs}
We used JWST NIRSpec IFS \citep{jakobsen2022,boker2023, boker2022} observations obtained as part of the NIRSpec Guarantee Observing Time program \href{https://ga-nifs.github.io/}{GA-NIFS}\footnote{Proposal ID 1262, cycle 1 (PI: Nora Luetzgendorf). The observations were acquired on 25th June 2023.} (Galaxy Assembly with NIRSpec IFS). Two disperser/filter spectral configurations were used: PRISM/CLEAR and G395H/F290LP with total exposure times of 3851 s and 36180 s, respectively. The PRISM setup provides a broad wavelength coverage, from 0.60 to 5.30 $\mu$m, corresponding to rest-frame wavelengths of approximately 620 to 5500~\AA\ at z = 8.683, with a spectral resolution of R $\sim$ 100. The G395H/F290LP configuration covers the redder part of the PRISM spectral range ($\lambda_{\rm rest}$ > 2900 \AA) at a significantly higher resolution, R $\sim$ 2700, allowing us to detect possible broad components in the emission lines associated with broad line regions or ionized gas outflows. Additionally, this configuration provides a higher signal-to-noise (S/N) ratio in rest-frame optical emission lines which are essential for deriving the physical conditions of the ionized gas, such as the electron density and chemical abundances.

We retrieved the raw data from the MAST (Barbara A. Mikulski Archive for Space Telescopes) archive. Then, we reduced them with a customized version of the JWST pipeline (version 1.11), using the Calibration Reference Data System (CRDS) context jwst\_1094.pmap. Some modifications to the pipeline allow us to improve the data quality and they are described in detail in \citet{perna2023}, but here we summarize the major changes. First, we applied the ``calwebb\_detector1'' step of the pipeline to account for detector level correction. Before calibrating the count-rate images through the ``calwebb\_spec2'' module of the pipeline, we corrected them by subtracting the 1/f noise through a polynomial fitting.
The final data cube was created by combining the individual calibrated 2-D exposures by using the ‘drizzle’ weighting obtaining a cube with a spaxel size of 0.05\arcsec.

Residual cosmic rays and other artifacts remaining after the standard calibration steps introduced a significant number of spectral spikes in the reduced data. To mitigate these effects, we applied a Fourier post-processing method (see Appendix \ref{app:clean}) after the data cube construction. We adopted the cleaning procedure described in \citet[][see their Section 3.4]{zamora2023fourier},
adapting the Tonry and Davis’ method \citep{TonryandDavis} for very high spectral resolution data. The main steps are (i) dividing the spectrum into logarithmic wavelength bins in order to get a uniform velocity shift 
, and (ii) filtering the high frequency variations below the spectral resolution of the grating. This filtering is performed by applying a band-pass filter to the Fourier-transformed spectrum using the most conservative set of parameters appropriate for the spectral resolution of the G395H/F290LP configuration ($\sigma$ $\sim$ 90 km/s, see Fig \ref{fig:fourier_}). The results of this filtering process are shown in Figure~\ref{fig:fourier}.

After the data reduction, we measured the redshift of the galaxy from the fitting of all emission lines in both spectral configurations, obtaining z$_{G395H}$ = 8.68022 $\pm$ 0.00003 and z$_{PRISM}$ = 8.6864 $\pm$ 0.0002 from the high and low resolution respectively, which results in a redshift offset of $\Delta z$ = z$_{PRISM}$ - z$_{G395H}$ = 0.006. This value is fully consistent with the previously reported offset between PRISM and medium-resolution gratings. A systematic redshift offset of $\Delta z = 0.0042$ has been identified in \citet{DEugenio2025}, increasing the scatter of the trend at lower redshifts due to the strong wavelength dependence of the PRISM’s spectral resolution. Similar values were found also in the IFU data for the same resolution modes \citep[i.e.][0.006 - 0.009 at z =8.5]{Perez-Gonzalez2025,Jones2025b}.

We registered the JWST NIRSpec astrometry to the F200W NIRCam observations available in the STScI MAST archive by comparing the positions of the 3 clumps of  EGSY8p7 determined from the PRISM IFS data at $\sim2~\mu$m  and those estimated from NIRCam image (Figure~\ref{fig:mosifs}). We thus applied a shift of $\Delta$RA = 0.17 arcsec and $\Delta$DEC = 0.054 arcsec to the IFS cube.

\section{Analysis}

\subsection{Detection of broad components in the oxygen lines}
\label{sec:broad}

The nature of the ionizing source in EGSY8p7 has been extensively discussed in the literature. \citet{Larson2023} reported a $2.5\,\sigma$ detection of a broad component in the H$\beta$ line, with a full width at half maximum (FWHM) of $\sim$ 1200 km/s, using the medium-resolution NIRSpec G395M/F290LP spectrum obtained with the MOS mode.
Notably, they did not detect a corresponding broad component in strong forbidden lines, such as [OIII]$\lambda$ 5007 \AA, which led them to associate the broad H$\beta$ feature with the BLR of an AGN. Based on this, they classified EGSY8p7 as a potential progenitor of massive z > 6 quasars hosting SMBH.

Subsequently, \citet{Marques-Chaves2024} analyzed the same dataset with a different data reduction approach. They reported a $2.2\,\sigma$ detection of the broad H$\beta$ component  with FWHM$^{broad}$ = 1156 $\pm$ 370 km/s, three times larger than the narrow component (FWHM$^{narrow}$ $\sim$ 370 km/s), and in line with \citet{Larson2023}. However, they also showed that the H$\beta$ line can be satisfactorily fitted with a single Gaussian profile with a FWHM$^{narrow}$ = 452 $\pm$ 68 km/s without  penalizing the residuals. Additionally, they analyzed the rest-frame UV emission lines commonly attributed to BLRs or high-density environments, such as NIV], CIV, and CIII]. If these emission lines originate from the BLR, they are expected to exhibit similarly broad Doppler and lower observed fluxes due to the dust extinction. However, the authors find that these lines appear narrow (FWHM < 90 km/s), making it unlikely that they are produced in the BLR. 

These discrepancies highlight the necessity of deeper observations of Balmer emission lines to unambiguously confirm the presence and nature of any broad components. Therefore, this issue could also be addressed by increasing the spectral resolution of the data, which improves the sampling of the line profile and allow to disentangle multiple components in each emission line. In this context, the significant increase in total exposure time in our observations (from 3107 s to 36180 s) combined with the highest spectral resolution of NIRSpec, places our dataset in an ideal position to solve this open question. Then, we analyzed the line profiles of H$\beta$ and [OIII] in our IFS data to investigate the presence of broad components that could be associated with gas originating from the BLR or potential outflows. 

\begin{figure}[h]
\centering
\includegraphics[width=\linewidth]{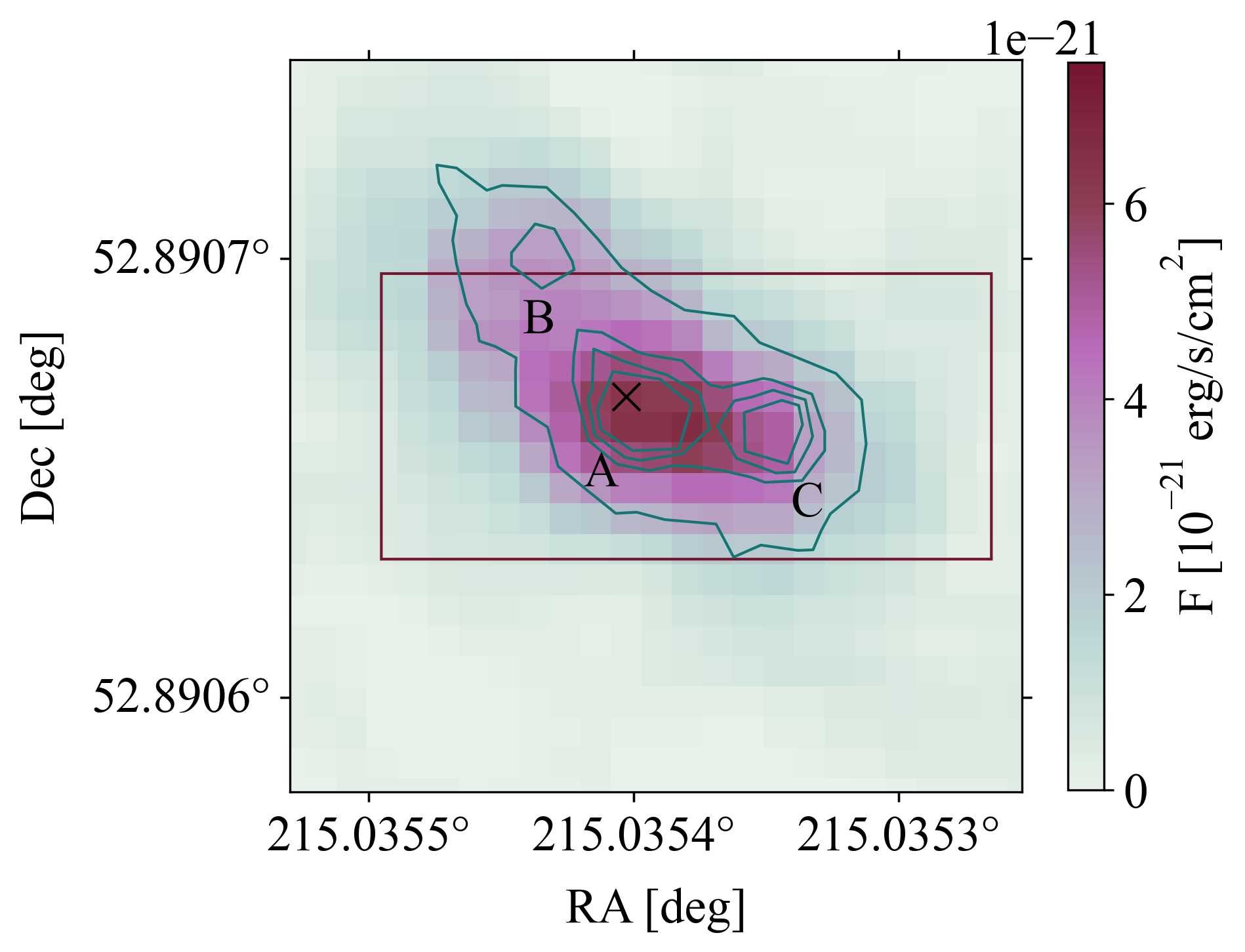}
\includegraphics[width=\linewidth]{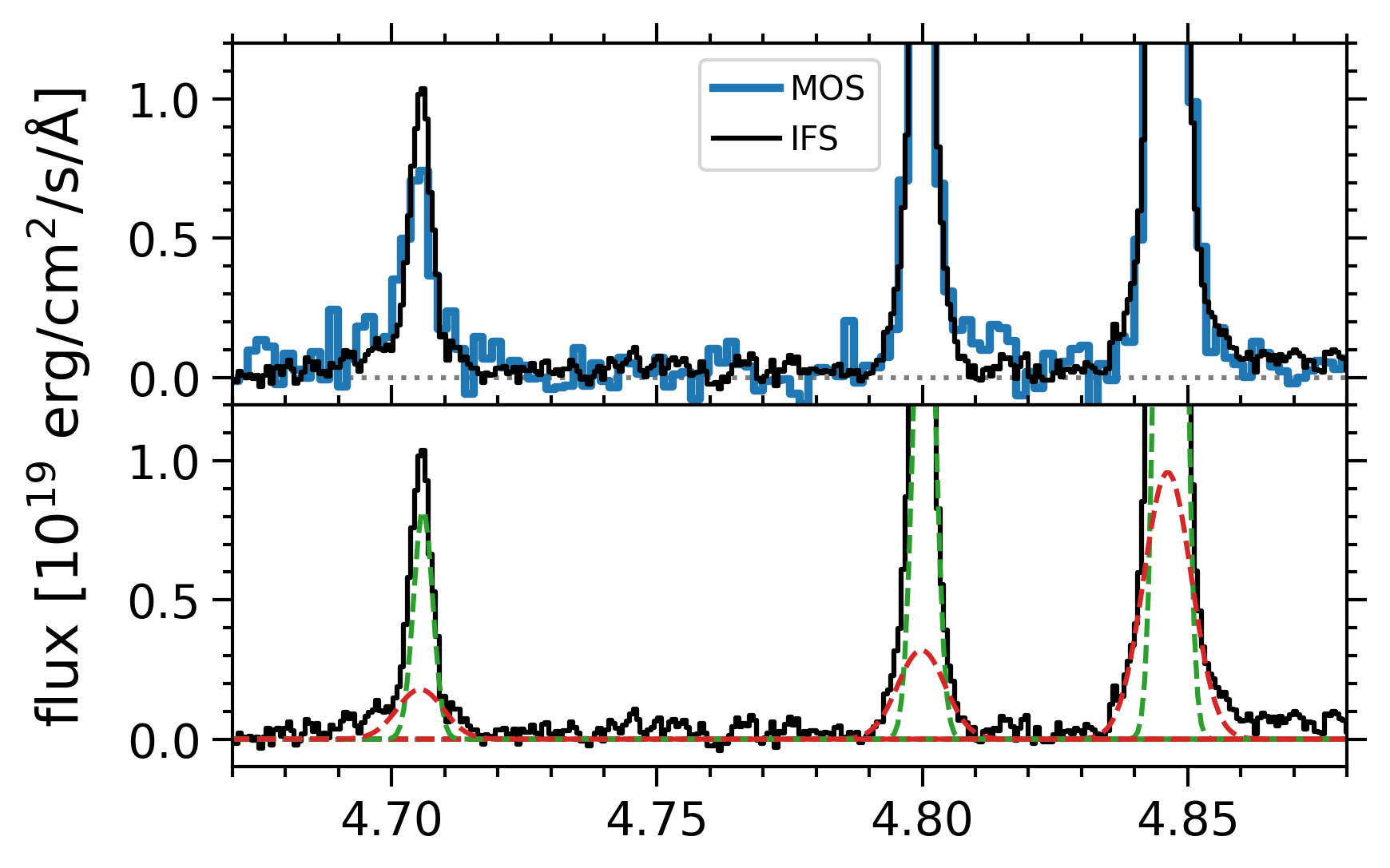}
\caption{Top: JWST NIRSpec images from collapsed continuum emission (2.2 $\mu$m < $\lambda$ < 2.4 $\mu$m) in the low-resolution PRISM data cube. The central NIRSpec/MOS shutter from program ERS 1345 used by \citet{Larson2023} and \citet{Marques-Chaves2024} is overlaid as a red solid rectangle, while the JWST/NIRCam F150W contours are shown with blue solid lines. The central point of the extraction aperture used in the following sections is marked with a black cross. Bottom: NIRSpec/MOS data (blue) and NIRSpec/IFS spectrum (black) extracted from a rectangle region as large as the MOS shutter (see text). A best IFS fit-model with two component (tied in velocity and velocity dispersion) for each line is shown in the bottom panel.}
\label{fig:mosifs}
\end{figure}

The top panel of Figure \ref{fig:mosifs} shows the continuum emission of the galaxy, obtained by collapsing the PRISM data cube in channels free from line emission. 
We extracted the integrated spectrum from our high-resolution data cube using this aperture (see NIRSPec/MOS shutter shown in Figure \ref{fig:mosifs}), to allow for a direct comparison with the medium-resolution spectra analyzed in those studies \citep{Larson2023,Marques-Chaves2024}. The line fluxes measured from the integrated MOS and IFS spectra are fully consistent within the errors, demonstrating the effectiveness of the pipeline's pathloss corrections for this object. The bottom panel of the same figure presents a comparison between the medium (R $\sim$ 1000) and high resolution (R$\sim$2700) spectra of the galaxy, focusing on the H$\beta$ and [OIII] optical emission lines. The broad component of H$\beta$ identified in previous studies is also present in our data. However, the higher resolution observations allow us to also distinguish a clear broad emission feature in the [OIII] forbidden lines.

We modeled the broad emission features using three different approaches: (i) a BLR component in H$\beta$, as suggested by the analysis of \citet{Larson2023}, and an independent broad component in [OIII] associated with outflowing gas; (ii) an outflow component in both H$\beta$ and [OIII] with the same width and velocity shift (see bottom panel of Fig. \ref{fig:mosifs}); and (iii) a combination of the previous two scenarios, incorporating both a BLR and outflow component in H$\beta$ (see the three panels of Fig. \ref{fig:bzoomin}). The fittings confirm the presence of a broad component in [OIII] with a FWHM of approximately 650 km/s in the three independent approaches (see Table \ref{tab:MOS_res}), suggesting the presence of an outflow in this galaxy. However, the best-fit results yield a reduced $\chi^2$ of 1.154–1.159 and a Bayesian Information Criterion (BIC) of 1447–1449, depending on the model. Therefore, it is not statistically possible to determine from the integrated spectrum whether the broad H$\beta$  emission originates from the same outflow, from the BLR of an AGN, or from a combination of both.

\subsection{Spatially integrated spectrum}\label{sec:int}
\label{sec:spectrum}

We extracted the spectra of the galaxy in PRISM and  grating data from a circular aperture with a radius of R = 0.25 arcsec, centered at the H$\alpha$ spatial peak (RA = 215.03525 deg, Dec = 52.89070 deg, see black cross in Figure \ref{fig:mosifs}). The aperture includes the three clumps and
we analyzed them as components of a single galactic system. The uncertainties in the integrated spectra were estimated by propagating in quadrature the errors provided in the error extension of the data cubes, calculated in the same spaxels used for the spectral extraction.

\begin{figure*}
\centering
\includegraphics[width=\textwidth]{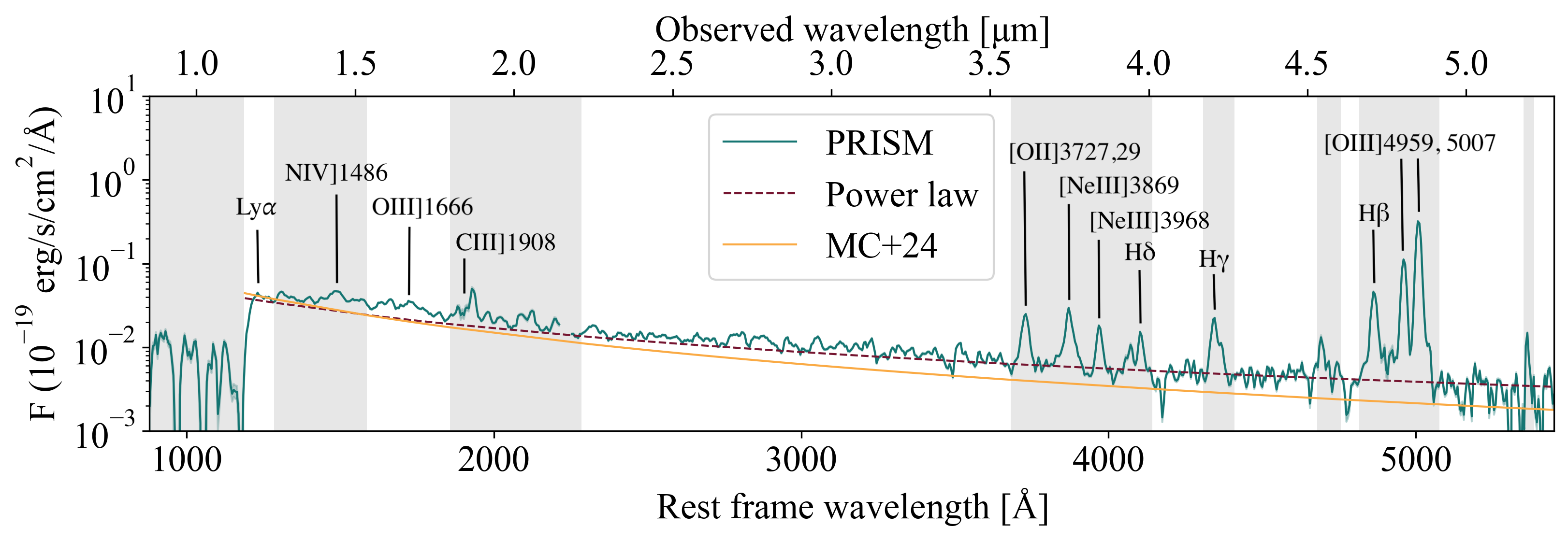}
\includegraphics[width=1\textwidth]{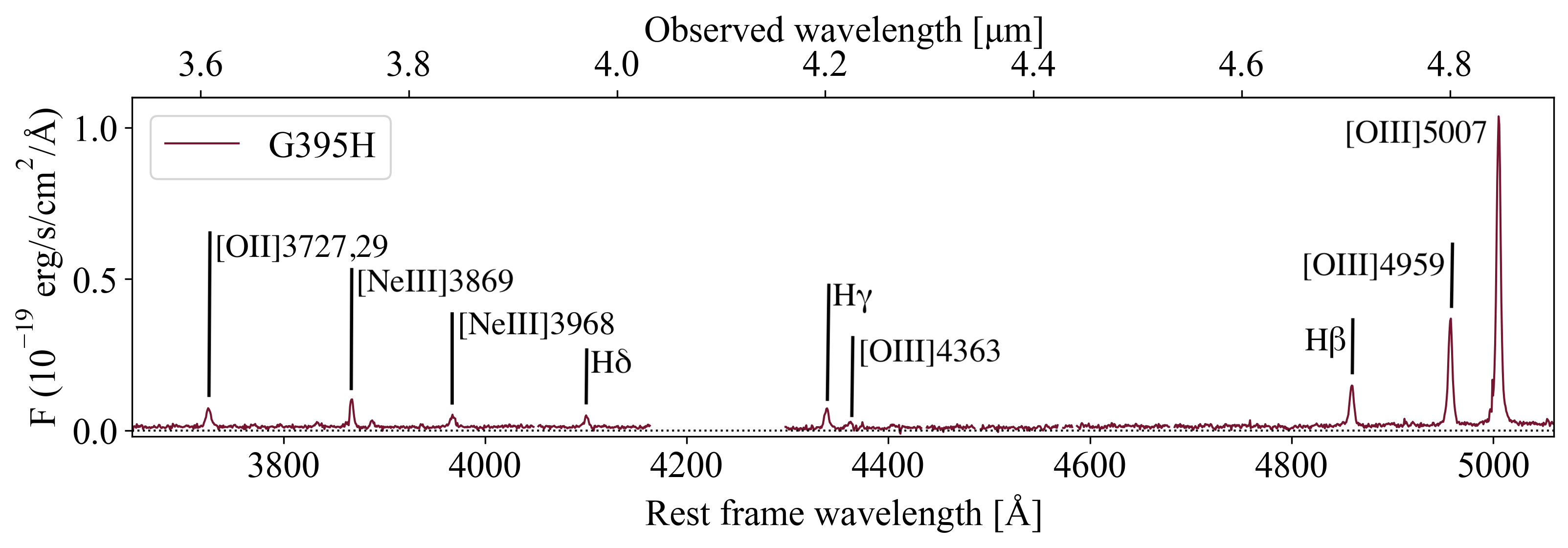}
\caption{JWST/NIRSpec PRISM and G395H/F290LP spectra of the galaxy (top and bottom panels respectively) extracted from an aperture of R= 0.25 arcsec. In the top panel, the gray shading areas are those excluded in the continuum fitting (see text), the red dashed line shows the power-law continuum model fitted to reproduce the continuum emission, and the solid orange line is the one proposed by \citet{Marques-Chaves2024}. The brightest nebular lines are marked. }
\label{fig:integrated spectrum}
\end{figure*}

The upper panel of Fig. \ref{fig:integrated spectrum} shows the low resolution spectrum from the PRISM data. 
The main remarkable feature is the non-detection of the NV]$\lambda$1242 \AA\ emission line, which probe energies.
This line was reported with a level of confidence 4.6 $\sigma$  by \citet{Mainali2018} from the analysis of the MOSFIRE spectrum. The authors report an integrated flux of (2.8 $\pm$ 0.6) $\times$ 10$^{-18}$ erg/s/cm$^{2}$. We note that this emission line was not detected in the medium-resolution MOS spectrum by \citet{Marques-Chaves2024} either (< 1.44 10$^{18}$ erg/cm$^2$). Nonetheless, the conclusion by \citet{Mainali2018} regarding the need for a hard ionizing spectrum to produce the energy required for the ionization of this species is still valid (larger than 77 and 98 eV for NIV and NV, respectively). Indeed we clearly detect the NIV] $\lambda$1486 \AA\ in the ultraviolet spectrum and this line
and this feature, together with the weak NIII] $\lambda$1750 \AA\ emission \citep[0.73 $\pm$ 0.30 $\times$ 10$^{-18}$ erg s$^{-1}$ cm$^{-2}$;][]{Marques-Chaves2024}, indicates that a significant fraction of nitrogen is in high ionization states. This line can originate in high-density star-forming regions or in the vicinity of an AGN, both supporting the presence of the extreme radiation fields observed in the Ly$\alpha$ emitting population \citep{Villar-Martin2004,Vanzella2010,Fosbury2003}.

A blue rest-frame UV continuum slope is clearly identified in this spectrum. We modeled the continuum by fitting a power-law function (f$_\lambda$ $\sim$ $\lambda^\beta$) to the PRISM spectrum, excluding emission lines and prominent spikes (highlighted in grey in the figure). The resulting fit, shown as a dashed red line, yields a UV slope of $\beta_{\rm UV} = -1.81 \pm 0.06$. For comparison, we also show the value reported by \citet[][]{Marques-Chaves2024} derived by fitting the continuum between 1300 - 2200 \AA\ in the rest-frame ($\beta_{\rm UV} = -2.11 \pm 0.09$, orange solid line). The discrepancy between the two results might be associated with the slit losses in the MOS observations, which would explain the fainter continuum emission at longer wavelengths.

The bottom panel of Fig. \ref{fig:integrated spectrum} shows the high resolution spectrum from the G395H/F290LP configuration. This  reveals the presence of the [OIII]$\lambda $4363 \AA\, [OIII]$\lambda \lambda$4959,5007 \AA\, [NeIII]$\lambda$3869 \AA\, and [NeIII]$\lambda$3968 \AA\ lines, emitted by the O$^{++}$ and Ne$^{++}$ ions, which are associated with  high-ionization states (35 eV and 41 eV respectively). However, the absence of emission lines such as [NIII]$\lambda$4640 \AA, [CIII]$\lambda$4650 \AA, or He II $\lambda$4686 \AA\ indicates that the hard ionizing spectrum is unlikely to originate from the presence of Wolf-Rayet stars \citep{Brinchmann2008}.

\begin{table}
\caption{Emission line fluxes and velocity dispersions from the high spectral resolution integrated spectrum of the galaxy (G395H grating).}\label{tab:fluxes}
\centering
\begin{tabular}{lcc }
\hline
Lines                           & F$_{g395h}^{\rm ~\dagger}$ & $\sigma$\\
                           & $[{\rm 10^{-19}~erg/s/cm^{2}}]$ & [${\rm km/s}$]\\

\hline
{[}OIII{]}$\lambda$4959         &  142 $\pm$ 4 & 112 $\pm$ 2  \\
{[}OIII{]}$\lambda$5007         &  432 $\pm$ 11 \\
H$\beta$                        &  57 $\pm$ 4  \\
{[}OII{]}$\lambda $3727  &  126 $\pm$ 19  \\
{[}OII{]}$\lambda $3729  &  46 $\pm$ 6  \\
{[}NeIII{]}$\lambda $3869       &  37 $\pm$ 1   \\
HeI$\lambda $3889               &  8 $\pm$ 1  \\
H$\delta$                       &  15 $\pm$ 1   \\
H$\gamma$                       &  28 $\pm$ 1  \\
{[}OIII{]}$\lambda $4363        &  7 $\pm$ 1  \\
\hline 
{[}OIII{]}$\lambda$4959$^{~\rm outflow}$     &  46 $\pm$ 6  &    318 $\pm$ 9  \\
{[}OIII{]}$\lambda$5007$^{~\rm outflow}$     &  126 $\pm$ 19  \\
H$\beta^{~\rm outflow}$                 &  15 $\pm$ 6       \\
\hline
\end{tabular}
\\Note. $\dagger~{\rm 10^{-19}~erg/s/cm^{2}}$; $\ddagger~{\rm km/s} $
\end{table}

We modeled the spectra by fitting the continuum with a second-order polynomial, combined with Gaussian profiles for the emission features. As mentioned in Section \ref{sec:broad}, the [OIII]$\lambda\lambda$4959, 5007 and H$\beta$ emission lines exhibit two kinematic components in the high-resolution configuration. In these two lines, the profiles were modeled using a combination of two Gaussian functions: (i) a narrow component with a velocity dispersion $\sigma$ < 150 km/s, and (ii) a broad component with 200 < $\sigma$ < 350 km/s. The remaining emission lines in the spectrum are well described by a single narrow Gaussian component, adopting the same velocity shift and width as those of the narrow components of [OIII] and H$\beta$.
The emission-line flux uncertainties were estimated using Monte Carlo realizations by adding Gaussian noise consistent with the local root mean square of each line and adopting the mean and standard deviation of the resulting distributions as the flux value and its uncertainty, respectively. The flux measurements are summarized in Table \ref{tab:fluxes}.

\subsection{Spatially resolved analysis}
\label{sec:kinematic}

The top panel of Figure \ref{fig:mosifs} shows the continuum emission of the galaxy, with the NIRCam/F150W contours superimposed with blue solid lines. The continuum reveals three distinct clumps whose properties have been studied in previous works. According to \citep[see][]{Marques-Chaves2024}, clumps A and B are star-forming, with stellar population ages of 4 and 5.7 Myr, stellar masses of log(M$_*$/M$_\odot$) = 8.76 and 8.66, and star formation rates (SFRs) of 148 and 83 M$_\odot$/yr, respectively. Clump C is more massive, with log(M$_*$/M$_\odot$) = 8.94, and hosts a non-ionising stellar population with an age of 15 Myr. 

The different properties obtained from the photometric analysis for the three clumps encourage us to construct two-dimensional (2D) maps of the various emission lines, enabling us to investigate the spatial distribution of the emission and the location of the outflow within the galaxy. We used the PRISM data cube, as this dataset provides a higher signal-to-noise ratio and includes the UV part of the spectrum. We subtracted the stellar continuum emission by creating a 2D map of the continuum at wavelengths close to the selected emission line. In particular we selected two sidebands around each line. The continuum maps were then removed from the flux maps of the emission lines. The selected wavelengths range adopted to generate the 2D maps of line and and continuum emission are reported in Table~\ref{tab:line_maps}. The final flux maps are shown in Figure~\ref{fig:line_maps}.

\begin{figure}[h]
\centering
\includegraphics[width=0.8\columnwidth]{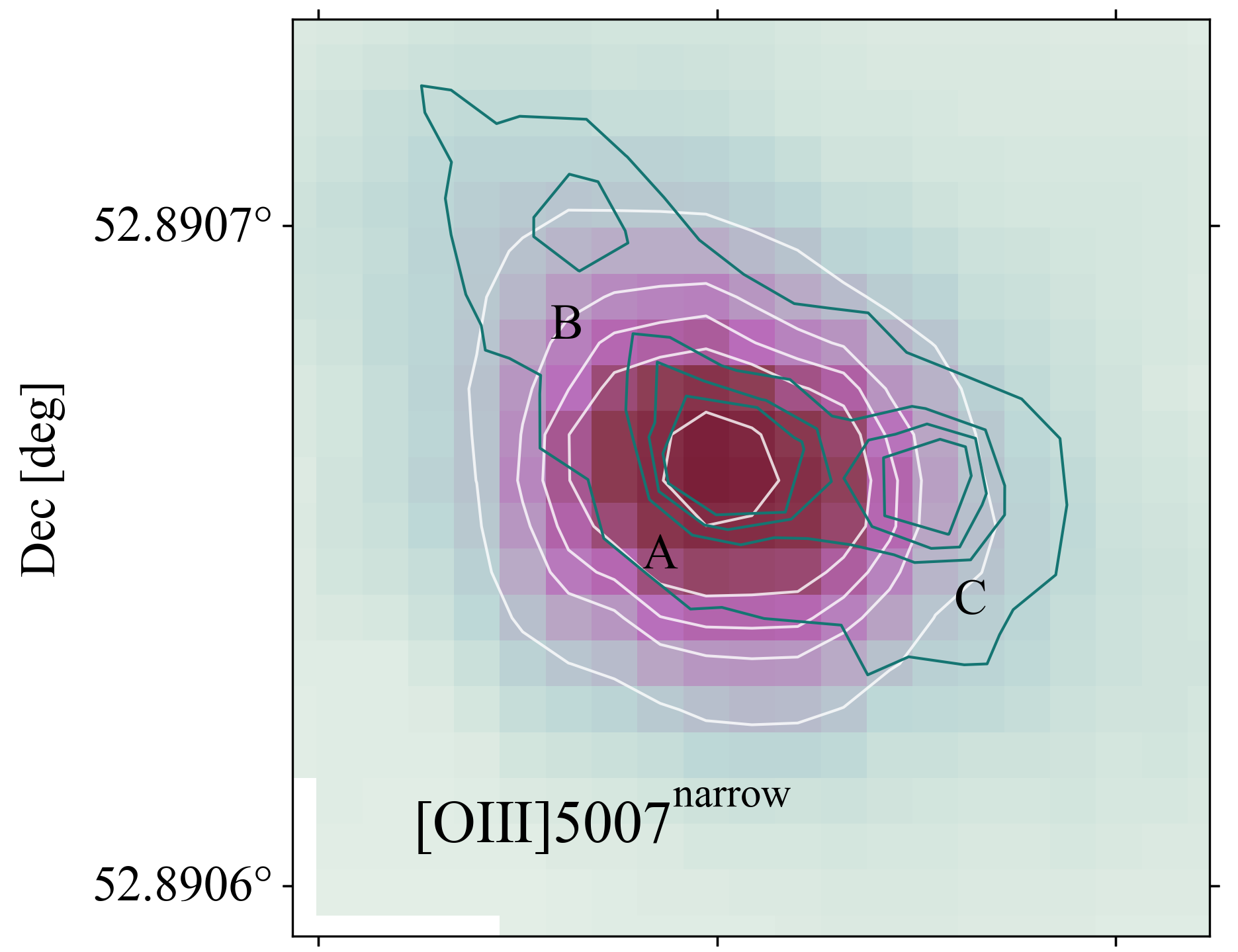}
\includegraphics[width=0.8\columnwidth]{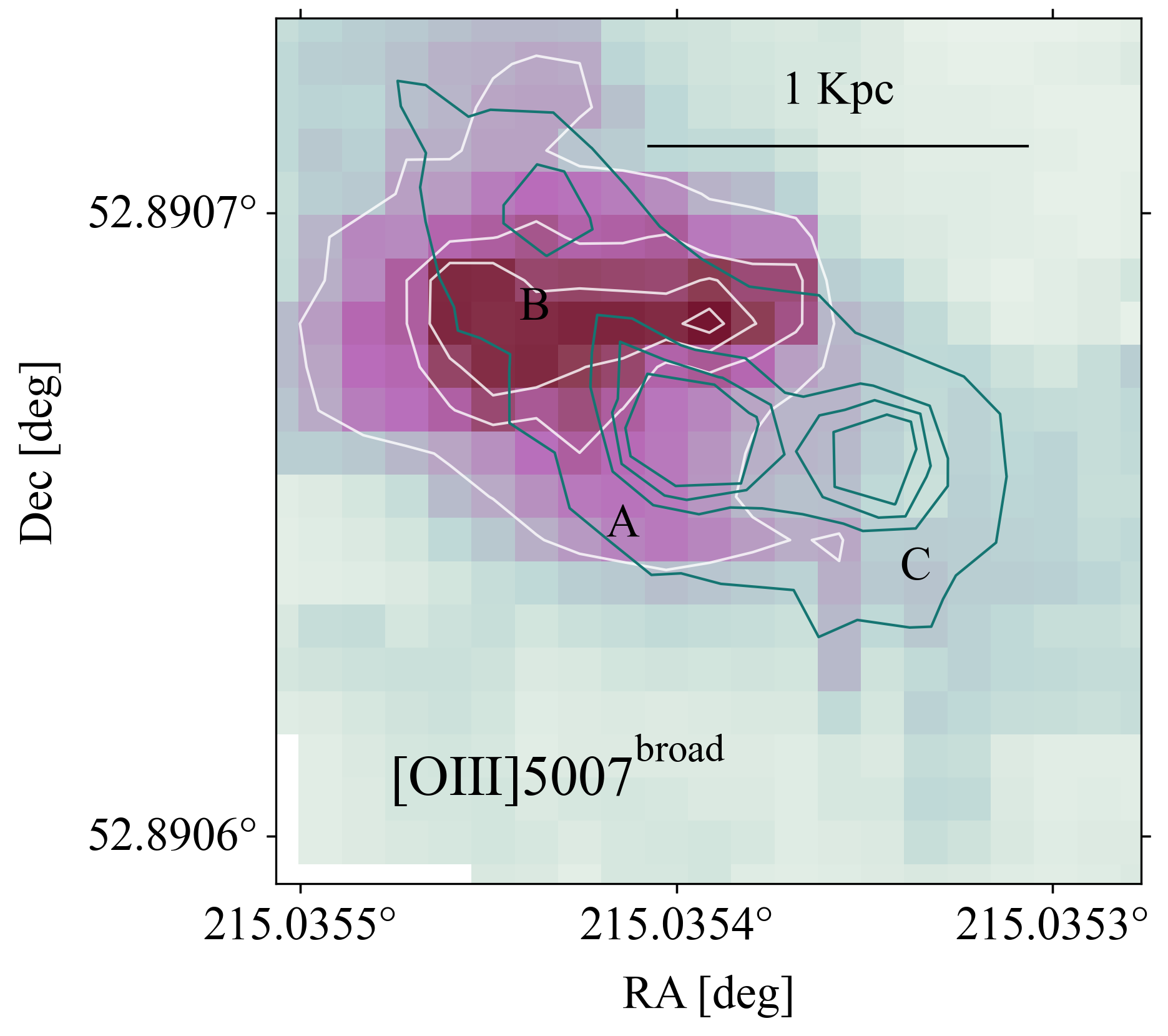}
\caption{Spatially resolved fitting of the [OIII]$\lambda$ 5007 \AA\ emission line in the high-resolution data cube (G395H). The narrow and broad emission line flux maps are shown in the top and bottom panels, respectively. The JWST/NIRCam F150W and the emmision map shown contours are shown with blue and white solid lines respectively. }
\label{fig:outflow}
\end{figure}

In order to reliably identify the origin of the broad component observed in the [OIII] emission line, we performed a
line fitting on a spaxel-by-spaxel basis, using the two Gaussian described in Section \ref{sec:int} in the high resolution data cube (G395H), and assuming two components when BIC > 4. Figure \ref{fig:outflow} presents the flux maps of the narrow and broad emission components, shown in the top and bottom panels respectively. We do not detect any significant velocity structure in the broad component [OIII] line.

The main narrow emission features observed in the galaxy originate from a compact region consistent with the primary star-forming clump A. However, the broad oxygen emission line [OIII]$\lambda$5007 \AA\ is spatially shifted toward clump B, approximately halfway between clumps A and B, with its centroid located at a projected distance of 0.5712 kpc from clump A. The projected extent of this outflow component is asymmetric, reaching a maximum extension of approximately 1 kpc in the direction perpendicular to the clumps.

Unfortunately, [OIII] is the only emission line for which we can perform a pixel by pixel analysis with a sufficiently high signal-to-noise ratio. However, to further investigate the origin of the broad component observed in H$\beta$, we constructed channel maps of this emission line (see Appendix \ref{ap:spectrophot}). Panels c and d of Figure \ref{fig:channel} show the channels in which the narrow component emission dominates the H$\beta$ profile, with the emission centered on the ionizing clump A. On the other hand, panels b, and e show the channels in which the broad component emission is comparable or stronger than the narrow component. In these last maps, a secondary emission clearly arises from a region that is spatially offset and consistent with the location of the broad [OIII] emission. This result supports the interpretation that both broad components originate from the same gas.

\section{Interstellar medium properties}
\label{sec:ism}

\subsection{Dust extinction}
We used the ratios of the Balmer emission lines detected in the integrated high spectral resolution spectrum, H$\beta$, H$\gamma$, and H$\delta$, to estimate the nebular dust reddening of the galaxy, assuming the extinction law of \citet{miller1972} with a specific attenuation of R$_v$ = 3.2. We adopted Case B recombination and we computed the intrinsic line ratios using PyNeb \citep{pyneb} for T = 10$^4$ K and $n_{\rm e}$ = 100 cm$^{-3}$ \citep{Osterbrock2006}, based on the atomic data from \citet{storey1995}.

We applied a linear regression using all the Balmer lines available in the spectrum to estimate the extinction coefficient and then reduce the uncertainties in its determination \citep[see][]{diaz2007,zamora2022}.
In the regression, the difference between the observed and theoretical logarithmic line ratios is the dependent variable, while the normalized logarithmic extinction at each wavelength (f (H$\beta$) = 0) acts as the independent variable. The slope of the fit corresponds to the logarithmic extinction coefficient, c(H$\beta$), which has a value of 0.06 $\pm$ 0.1, corresponding to a visual attenuation of A$_v$ = 0.1 $\pm$ 0.4 mag and consistent with no dust attenuation. Previous measurements in the literature report similarly low extinction values, also compatible with our results \citep[see][]{Marques-Chaves2024}. Therefore, assuming no dust attenuation is an appropriate approximation for this work, and we do not propagate the extinction uncertainties into the subsequent measurements.
  
The \citet{cardelli1989} curve is commonly used as an approximation for nebular attenuation in high-redshift star-forming galaxies \citep[see][]{reddy2020}. However, the actual attenuation law likely varies across different galaxies \citep{sanders2024}. Moreover, the assumption of Case B recombination may not be valid for hot ionizing sources in extremely low-metallicity nebulae \citep[i.e.][]{Scarlata2024,McClymont2025}. Nevertheless, the resulting variation in attenuation value is small enough that it does not significantly affect our results. Indeed, \citet{Marques-Chaves2024} obtained a dust attenuation of A$_v$ = 0.12 $\pm$ 0.11 mag using a far-UV dust attenuation curve suitable for high redshift \citep[][$\lambda$ = 950 -- 1500 \AA, z $\sim$ 3]{Reddy2016}, which is compatible with our result within the errors.

\subsection{UV continuum slope and photon escape fraction}

The dust extinction is known to correlate well with the UV slope, as demonstrated by several previous studies across a wide range of redshifts. This correlation has been observed at $z \sim 0$ \citep{Meurer1995}, $z \sim 1$ \citep{Laird2005}, $z \sim 2$ \citep{Daddi2004}, $z = 3$ \citep{Meurer1999}, and $z = 4$ - 6 \citep{Lehnert2003}. As a result, the UV slope can be reliably used to estimate dust attenuation even at higher redshifts. Recent JWST observations have revealed an increasingly steep UV-continuum slope, reaching $\beta \simeq -2.6$ at $z \simeq 12$, which suggests a transition in the galaxy population toward systems with minimal dust content \citep[e.g.,][]{Topping2023}. 

The UV slope derived from the integrated spectrum of our galaxy is $\beta_{\rm UV} = -1.81 \pm 0.06$ (see Sec. \ref{sec:int}). Assuming that the underlying stellar population has a well-defined intrinsic UV slope for young populations (< 100 Myr) with a continuous star-forming and solar metallicities, we used the relation proposed by \citet{Meurer1999} and adopted the same extinction law as for the gas \citep[][with R$_V$ = 3.2]{miller1972} to calculate the visual extinction. This yields a value of A$_v$ = 0.033 $\pm$ 0.005 mag which, similarly to the gas extinction, is effectively negligible.

The UV slope derived in this work is consistent with the value obtained by \citet{Larson2023}, of $-1.77^{+0.11}_{-0.12}$ and $-1.80^{+0.09}_{-0.10}$ for a continuous and busty SFH respectively. This value is also consistent with typical measurements reported for galaxies at similar redshifts \citep[see][]{Cullen2024}. However, our slope is slightly flatter than the one derived by \citet[][$\beta_{\rm UV} = -2.11 \pm 0.09$]{Marques-Chaves2024}, and also  than the value inferred from the F150W and F200W photometry ($\beta_{\rm UV}^{\rm phot} = -2.11 \pm 0.11$). We tried to perform the same fitting over the wavelength range 1350 - 2600 \AA\ to avoid the effect of damped Ly$\alpha$ absorption \citep[see][]{Heintz2025}, but the obtained UV-slope is even flatter.

In order to investigate the origin of this discrepancy, we extracted the integrated spectra of the three individual clumps, centered at the coordinates reported by \citet{Marques-Chaves2024} and using a circular aperture with a radius of 0.025 arcsec. We fitted the continuum following the same procedure as for the integrated spectrum (see Section \ref{sec:int}). The resulting power-law UV slopes are $\beta$ = --1.94, --1.54, and --2.03 for clumps A, B, and C, respectively. This would explain the discrepancy because the aperture used by \citet{Marques-Chaves2024} does not include clump B, which has the flattest UV slope. The redder color of clump B may also indicate the presence of moderate dust attenuation in that region of the system. Unfortunately, we cannot provide further information on this issue because the H$\gamma$ emission map from PRISM does not have sufficient signal-to-noise to perform a pixel-by-pixel reddening analysis, and the H$\beta$ emission in the G395H/F290LP data cube is contaminated by outflow emission. 

The UV continuum slope is also correlated with some high-redshift galaxy properties relevant to their contribution to cosmic reionization. 
In particular, blue UV slopes are indicative of elevated ionizing photon production efficiencies \citep{Cullen2024, Topping2024} and are frequently associated with significant ionizing photon escape fractions (f$_{esc}$) \citep{Begley2022, Chisholm2022, Dottorini2025}. We obtained a value of f$_{esc}$ = 0.021 $\pm$ 0.014, which increases to $\sim$ 0.03 when considering only clump A. Such low escape fraction remains insufficient for the galaxy to be considered a primary contributor to cosmic IGM reionisation. Galaxies that play a dominant role on it are required to have significantly higher escape fractions of ionizing photons, typically in the range of $\sim$10\%–20\% \citep[see for example][]{Robertson2013}.

\subsection{Ionic and total oxygen abundances}\label{sec:ion_abundances}

The high spectral resolution of our new dataset enables the detection of the [OIII] auroral line in the integrated spectrum with a S/N ratio of 4.6, higher than that obtained in previous studies of this galaxy \citep[i.e.][]{Marques-Chaves2024}. This yields  R$_\mathrm{O3}$ = I ($\lambda$4959 \AA+ $\lambda$5007 \AA)/ I ($\lambda$4363 \AA) of  $78.10 \pm 0.17$. We then used  R$_\mathrm{O3}$ to determine the electron temperature for  O$^{+}$ and O$^{++}$ gas. In particular, we inferred an an electron temperature of T$_e$([OIII]) = (1.4179  $\pm$ 0.0014) $\times$ 10$^4$ K and  T$_e$([OII]) = (1.2636 $\pm$ 0.0010) $\times$ 10$^4$ K by adopting the empirical calibration by \citet{hagele2008} and \citet{Izotov2006}, respectively. We note that various calibrations have been reported in the literature for deriving T$_e$([OIII]), but the electron temperatures resulting from them are consistent within the uncertainties (see Table~\ref{tab:temperatures}). The electron temperature of the intermediate-ionization zone, T$_e$([SIII]), can also be estimated from T$_e$([OIII]) using the relation provided by \citet{garnett1992}, yielding a value of (1.3469 $\pm$ 0.0011) $\times$ 10$^4$ \citep[using ][ we obtained (1.56 $\pm$  0.24) $\times$ 10$^4$ K, totally compatible within the errors]{croxall2016}.

The ionic abundances of O$^+$ and O$^{++}$ were calculated using T$_e$([OII]) and T$_e$([OIII]), respectively, based on the expressions provided by \citet{diaz2022}, which were derived under the assumption of a uniform density structure with $n_{\rm e}$ = 100 cm$^{-3}$. The resulting ionic abundances, expressed in the form 12 + log(X$^i$/H$^+$), are 6.901 $\pm$ 0.075 for O$^+$ and 7.983 $\pm$ 0.071 for O$^{++}$ providing a total oxygen abundance of 12 + log(O/H) = 8.017 $\pm$ 0.066. In high-excitation objects, such as those exhibiting He II emission, a small contribution from O$^{3+}$ may also be present. However, we did not detect the He II $\lambda4686$ \AA\ emission line, so we considered the contribution from O$^{3+}$ negligible.

The uncertainties in the derived abundances are mainly associated with the electron temperature determinations, as the observational errors in the strong recombination and collisionally excited line fluxes are relatively small. The relation between T$_e$([OII]) and T$_e$([OIII]) is not straightforward and depends strongly on the electron density \citep[see][]{hagele2006}. This fact introduces significant uncertainties in the determination of O$^+$/H$^+$. However, since the majority of the oxygen in the studied galaxy is in the form of O$^{++}$ ($\sim$ 90\%), the calibration used to estimate the temperature of the low-ionization zone does not significantly affect the total O/H abundance determination. Despite the uncertainties, our oxygen abundance measurements is in agreement with those already reported in previous studies and calculated with auroral line or different calibration \citep{Nakajima2023,Isobeb2023,Larson2023,Tang2023,Marques-Chaves2024}.

\subsection{Ionization parameter}\label{sec:ionization_parameter}
The gas ionization degree is represented by the ionization parameter, defined as the ratio of the ionizing photon flux and the gas atom density, q = Q(H$_0$)/(4$\pi$R$^2$$n_{\rm e}$) where Q(H$_0$) is the rate of hydrogen-ionizing photons, R is the characteristic radius of the ionized region, and $n_{\rm e}$ is the electron density. Physically, it corresponds to the velocity at which the ionization front propagates through the medium under the influence of the radiation field.  It is often expressed in dimensionless units as  $u = q/c$, where $c$ is the speed of light.

The ionization parameter is commonly estimated using two lines of the same element corresponding to two contiguous ionization states, i. e. [OIII]$\lambda$5007 \AA /[OII]$\lambda$3727\AA\ as originally proposed by \citet{alloin1978}. In our galaxy, this ratio yields a value of 1.300 $\pm$ 0.033. Then, using the empirical calibration of \citet{diaz2000} we obtained an ionization parameter log(u) = --1.980 $\pm$ 0.066.

However, the large wavelength separation between the [OIII] and [OII] emission lines in the red part of the rest-frame optical spectrum makes the [OIII]/[OII] ratio susceptible to dust extinction effects. Additionally, this ratio depends on the gas metallicity, which introduces further degeneracies when interpreting ionization conditions. To mitigate these issues, alternative diagnostics for the ionization parameter have been proposed. One such diagnostic is the [NeIII]$\lambda$3869 \AA / [OII]$\lambda$3727 \AA\ ratio. Due to the close proximity in wavelength of these lines, this ratio is significantly less affected by reddening and can be observed at higher redshifts. Moreover, since neon is produced in massive stars, its abundance is expected to closely trace that of oxygen, making it a reliable tracer in chemical and ionization studies. We obtained a value of log([NeIII]$\lambda$3869 \AA /[OII] ]$\lambda$3727 \AA) = 0.234 $\pm$ 0.034, which, using the calibration proposed by \citet{levesque2014}, corresponds to an ionization parameter of log(u) = --2.216 $\pm$ 0.035, lower than the value derived from the [OIII]/[OII] ratio.

\subsection{Electron density}\label{sec:density}
Electron density can be estimated by measuring emission-line doublets of the same ionic species whose lines have different critical densities that trace electron densities in their respective ionization zones. Examples include [SII]$\lambda\lambda$6717,6731 \AA\ and [OII]$\lambda\lambda$3726,3729 \AA\ for the low-ionization zone, and [ArIV]$\lambda\lambda$4713,4741 \AA\ for the high-ionization zone. We used the [OII] doublet as our electron density diagnostic because the [SII] lines fall outside our wavelength coverage, and the [ArIV] lines are unresolved at the resolution of our PRISM data. 

We calculated the electron density using PyNeb \citep{pyneb}, assuming that the electron temperature derived from [SIII] corresponds to the average electron temperature across the galaxy. For the calculations, we adopted atomic data from \citet{zeippen1982} and \citet{wiese1996}, and collision strengths from \citet{pradhan2006}. Uncertainties were estimated by propagating the errors in both the electron temperature and the emission line ratio, with the latter being the dominant one. 
From the oxygen lines, we inferred an intensity ratio I($\lambda$3729\AA)/I($\lambda$3727\AA) of 0.54 $\pm$ 0.09 which corresponds to an electron density of  log($n_{\rm e}$/cm$^{-3}$) = 3.56$^{+0.31}_{-0.21}$. The intensity ratio calculated in this work is consistent with the value reported by \citet[][0.639 ± 0.255]{Larson2023} and slightly different from that reported by \citet[][0.98 $\pm$ 0.23]{Marques-Chaves2024}. This discrepancy depends on the assumptions about the line fitting, as the [OII] doublet is spectrally blended due to kinematic broadening in the medium spectral resolution data used in previous works. Therefore, the ratio between the two emission lines suffers from a large degeneracy. 
It is important to emphasize that our dataset, with higher spectral resolution and better signal-to-noise ratio than previous observations, yields more reliable density measurements (see Fig. \ref{fig:lines-high}).

It should be noted that we assumed a uniform density structure with $n_{\rm e}$ = 100 cm$^{-3}$ to calculate the ionic abundances. Although we found a relatively high electron density in this galaxy, this assumption is not expected to significantly affect our results. The O$^{++}$ abundance is largely insensitive to density, varying by only 0.004 dex over the range 100 cm$^{-3}$ < $n_{\rm e}$ < 1000 cm$^{-3}$ \citep{stanton2025}, a factor more than an order of magnitude smaller than the uncertainties in the abundances. The impact on the O$^+$ abundance is more significant. However, this is not a major concern since O$^{++}$ dominates the total oxygen abundance in this galaxy, as discussed in Section~\ref{sec:ion_abundances}.

It is also important to highlight that the derived electron density, while relatively high, remains below the critical densities of all the key emission lines used for the O$^{2+}$ electron temperature determination. The [OIII]$\lambda \lambda$4959,5007 \AA\ lines have critical densities of n$_{crit}$ $\sim$ 6.8 $\times$ 10$^5$ cm$^{-3}$, similar to that of the [OIII]$\lambda $4363 \AA\ auroral line. In contrast, the [OII]$\lambda \lambda$3727,29 \AA\ doublet has lower critical densities of approximately 1.5 $\times$ 10$^4$ cm$^{-3}$ and 3.4 $\times$ 10$^3$ cm$^{-3}$, respectively \citep{Osterbrock2006,zeippen1982}. The latter value is close to the electron density derived for our galaxy, indicating that [OII] emission may be partially affected by collisional de-excitation, and n$_e$ could be overestimated. This line has been used to estimate the electron density, the ionization parameter, and the O$^+$ ionic abundance. However, for the latter it is less critical, as O$^{++}$ dominates the total oxygen abundance, as discussed previously.

\section{Gas excitation conditions}
\label{sec:diagram}
\subsection{Photoionizatio and radiative shock modeling}\label{sec:models}

To investigate the physical nature of the ionized gas in our source, we compared our measurements with a set of photoionization and shock models.

The photoionization models are obtained by using input radiation fields of star-forming galaxy spectral energy distribution templates with two different initial mass functions (IMFs), as well as models in which the ionizing source is an active galactic nucleus (AGN). They were computed using the CLOUDY \citep{cloudy} code, assuming an ionization-bounded nebulae and a plane-parallel geometry. We adopted ionization parameters of $\log(u) = -2.0$ and $-2.5$, consistent with the values derived from the [OIII]/[OII] and [NeIII]/[OII] emission line ratios (see Section \ref{sec:ionization_parameter}). The gas metallicity was fixed to the measured value, 12 + log(O/H)= 8.0 (see Sec. \ref{sec:ion_abundances}), with the rest of the elements in solar proportions \citep{asplund2009} and a constant value of the electron density, log($n_{\rm e}$) = 3.56 (see Sec. \ref{sec:density}). We applied a depletion factor of 0.1 to the refractory elements to account for their incorporation into dust grains.

For the star-forming galaxy models (hereafter referred to as \textit{SF}), the nebula is ionized by a young star cluster synthesised using PopStar \citep{popstar1} with the \citet{salpeter1955} IMF, with a lower and upper mass limit of 0.15 and 100 M$_\odot$, respectively. We also included a top-heavy IMF (hereafter referred to as \textit{Top-heavy IMF}), as described in \citet{kauffmann2024}, characterized by a slope of –0.8 for stellar masses above 10 M$_\odot$, and implemented using the PopStar models \citep{millan2021}. We selected ages from 2 to 6 Myr in steps of 1 Myr to represent the simulated clusters, and we included the nebular continuum in a self-consistent way. 

For AGN models we used (i) power laws with index between --1.5 and --2, (ii)  a ``typical'' radio-quiet AGN \citep[][with the continuum having a sub-millimeter break at 10 microns]{mathews1987}, and (iii) a low accretion AGN with an Eddington ratio $\lambda_{\rm edd}$ = L$_{\rm bol}$/L$_{\rm edd}$ = 0.07 \citep{jin2012}.

The scenario of an obscured QSO (hereafter referred to as \textit{Absorbed AGN}) has also been explored. In this case, the UV radiation from a powerful QSO is assumed to emerge from a dense, dusty medium where the QSO is embedded. Strong bipolar outflows would disrupt the surrounding dust. However, if the ionization cone is not aligned with respect to our line of sight, the scattered continuum from the inner regions is obscured. As a result, the emergent radiation is filtered by the gas, modifying the intrinsic spectral energy distribution. Specifically, photons in the 54 – 80 eV energy range are preferentially absorbed, altering the power law spectrum that escapes from the nucleus. For our analysis, we adopted the filtered SEDs proposed by \citet{binette2003}, and we employed the same Cloudy photoionization setup as in the unfiltered power-law models.

We also used the MAPPINGS code \citep{sutherland1993,mappings2008} to simulate radiative shocks. This code self-consistently computes the emission from both the shock and its associated photoionized precursor. The main parameters which affect the shock properties are: the shock velocity, which we varied between 100 and 1000 km s$^{-1}$, the transverse component of the pre-shock magnetic field, B = 0.1 and 10 $\mu$G, and the metallicity of the gas. The gas abundance controls the cooling efficiency, influencing the spatial and temporal evolution of the system. If the observed emission lines are significantly affected by shocks, the metallicity estimate derived in Section \ref{sec:ion_abundances} may be unreliable. Therefore, we adopted solar metallicity for simplicity. The pre-shock density, which cannot be directly constrained from our data, was assumed to be 100 cm$^{-3}$, consistent with the low-density regime typical of ionized nebulae.

\subsection{Ionization nature: AGN vs star formation}\label{sec:bpts}
We used different emission line diagnostic diagrams to investigate the physical conditions of the gas and to infer the ionization mechanism of our source. At the redshift of our target, H$\alpha$ and its nearby nebular lines, such as [SII]$\lambda \lambda$6717,31 \AA\ or [NII]$\lambda \lambda$6548,84 \AA , are redshifted beyond the spectral coverage of our observations. Then, the classical BPT and other diagnostic diagrams \citep{bpt,Veilleux1987} cannot be applied. To overcome this limitation, alternative diagnostics using rest-frame blue optical emission lines have been proposed in the literature, commonly referred as “blue” BPT diagrams \citep[i.e.][]{Backhaus2022,Mazzolari2024}. They typically use combinations of [NeIII]$\lambda$3869, [OII]$\lambda \lambda$3727,29 \AA, [OIII]$\lambda$4363 \AA , [OIII]$\lambda$5007 \AA\ and H$\beta$, providing a robust method for distinguishing between star-forming galaxies and active galactic nuclei (AGN) at high redshifts, where the traditional line ratios are inaccessible. 

\begin{figure*}[h]
\centering
\includegraphics[width=0.9\textwidth]{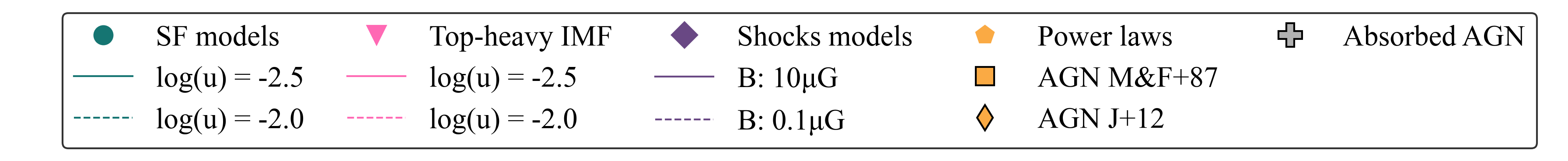}
\includegraphics[width=0.8\columnwidth]{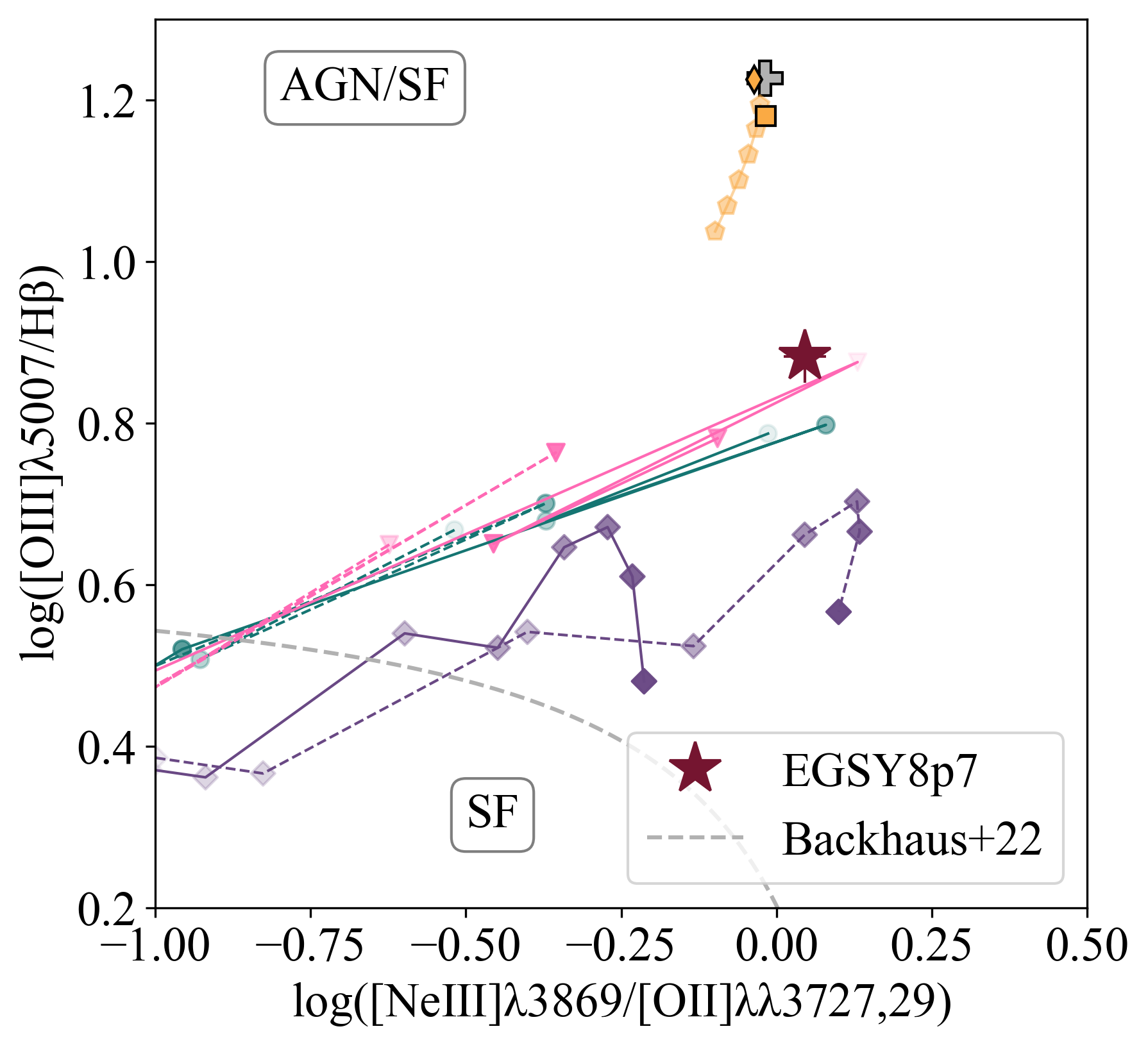}
\includegraphics[width=0.8\columnwidth]{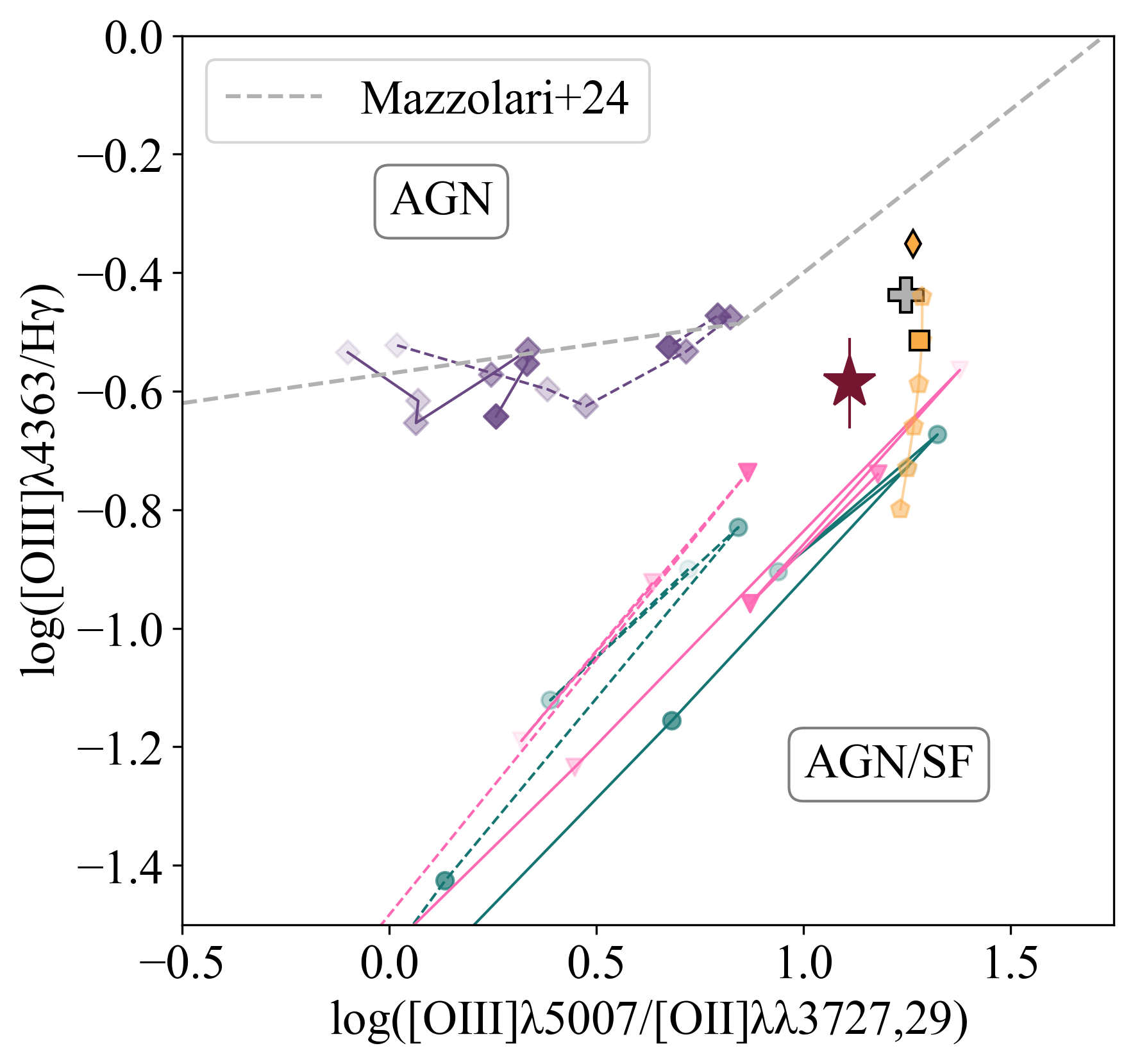}
\includegraphics[width=0.8\columnwidth]{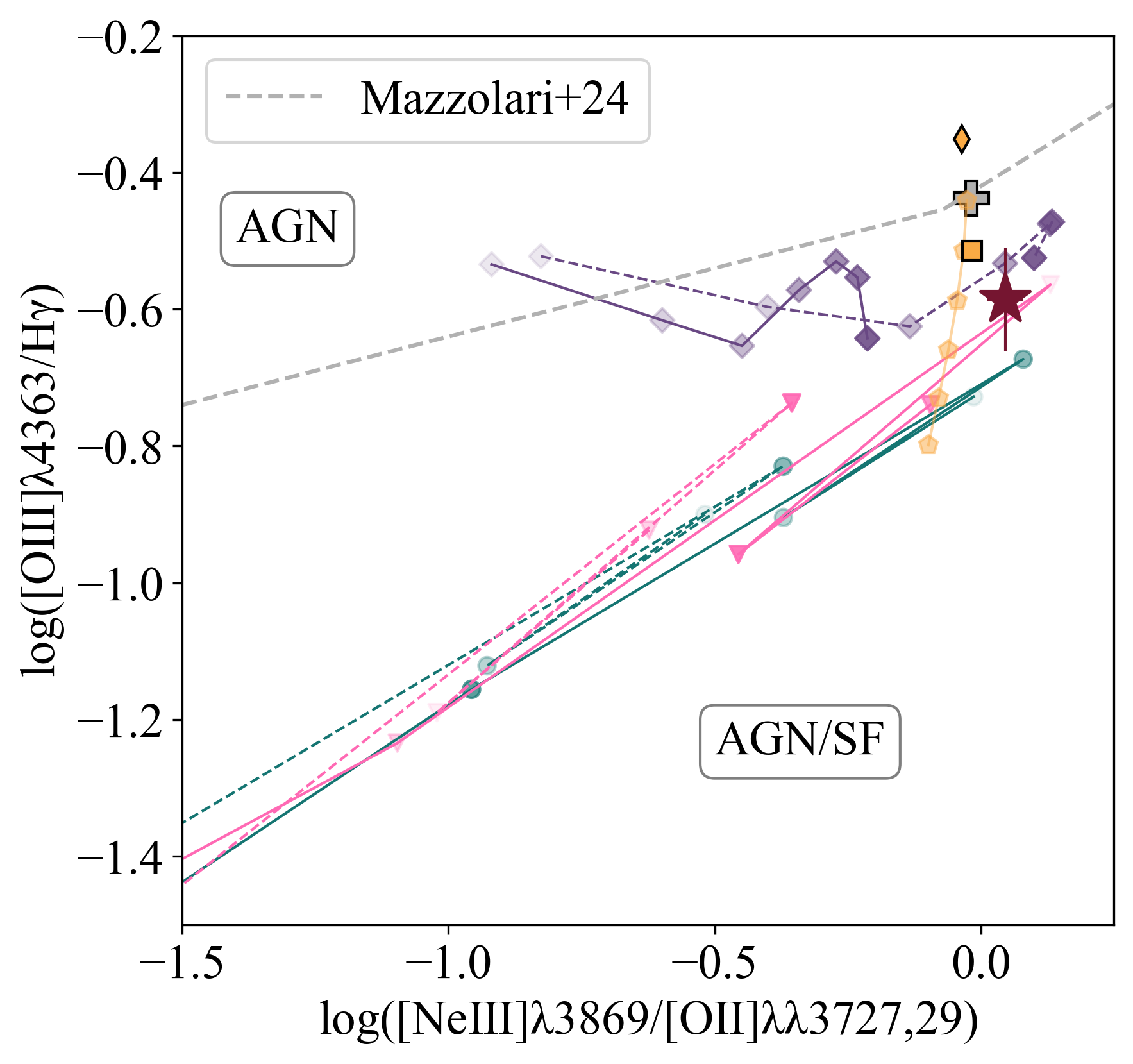}
\includegraphics[width=0.8\columnwidth]{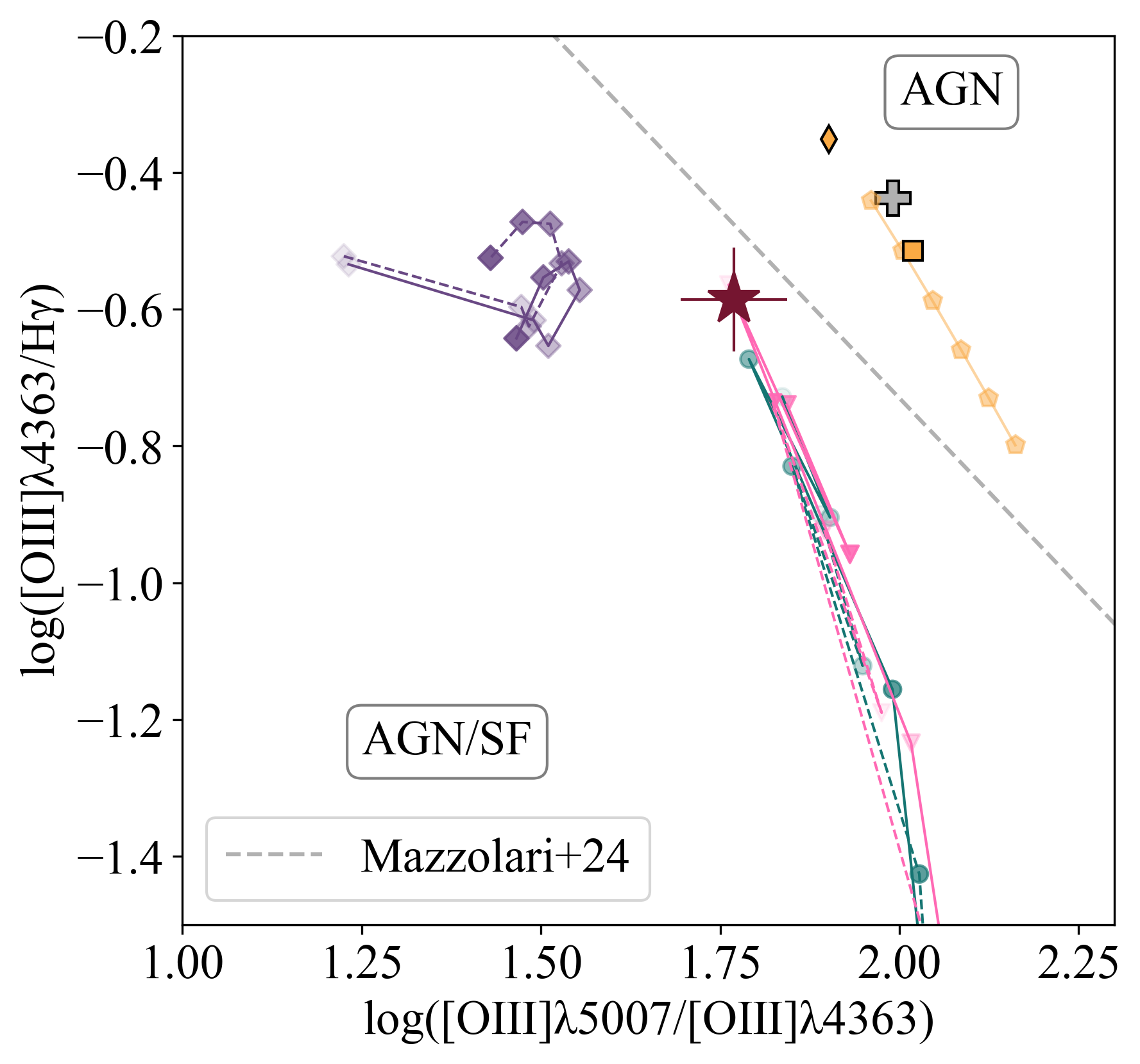}
\caption{Emission-line diagnostic diagrams together with theoretical models specifically computed for the nebular properties derived for this source. Photoionization models with a \citet{salpeter1955} and a top-heavy IMF \citep{kauffmann2024} are shown with blue dots and pink triangles, respectively, for two ionization parameters. Shock models for two different magnetic fields are shown with purple diamonds. The transparency of the colors decreases with age and velocity in the SF and shock models, respectively. Finally, different ionization models with power laws and absorbed AGNs are shown with yellow and grey symbols. Division lines between AGN and SF photoionization from \citet{Backhaus2022} and \citet{Mazzolari2024} are shown with dashed lines in the panels.}
\label{fig:blue_bpts}
\end{figure*}

The so called ``OHNO'' diagram, shown in the upper left panel of Figure \ref{fig:blue_bpts}, was previously presented by \citet{Larson2023}. They concluded that the source is powered by a high-ionization field and that the measurements fall within a region of this diagnostic diagram that makes it difficult to distinguish between an AGN and metal-poor, high-ionization star-forming HII regions. The authors found that the position of the galaxy is compatible with CLOUDY star-forming models (log(u) = --1.9 to --1.5), AGN models (log(u) $\sim$ --2.1), and MAPPINGS shock models (log(U) = --2.5 to --1.5). 

Given that they are not able to fully distinguish the nature of the emission source in this galaxy, we present the same diagram and three additional ones using different emission line ratios (see Figure \ref{fig:blue_bpts}). We also include the theoretical models described in Section \ref{sec:models}, which are specifically computed for the nebular properties derived for this source. 

The [OIII]$\lambda$4363/H$\gamma$ and [NeIII]$\lambda$3869/[OII]$\lambda \lambda$3727,29 ratios \citep{Mazzolari2024} are consistent with any of the photoionization and shock models computed for this source. In contrast, other diagnostic ratios, such as [OIII]$\lambda$5007/H$\beta$, [OIII]$\lambda$5007/[OII]$\lambda \lambda$3727,29, and [OIII]$\lambda$5007/ [OIII]$\lambda$4363 ratios, are well reproduced by the SF models only. According to these results, the nature of this source is consistent with a star-forming scenario. However, the emission line ratios used are primarily sensitive to ionization conditions and metallicity, rather than to variations in the upper end of the IMF. Then, we are not able to constrain the shape of the IMF.

Other UV emission line ratios were used in \citet{Marques-Chaves2024} to distinguish the nature of this source, including CIII]/He II, CIV/He II, [OIII]/He II, and NV/NIV]. They also compare them with star-forming and AGN models compatible with the inferred gas properties. They used upper limits for He II and NV, which were not detected, as in the present study. The absence of these emission lines, probing energies >54 eV and >77 eV respectively, places the galaxy outside the region occupied by AGN models. Then, they conclude that the high-ionization lines observed are also consistent with stellar photoionization.

It is also important to emphasize that the photoionization models support the ionization structure assumed in our analysis (see Section \ref{sec:ion_abundances}). Consequently, the ionic abundances of both low and high ionization species provide a good approximation of the total oxygen abundance in the galaxy, with O$^{+}$/H$^{+}$ + O$^{++}$/H$^{+}$ $\sim$ O/H.

\section{Stellar populations}
\label{sec:star}
\subsection{Star formation rate}

We estimated the star-formation rate from the extinction-corrected luminosity of the Balmer lines, which are sensitive to the star formation activity over the past 10 Myr. We  estimated the luminosity of H$\alpha$ from the extinction-corrected luminosity of H$\beta$, assuming the theoretical  H$\alpha$/H$\beta$ ratio of 2.86 \citep[for $T = 10^4$~K and $n_e = 100$~cm$^{-3}$;][]{pyneb}. We then employed the  H$\alpha$ luminosity-to-SFR conversion factor of $3.236 \times 10^{-42}$ (M$_\odot$/yr)/(erg/s) from \citet{reddy2018}, obtaining SFR$_{H\beta}$ = 49 $\pm$ 3 M$_\odot$/yr.

This conversion factor is lower than those adopted in other works. For instance, the factor is $4.165 \times 10^{-42}$ (M$_\odot$/yr)/(erg/s) for the calibration of \citet{kennicutt1998}, assuming a \citet{chabrier2003} initial mass function (IMF), and $4.634 \times 10^{-42}$ (M$_\odot$/yr)/(erg/s) in \citet{hao2011} using the same IMF. However, the \citet{reddy2018} calibration is more suitable for low-metallicity galaxies at high redshifts (stellar population modeling assuming \citet{Bruzual2003}, and Z = 0.28 Z$_\odot$).

Young, massive stars also emit strongly at UV wavelengths, particularly between 1500 and 2800 \AA, significantly contributing to the UV continuum. As a result, the star formation rate can also be reliably traced using this emission. We used the relation from \citet{kennicutt1998}, which provides a UV luminosity-to-SFR conversion factor of $1.4 \times 10^{-28}$, with the luminosity expressed in units of erg\,s$^{-1}$\,Hz$^{-1}$. We obtained SFR$_{UV}$ = 4.2 $\pm$ 0.7 M$_\odot$/yr.

It is important to note that the UV-based SFR tracer relies on the assumption that the UV continuum is dominated by young stellar populations. However, at these wavelengths, the continuum can be significantly contaminated by nebular emission, particularly by the two-photon continuum. This contamination can introduce systematic offsets in the SFR calibration of approximately 0.05 to 0.15 dex, with larger deviations expected at lower metallicities \citep{raiter2010}. Then, UV-based SFR may be lower than the one calculated from the Balmer emission. However, since there is no evidence of a Balmer jump in our integrated spectrum and continuum features near Ly$\alpha$, a significant contribution from the nebular continuum or two-photon emission appears unlikely for this particular source. It is important also to emphasize that the SFR calibration by \citet{kennicutt1998} may underestimate the number of ionizing photons in galaxies with metal-poor stars, so a UV-based SFR could be lower than the value derived for our sources. Nevertheless, this does not affect the conclusions regarding the star formation history of the galaxy presented here.

The young, massive, and ionizing O and B stars responsible for the H$\beta$ emission lines have lifespans shorter than 10 Myr. On the contrary, the stars contributing to the UV continuum emission have longer lifespans, typically between 10 and 100 Myr. Therefore, comparing the star formation rates derived from the H$\beta$ emission and the UV continuum provides valuable information about the recent star formation history of the galaxy. The higher H$\beta$-based SFR relative to the UV-based one suggests that this galaxy has not experienced a constant star formation history over the past 100 Myr; instead, it experienced a strong burst of star formation within the last 10 Myr.

\subsection{Temperature of the ionizing stars}
The number of hydrogen-ionizing photons, Q(H$_0$), can be calculated from their extinction-corrected H$\beta$ flux once translated into luminosities. We used the equation derived using the recombination coefficient of the H$\beta$ line assuming a constant value of electron density of 100 cm$^{-3}$, a temperature of 10$^4$ K and case B recombination \citep{Osterbrock2006}. This quantity is (1.110 $\pm$ 0.078) $\times$ 10$^{55}$ photons/s, corresponding to H$\alpha$ luminosity of (1.52 $\pm$ 0.11) $\times$ 10$^{43}$ erg/s. We also calculated the number of helium-ionizing photons, Q(He$_0$), from the observed luminosity of the HeI $\lambda$3889 \AA\ emission line, assuming the same physical conditions described above. We obtained (2.61 $\pm$ 0.37) $\times$ 10$^{52}$ photons/s.

The hardness of the ionizing radiation can be inferred from the spectral energy distribution of the ionizing source, which is commonly parameterized by the $\eta$ parameter. Defined by \citet{vilchez1988}, this is the ratio of oxygen and sulphur ionic abundance ratios, (O$^{+}$/O$^{++}$) and (S$^{+}$/S$^{++}$). This parameter serves as a proxy for the effective temperature of the ionizing stars, as it reflects the ionization structure of the nebula which is determined by the slope of the spectral ionizing radiation. High values of the $\eta$ parameter correspond to low stellar effective temperatures \citep[see Figure 14 of][]{zamora2023}. However, due to the wavelength coverage of our data, we were unable to determine the sulfur ionic abundances and therefore could not apply this approach.

The ionization structure of the nebula can also be inferred from the ratio between the number of helium and hydrogen ionizing photons, Q(He$_0$)/Q(H$_0$), since for effective temperatures below $\sim$ 40000 K, the He$^+$ Strömgren zone is smaller than the H$^+$ one due to the higher ionization potential of helium \citep{Osterbrock2006}. Then, by comparing the ionizing photon ratio with predictions from photoionization models, we can infer the equivalent effective temperature of the ionizing stellar population. 

We estimated the effective temperature of the ionising stellar population in our galaxy by applying the relation proposed by \citet{zamora2023}, which is based on Cloudy models \citep[][log(u) = --4.0 to --2.5, $n_{\rm e}$ = 100 cm$^{-3}$ and solar metallicity]{cloudy} and stellar atmospheres from \citet[][non-LTE models, log(g) = 4 and T$_{eff}$ = 30000 - 55000 K]{mihalas1978}. This relation can be used only for log(Q(H$_0$)/Q(He$_0$)) $>$ 2.8 because below this threshold the logarithmic ratio becomes insensitive to T$_{eff}$, because the ionization zones since the ionization zones of helium and hydrogen coincide \citep{zamora2023}. We obtained log(Q(H$_0$)/Q(He$_0$)) $\sim$ 2.63, which corresponds to T$_{eff}$ > 40.000K. The effective temperature of the ionizing stellar population in the galaxy is consistent with the low chemical abundance derived for the gas (see Sec. \ref{sec:ion_abundances}), providing a coherent picture between the stellar radiation field and the nebular conditions.

It is also important to emphasize that the photoionization models calculated in Section \ref{sec:models} are consistent with the hydrogen and helium ionization structure derived from the Q(H$_0$)/Q(He$_0$) ratio: both ionization zones coincide. This further supports a star-forming origin for the ionization in the galaxy.

\section{Discussion}\label{sec:disc}

\subsection{Electron density evolution with z}
\begin{figure}[h!]
\centering
\includegraphics[width=\linewidth]{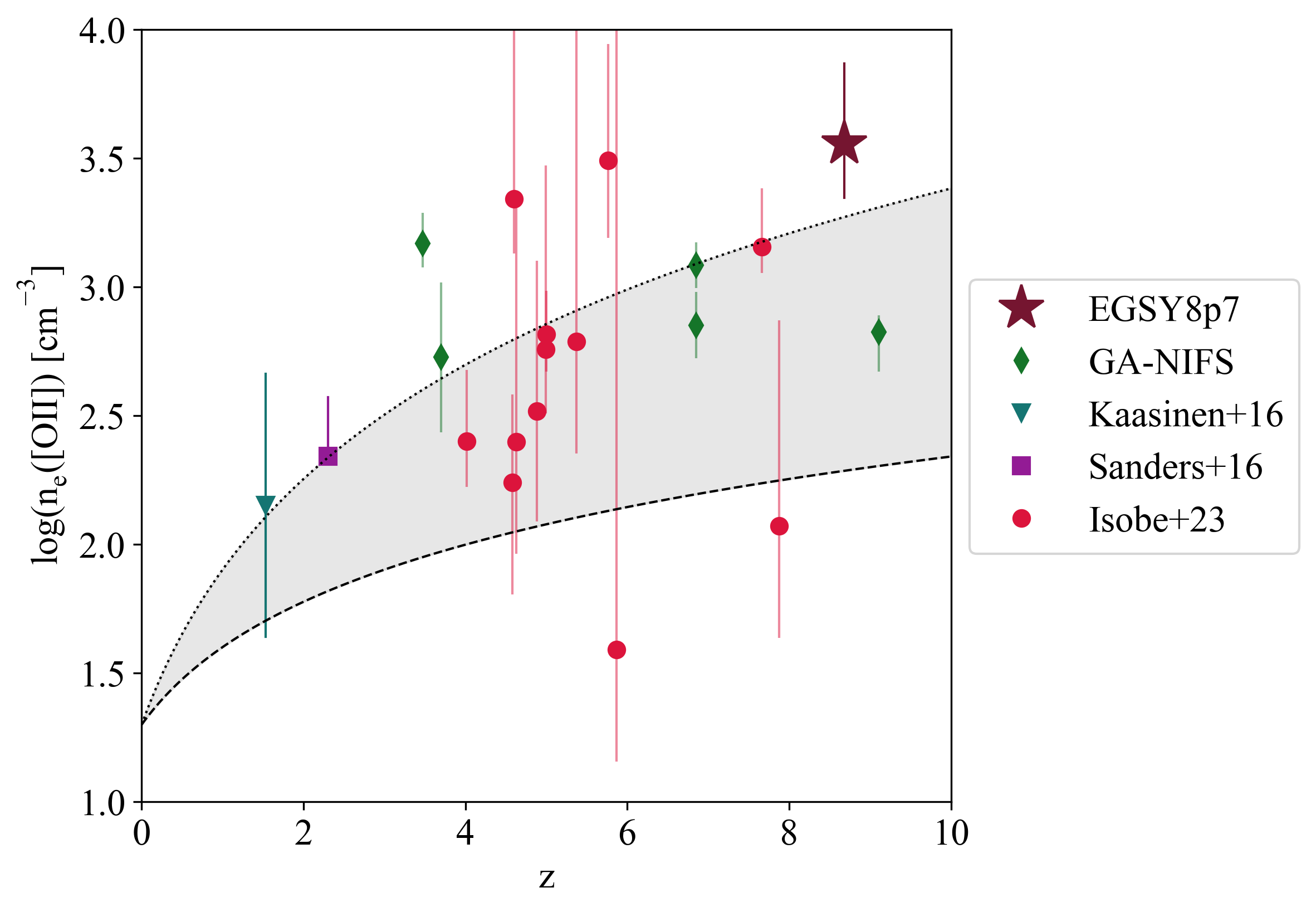}
\caption{Electron density $n_{\rm e}$ as a function of redshift. The dotted and dashed curves represent $n_{\rm e}$ $\sim$ (1+z)$^{1}$ and $\sim$ (1+z)$^{2}$ respectively. A sample of star-forming galaxies at z = 4–9 observed by the JWST/NIRSpec GLASS, Early Release Observations, and CEERS programs are shown for comparison purposes \citep{Isobe2023}. Also the means values from \citet{Kaasinen2017} and \cite{Sanders2016} at lower redshifts. Finally, a sample of GA-NIFS star-forming galaxies is shown \citep{Scholtz2025, Lamperti2024, Marconcini2024,delpino2024}.}
\label{fig:z-ne}
\end{figure}

The relationship between the physical conditions of the ISM, such as  ionization state, metallicity, density, and kinematics, across cosmic time remains an open question that requires further in-depth investigation. However, there is a clear trend of increasing electron density with redshift \citep{Davies2021,Isobe2023,topping2025}. 

High redshift star-forming galaxies are generally more compact at fixed stellar mass, showing smaller effective radii than their low-redshift counterparts. If the size evolution follows r $\sim$ (1+z)$^{-1}$ \citep{mo2002, shibuya2015, ono2023}, then the densities are expected to be proportional to $n_{\rm e}$ $\sim$ (1+z)$^{2}$ \citep[e.g.,][]{topping2025, Isobe2023,Davies2021}. However, the electron density measured in the studied galaxy, $\log(n_e) = 3.56^{+0.31}_{-0.21}$, is slightly higher than predictions for galaxies at z = 8–9 (see Fig. \ref{fig:z-ne}), assuming an extrapolation of the trends inferred at lower redshift \citep[see][]{Isobe2023}.
Our galaxy is not an isolated case; other galaxies reported in the literature exhibit a similar behavior, for example J0808+3948 in the local Universe, with $n_{\rm e}$ $\sim$ 1000 cm$^{-3}$ \citep{berg2022}, GS-5001 at redshift z = 3.47, with $n_{\rm e}$ $\sim$ 1480 cm$^{-3}$ \citep{Lamperti2024}, or CEERS-01658, having $n_{\rm e}$ $\sim$ 2200 cm$^{-3}$ at z = 4.6 \citep{Isobe2023}. Despite the large uncertainties in our measurement, this may suggest that the electron temperature in this object is not particularly high. Consequently, any additional pressure that could drive thermal expansion does not appear to be reducing the electron density. 

\subsection{Gas excitation properties of the galaxy}\label{sec:exc}

In Section~\ref{sec:bpts}, we establish that the ionization is not driven by an AGN but is instead powered by star-formation processes. Then, in order to study the excitation properties of the gas, we constructed several diagrams to characterize its degree of ionization. To place our results in the context of high-redshift galaxies, we compiled a sample of sources with high-quality data and the necessary emission lines for this analysis \citep{Curti2023,Schaerer2024,arellano-cordova2022, Topping2024,cameron2023,Heintz2024,Curti2025,sanders2024}, with direct measurements of the auroral [OIII]$\lambda$ 4364 \AA\ emission line and similar abundances as our source. We also include additional samples of low-metallicity HII galaxies, Green Pea galaxies, and Extreme Emission Line Galaxies (EELGs), which are considered analogs of high-redshift galaxies in the local Universe \citep{Izotov2006,Jaskot2013,Amorin2015}. 

\begin{figure*}[h]
\centering
\includegraphics[width=0.8\columnwidth]{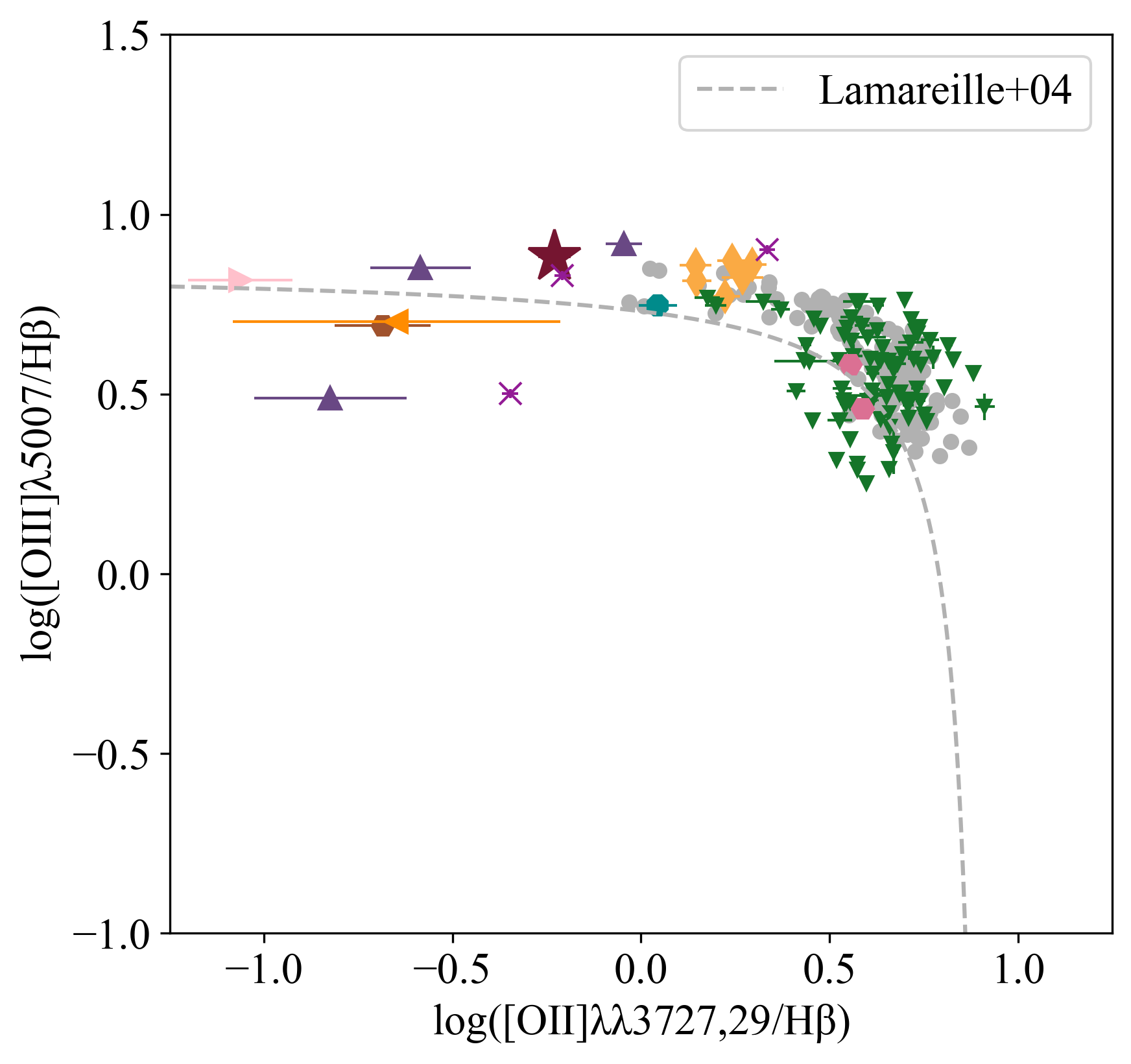}
\includegraphics[width=0.83\columnwidth]{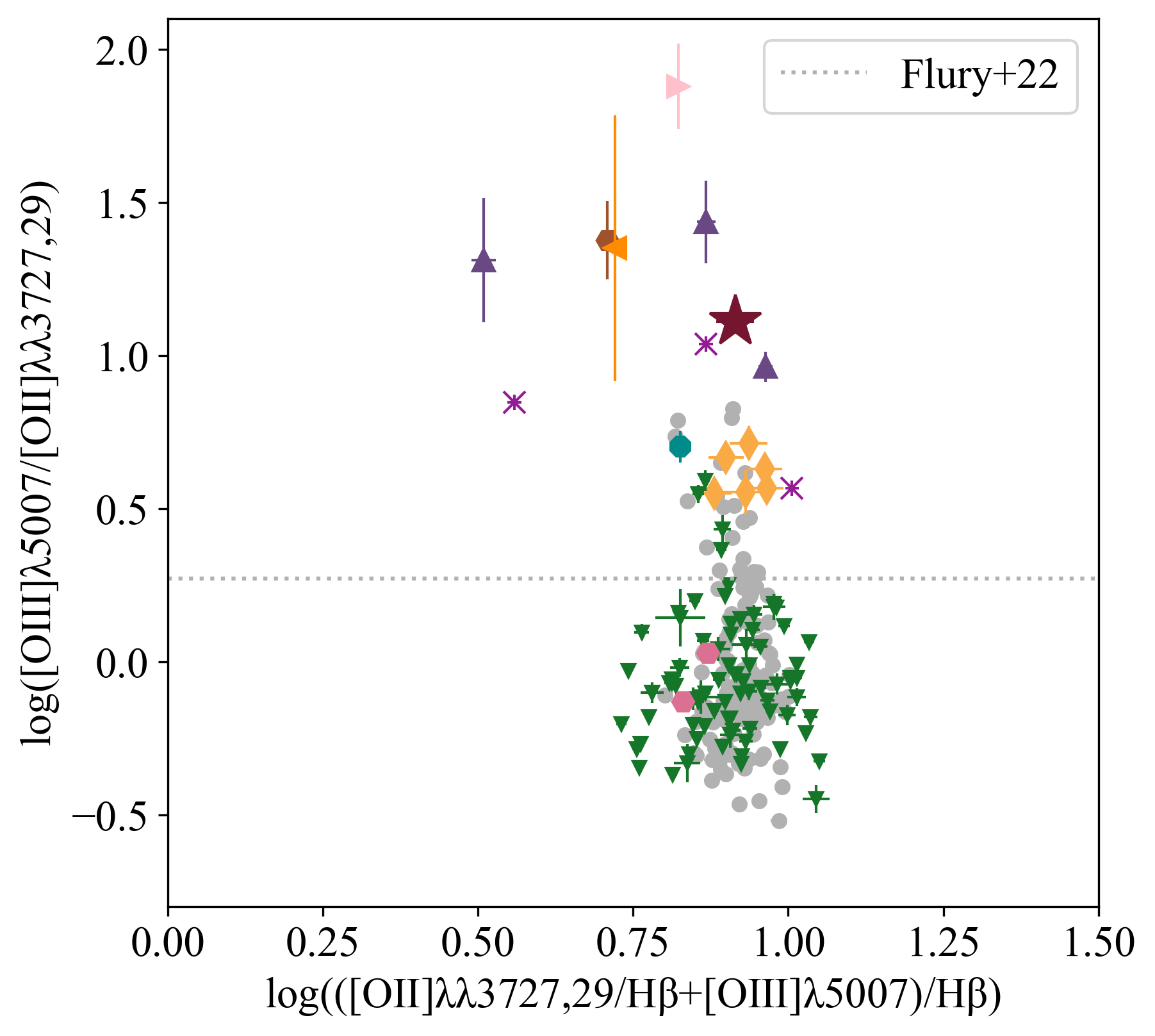}
\includegraphics[width=0.74\linewidth]{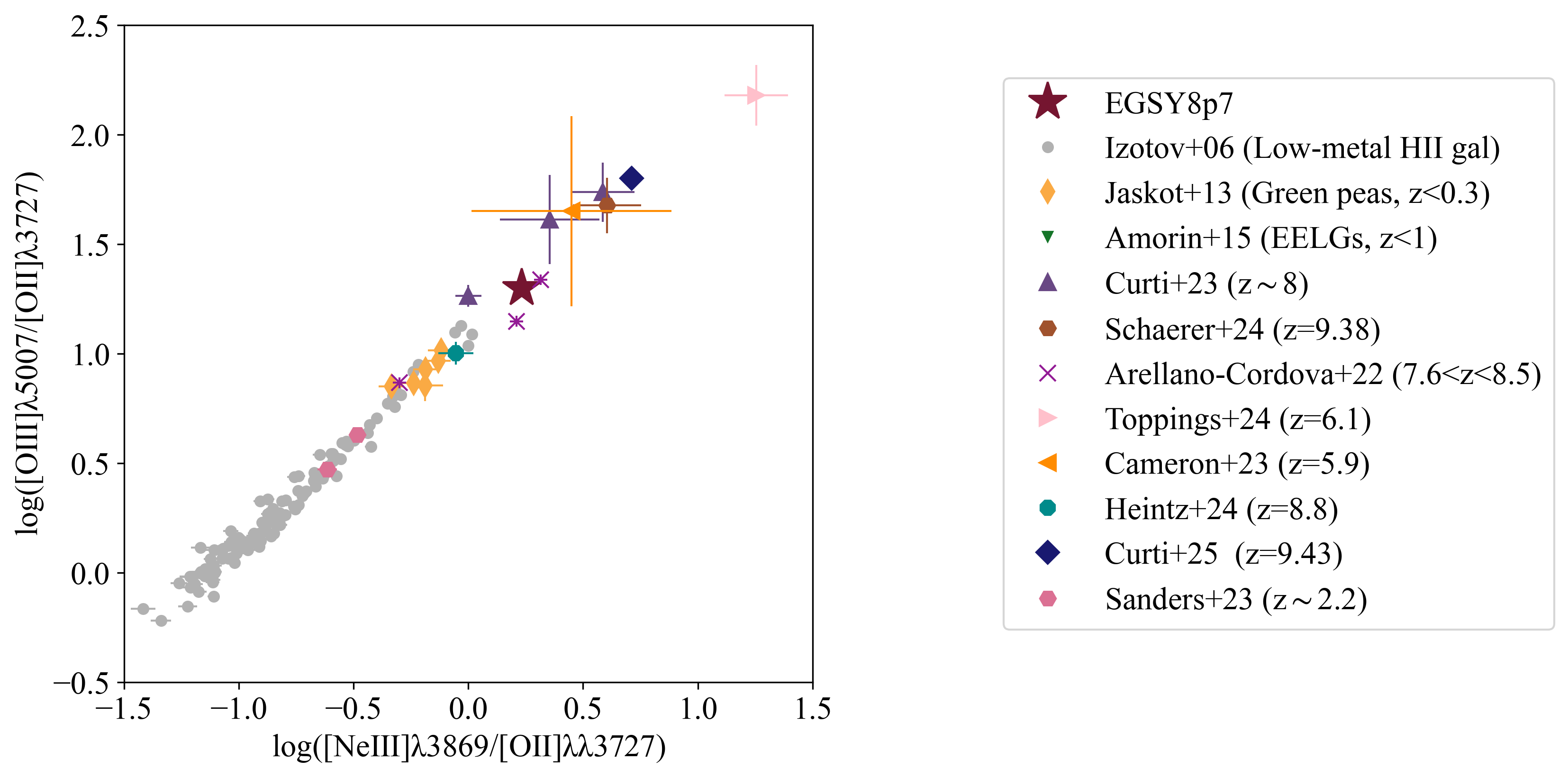}
\caption{Excitation diagrams. A compilation of high redshift galaxies with direct measurements of the auroral [OIII]$\lambda$ 4364 \AA\ emission line and similar abundances as our source are presented as a comparison \citep{Curti2023,Schaerer2024,arellano-cordova2022, Topping2024,cameron2023,Heintz2024,Curti2025,sanders2024}. Samples of local low-metal HII galaxies, Green peas and EELGs \citep{Izotov2006,Jaskot2013,Amorin2015} are also shown. The empirical divisions defined by \citet{Lamareille2004} in the local Universe to separate star-forming galaxies from AGNs, as well as the LyC leakage threshold proposed by \citet{Flury2022}, are shown with grey dashed and dotted lines, respectively.}\label{fig:exc_properties}
\end{figure*}

The upper left panel of Figure \ref{fig:exc_properties} shows the [OIII]$\lambda$5007/H$\beta$ versus [OII]$\lambda$3727,29/H$\beta$ diagram, used to infer the excitation properties of the galaxy. The dashed grey line marks the empirical boundary between star-forming galaxies and AGNs in the local Universe \citep{Lamareille2004,lamareille2010}. However, the higher ionization parameters and harder radiation fields that characterize high-redshift galaxies often result in enhanced [OIII] emission, which causes these systems to lie above the dividing line established for nearby galaxies \citep[see][]{maiolino2019}. This trend is further reinforced by the location of other high-redshift star-forming galaxies in the same diagram. In this context, although the studied galaxy does not fall within the local star-forming area, with a measured value of log([OII]$\lambda$3727,29/H$\beta$) = 0.882 $\pm$ 0.033, this  does not necessarily indicate the presence of an AGN. Instead, it likely reflects the elevated excitation conditions common among galaxies in the early Universe.

The lower left panel of the same figure shows [OIII]$\lambda$5007/[OII]$\lambda$3727 versus [NeIII]$\lambda$3869/[OII]$\lambda$3727, two proxies for the ionization parameter of the sources (see Section \ref{sec:ionization_parameter}). All the measurements lie along a linear regression, which is expected due to the similar ionization structures of Ne$^{2+}$ and O$^{2+}$. This diagram also confirms that the source is powered by a high ionization field, as found by \citet{Larson2023}.

Finally, the right panel of Fig.~\ref{fig:exc_properties} shows [OIII]$\lambda$5007/[OII]$\lambda$3727,29 versus ([OIII]$\lambda$5007+[OII]$\lambda$3727,29)/H$\beta$, also referred to as O${32}$ and R${23}$, respectively, which are proxies for the ionization parameter and metallicity of the sources. Our galaxy has a value of log([OIII]$\lambda$5007/[OII]$\lambda$3727,29) = 1.037 $\pm$ 0.048. Such high values are sometimes interpreted as indicative of density bounded HII regions, which could be associated with high escape fractions of Lyman continuum (LyC) photons. In the same figure, the dotted grey line marks the threshold for LyC leakers (LCEs), with f$_{\rm esc}$ > 0.05, proposed by \citet[][]{Flury2022}. Most high redshift sources lie above this threshold, although our galaxy has an escape fraction lower than this limit. Therefore, the high O$_{32}$ ratio in our galaxy is more likely associated with very high excitation conditions \citep[see][]{Chisholm2022}.

\subsection{MZR and SFR main sequence}
\begin{figure}[h]
\centering
\includegraphics[width=0.8\columnwidth]{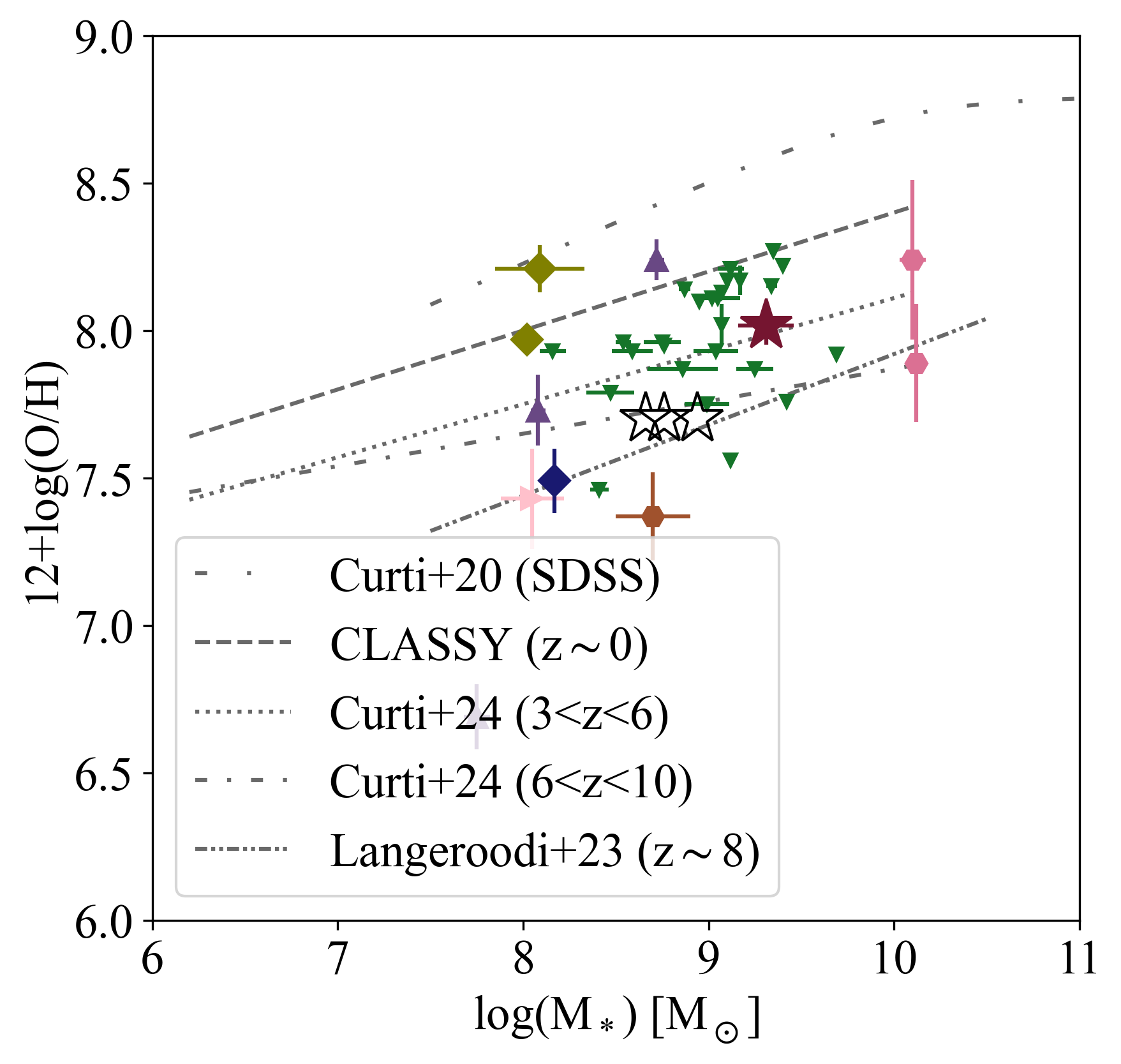}
\includegraphics[width=0.8\columnwidth]{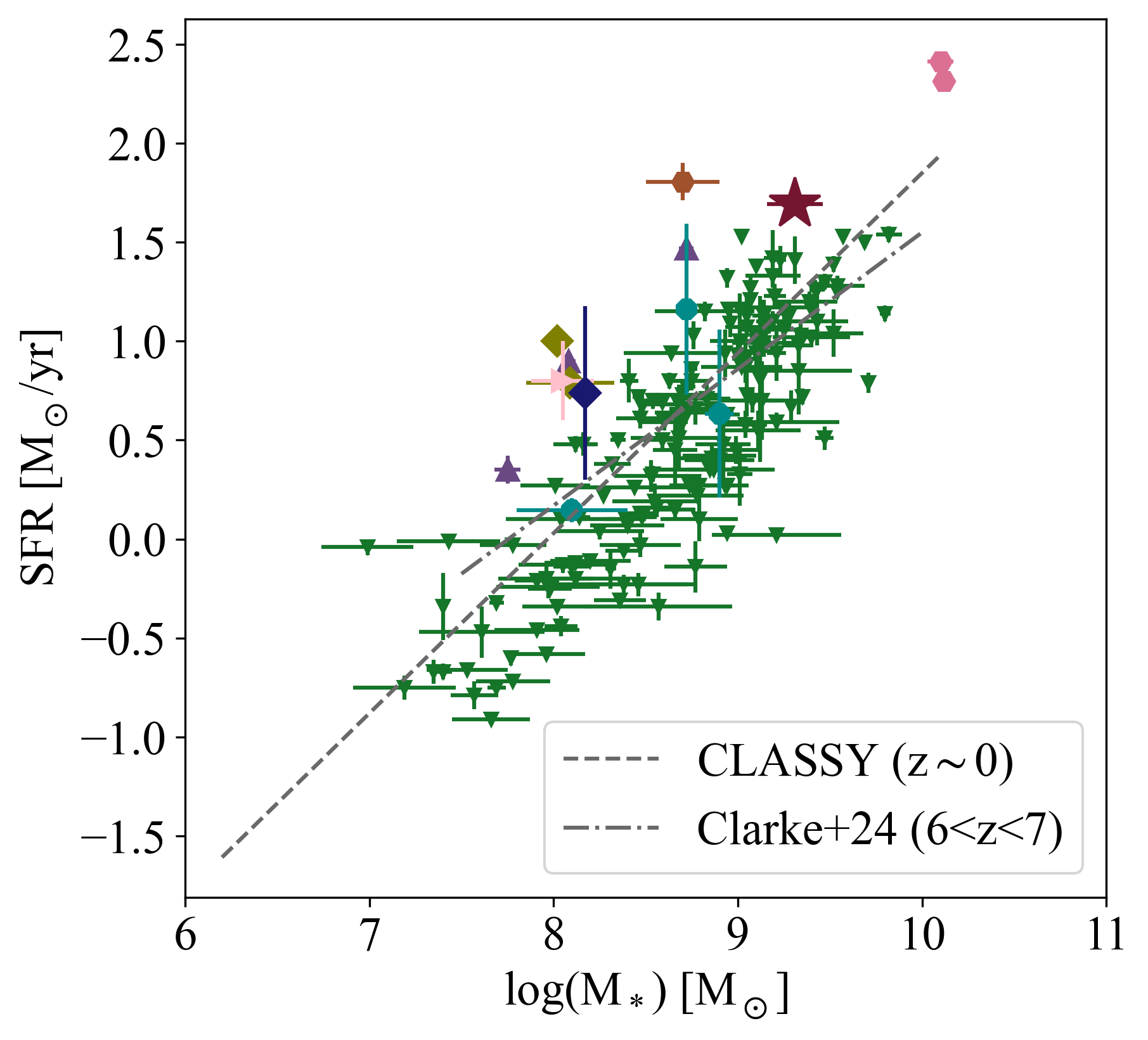}
\caption{The upper panel shows the mass metallicity relation with calibrations at different redshifts \citep{berg2022,Curti2024,Curti2020,Langeroodi2023}. The lower panel shows the SFR versus stellar mass relations found at low and high redshift \citep{Clarke2024,berg2022}. The high-redshift galaxy compilation used in Fig. \ref{fig:exc_properties} is also shown using the same symbols. The values for the individual clumps in our galaxy (open stars), as well as their stellar masses, are taken from \citet{Marques-Chaves2024}.}
\label{fig:MRZ}
\end{figure}

The top panel of Figure \ref{fig:MRZ} shows the position of our galaxy in the mass–metallicity relation (MZR). The additional galaxy sample described in Section \ref{sec:exc} is also included, together with the SFR–M$_*$ relations from other works in the literature \citep{berg2022, Curti2020, Curti2024, Langeroodi2023}. The integrated values reported in this work are consistent with the mass–metallicity relation proposed by \citet{Curti2024} at $3 < z < 6$, showing similar metallicity and stellar mass to the $z<1$ EELGs from \citet{Amorin2015}. Instead, this galaxy has a higher metallicity than the relation found for galaxies at z = 6–10  at fixed stellar mass \citep{Curti2024}.
On the other hand, the individual clumps of the galaxy studied in this work analyzed in \citet{Marques-Chaves2024} are compatible with the relation found by \citet{Curti2024} at $6 < z < 10$ and by \citet{Langeroodi2023} at z $\sim$ 8. However, in both cases at fixed stellar mass, the metallicity is lower than in galaxies in the local Universe.

The lower panel of Figure \ref{fig:MRZ} shows the position of our galaxy in the star-forming main sequence, compared with SFR–stellar mass relations derived from both local analogs and high-redshift galaxy populations \citep{berg2022, Clarke2024,Popesso2023}. Our galaxy lies significantly above the main sequence, indicating that it is undergoing a starburst phase, with a star formation rate notably higher than the average for galaxies of similar stellar mass. This suggests an episode of enhanced star formation, possibly driven by processes such as gas accretion, mergers, or internal dynamical instabilities. A similar offset is observed for the other high-redshift galaxies in our comparison sample, reinforcing the idea that intense star-forming activity is a common feature among galaxies in the early Universe. 

\subsection{Spatially structure of the ISM emission}
The main emission features observed in the integrated spectrum originate from a compact region consistent with the primary star-forming clump A, as shown in Figure \ref{fig:line_maps}. Previous studies have shown that low mass galaxies with strong nebular emission lines often exhibit compact, centrally concentrated, and bursty star formation, which is consistent with our findings \citep[see][]{Lyu2025}. The inner regions of these galaxies show younger stellar ages and elevated ionizing photon production efficiencies, suggesting the presence of younger stellar populations and higher specific star formation rates in the core compared to the outer regions. This supports an ``outside-in'' stellar mass assembly mode at $z \sim 7$ \citep[i.e.][]{Pan2015}, in which compaction processes driven by mechanisms such as major mergers or the accretion of tidal streams enhance central gas densities and trigger intense star formation in the early stages of galaxy evolution.

Notably, the oxygen auroral emission line [OIII]$\lambda$4363~\AA, as well as the [OII]$\lambda \lambda$3727,39~\AA\ doublet, are spatially shifted toward clump B (see Fig. \ref{fig:line_maps}). This may reflect complex ionization structures in the gas located between the two ionized regions.
These conditions may result from shock heating caused by interactions between the clumps, or from a scenario in which clump B is in an earlier evolutionary stage or it is experiencing feedback processes that modify its ionization balance.

Regarding the electron density of the galaxy, it is important to note that different ions trace different ionization zones. \citet{Marques-Chaves2024} measured the NIV]$\lambda \lambda$1483,87 \AA\ doublet line ratio, I(1483)/I(1487) = 0.50 $\pm$ 0.22, which corresponds to an electron density of $n_{\rm e}$ = 10$^4$ -- 10$^5$ cm$^{-3}$. Similarly, the non-detection of the redder line of the CIII]$\lambda \lambda$1907,1909 \AA\ doublet suggests densities above 10$^4$ cm$^{-3}$. However, these values are significantly higher than the density derived from the [OII] doublet, of $n_{\rm e}$ $\sim$ 3600 cm$^{-3}$ (see Section~\ref{sec:density}). The nitrogen and carbon lines originate from the primary star-forming clump (A), as shown in Figure \ref{fig:line_maps}, indicating that this region hosts gas at very high densities. In contrast, the [OII] emission primarily arises from the region between clumps A and B, which is characterized by lower excitation and lower gas densities. This spatial variation highlights the complex structure of the interstellar medium in this system, with localized high-density star-forming regions with a more diffuse, lower-density environment between the two principal star-forming clumps.

The outflow, as traced buy the broad components in [OIII]$\lambda\lambda$4959,5007 \AA\ and H$\beta$ (see Sect. \ref{sec:broad}), is spatially resolved and located between the two main star forming clumps, A and B. It appears to be directed toward clump B, close to the peak emission of the oxygen auroral line [OII]$\lambda $ 4363 \AA\  and the doublet [OII]$\lambda \lambda$ 3727,29 \AA\ (see Figure \ref{fig:outflow}). However, these emmision lines remains narrow, with a velocity dispersion of $\sigma$ $\sim$ 100 km/s, significantly lower than that measured for the outflow component. This suggests that the gas responsible for the [OII] emission is different from the gas associated with the outflow. 

\subsection{Outflow and feedback}
\label{sec:outflow}
The integrated spectrum of the galaxy exhibits clear evidence of broad components in both the [OIII] and H$\beta$ emission lines (see Fig.~\ref{fig:bzoomin}). As discussed in Section~\ref{sec:broad}, we are not able to statistically distinguish whether the broad H$\beta$ component originates from the BLR of an AGN or from an outflow. 
However, the broad [OIII] emission cannot stem from dense BLR gas and we ascribe it to galactic outflows. In the following, we estimate the properties of this outflowing gas. 
The luminosity of the [OIII] outflowing broad component is (1.18 $\pm$ 0.18) $\times$ 10$^{43}$ erg/s, with a $\sigma$ of approximately 300 km/s and a velocity offset relative to the systemic redshift of the galaxy of $\Delta $v$^{\rm out}$ = 78 $\pm$ 11 km/s. We estimated the outflow velocity using the relation v$^{\rm out}$ = |$\Delta $v$^{\rm out}$|+ 2$\sigma^{\rm out}$ \citep[see][]{Fiore2017}, yielding a velocity of 714 $\pm$ 21 km/s. The outflow radius was calculated as the projected distance between the center of clump A and the centroid of the broad [OIII] emission, giving a value of R$^{\rm out}$ = 0.6 kpc.

We determined the mass of the ionized outflowing gas using the [OIII] luminosity from the broad component, following \citet{carniani2015}. 
We fixed the parameters to the values obtained from the analysis of the integrated spectrum (see Sects. \ref{sec:ion_abundances} and \ref{sec:density}): the gas-phase metallicity to 12+log(O/H) = 8.017, and the electron density to log($n_{\rm e}$) = 3.56, as this value could not be estimated for the broad component and we expect high densities in the outflow \citep[see][]{Forster2019}.
We thus inferred a  mass of the ionized outflowing gas of  $M_{\rm ion}$ = (6.1 $\pm$ 1.3) $\times$ 10$^6$ M$_{\odot}$. Assuming a uniformly filled conical geometry, the mass outflow rate of the ionized gas can be determined as  $\dot{M}_{\rm ion}$ = $M_{\rm ion}$ $v^{\rm out}$/R$^{\rm out}$, where $M_{\mathrm{ion}}$ is the mass of the ionized gas, $v^{\mathrm{out}}$ is the outflow velocity, and $R^{\mathrm{out}}$ is the radial extent of the outflow. We obtained a mass outflow rate of 7.8 $\pm$ 1.7 M$_\odot$/yr. 

Using this result, we computed the mass-loading factor, defined as $\eta$ = $\dot{M}_{\rm ion}$/SFR, and obtained a value of approximately 0.16 (16\%). This suggests that star formation is more efficient in converting gas into stars relative to the mass loss through the outflow. 

The low mass loading factor is consistent with the low ionizing photon escape fractions measured in this galaxy, implying that most of the gas is likely retained within the system. This may indicate that the outflowing material does not escape the gravitational potential, or that a significant fraction of the expelled gas could eventually fall back and be recycled into future star formation. 
These conclusions assumes that the system has typically been experiencing outflows similar to the one currently observed. However, other temporal uncertainties should also be taken into account, such as the possibility that the current outflow has only recently begun or that it represents the first significant outflow in the system.
Also, it is important to note that our analysis probes only the ionized gas phase, and the measured ionized mass outflow rate likely represents a lower limit to the total mass outflow rate \citep{Fluetsch2021,Parlanti2025}. We also stress that the dust content of the outflow is unknown and therefore the [OIII] luminosity was not corrected for any dust attenuation. Consequently, the intrinsic luminosity associated with the outflow may be higher than the observed value, and the actual mass loading factor could also be higher.

The kinetic energy rate and the momentum rate of the outflow are also key quantities used to characterize its dynamical impact on the surrounding medium. The kinetic energy rate, defined as $\dot{E}_{\rm kin}$ = 1/2 $\dot{M}_{\rm out}$ v$_{\rm out}^2$, is (1.26 $\pm$ 0.28) $\times$ 10$^{42}$ erg/s. The kinetic energy injection rate from stellar feedback, estimated as $\dot{E}_{\rm kin}^{\rm stellar}$ $\sim$ 7 $\times$ 10$^{41}$ SFR \citep{Leitherer1999}, is approximately 3.43 $\times$ 10$^{43}$ erg/s. This value exceeds the kinetic energy of the observed outflow by one order of magnitude, suggesting that stellar feedback alone could be responsible for driving the outflow \citep[see][]{Veilleux2005,Heckman2015}.

The momentum rate of the outflow, defined as $\dot{P}$ = $\dot{M}_{\rm out}$ v$_{\rm out}$, is (3.52 $\pm$ 0.76) $\times$ 10$^{34}$ g cm/s$^2$. We compare this with the radiation momentum rate, calculated as $\dot{P}^{\rm rad}$ = L$^{\rm ion}$/c. The ionizing luminosity, $L_{\mathrm{\rm ion}}$, can be approximated from the H$\alpha$ luminosity using a scaling factor, which is found to be $\sim$20 for local galaxies \citep{kennicutt1998}, and ranges from 20 to 40 for high redshift systems \citep{Chevallard2018, Stark2015, Stark2017}. Based on our H$\alpha$ luminosity, this latter range yields a radiation momentum rate between 5.1 $\times$ 10$^{33}$ and  2.0 $\times$ 10$^{34}$ g cm/s$^2$. This range is comparable or lower than the outflow momentum rate, suggesting that the total energy from star formation is sufficient to power the outflow and the radiation pressure alone may account for the required momentum transfer. This implies that additional stellar feedback processes, such as supernovae explosions and stellar winds, are not needed to the driving mechanism of the outflow \citep{Ferrara2024}.

\begin{table}[h]
\caption{Outflow properties inferred from the \oiii\ broad component.}
\centering
\begin{tabular}{lc}
\hline \hline
$\log(M_{\rm out}/{M_\odot)}$                 &  6.787 $\pm$ 0.093 \\ 
$v_{\rm out}~{\rm [\kms]}$                     &  714 $\pm$ 21   \\ 
$\log(\dot{M}_{\rm out}/{[M_\odot~\rm yr^{-1}]})$   & 0.894 $\pm$ 0.093 \\ 
$\log(\dot{E}_{\rm out}/{\rm [erg~s^{-1}])}$        & 42.100 $\pm$ 0.097\\  
$\log(\dot{P}_{\rm out}/{\rm [g~cm~s^{-2}]})$     &  34.547 $\pm$ 0.094 \\ 
$\eta$        &   0.160 $\pm$ 0.036 \\  

\hline \hline 
\end{tabular}
\label{tab:OUTprop}  
\end{table}

Finally, to address the uncertainty regarding the origin of the broad H$\beta$ component (outflow, a BLR, or a combination of both; see Section \ref{sec:broad}), we compared the ionized outflow mass estimated independently from the H$\beta$ and [OIII] broad components. According to \citet{carniani2015}, if the broad components of both emission lines originate from the same outflow, their fluxes are expected to follow the relation  $F^{outflow}_{[OIII]}/F^{outflow}_{H\beta} = 21.25 \cdot (Z/Z_\odot)$, where $Z/Z_\odot$ is the gas-phase metallicity relative to solar. In the scenario where all of the broad H$\beta$ emission arises from the outflow (medium panel of Fig. \ref{fig:bzoomin}), we measure a flux ratio of F$^{outflow}_{[OIII]}$/F$^{outflow}_{H\beta}$ = 0.176 $\pm$ 0.034. This corresponds to a metallicity of Z/Z$\odot$ = 0.27 $\pm$ 0.05, which is fully consistent with the metallicity derived for the star-forming gas in the galaxy (Sect. \ref{sec:ion_abundances}). On the other hand, assuming a combined contribution from both a BLR and an outflow (right panel of Fig. \ref{fig:bzoomin}), the observed flux ratio would imply a higher metallicity of Z/Z$_\odot$ = 0.6 $\pm$ 0.1, approximately twice the metallicity derived for the gas in the galaxy. This further supports the scenario in which the broad H$\beta$ emission is not associated with a BLR. This result, combined with the analysis of the ionisation conditions (see Section \ref{sec:bpts}),  reinforces the conclusion that the presence of an AGN  in this system is unlikely.

\section{Conclusions}\label{sec:conc}
In this study, we present new JWST/NIRSpec IFS observations of EGSY8p7, a galaxy at z = 8.6782 confirmed by its strong Ly$\alpha$ emission. The higher spectral resolution and deeper observations compared to previous studies provide an opportunity to address two open questions about this source: (i) the nature of its ionization mechanism and (ii) the origin of its enhanced N/O ratio. Furthermore, we investigate, for the first time, the spatially resolved ionization properties of this galaxy.

\begin{enumerate}[label=(\roman*)]
\item This galaxy was identified as a potential progenitor of massive $z > 6$ quasars hosting supermassive black holes, based on the detection of a broad component in the H$\beta$ line with MSA-JWST using the medium resolution grating \citep[][2.5 $\sigma$ and 2.2 $\sigma$ detection respectively]{Larson2023,Marques-Chaves2024}. We analyzed the line profiles of H$\beta$ and [OIII] in our IFS data to investigate the presence of broad components, confirming the H$\beta$ feature reported in previous studies. In addition, we identified a broad [OIII] emission feature (FWHM $\sim$ 650 km/s), which confirms the presence of an outflow in this galaxy. This emission is also spatially resolved, showing an asymmetric extension reaching $\sim$ 1 kpc in the direction perpendicular to two of the clumps A and B, with its centroid located at a projected distance of 0.6 kpc from the clump responsible for ionizing the ISM.

The luminosity of the outflow is $(1.18 \pm 0.18) \times 10^{43}$ erg/s. It is expelling $(6.1 \pm 1.3) \times 10^{6}$ M$_{\odot}$ at a velocity of $714 \pm 21$ km/s. The ionized outflow mass calculated from the H$\beta$ and [OIII] broad components is fully consistent when assuming the metallicity derived for the ionized gas. The mass loss through the ionised outflow is not efficient, yielding a mass loading factor of approximately 0.16, which is consistent with the low ionizing photon escape fraction measured ($f_{\rm esc} = 0.021 \pm 0.014$) and implies that most of the gas is likely retained within the system. We compared the kinetic energy of the observed outflow with the kinetic energy injection rate from stellar feedback, finding the latter to be larger by an order of magnitude. Additionally, the radiation pressure rate is comparable to the outflow momentum rate. Therefore, stellar feedback alone could power the outflow, while radiation pressure may account for the required momentum transfer.

We also compared several diagnostic emission line ratios measured for this galaxy with the predictions from a set of theoretical photoionization and shock models computed specifically for the physical properties of this source. The results suggest that the nature of this source is consistent with star-forming photoionization, although we are not able to constrain the shape of the IMF. We find no evidence for an accreting supermassive black hole in this galaxy.

\item This galaxy was identified by its strongly enhanced N/O ratio (log(N/O) = 0.18 $\pm$ 0.11) and a C/O ratio comparable to other galaxies with similar abundances. \citet{Marques-Chaves2024} suggested that the high N/O abundance in this galaxy can be explained either by massive-star winds from WR stars, assuming a special timing and essentially no dilution with the ambient ISM, or by mixing the ejecta from supermassive stars with comparable amounts of unenriched ISM. In the first scenario, variations in the C/O ratio would be expected, while in the second scenario only small changes are predicted. 
Our deep, high–spectral resolution observations do not reveal any WR spectroscopic features, thus supporting the enrichment scenario driven by supermassive stars. However we also note that new scenarios have been recently proposed to explaining the high N/O abundance: a) gas ejected by diﬀerential galactic winds in galaxies with a bursty star-formation history \citep{Rizzuti:2025}; b) ejecta for the first AGB stars \citep{DAntona:2023, DAntona:2025}.

\end{enumerate}
Additionally, we derived other physical properties of the galaxy. The ionization degree of the gas is high, comparable to values found in other galaxies in the early Universe. This galaxy has not experienced a constant star-formation history over the past 100 Myr; instead, it experienced a strong burst of star formation within the last 10 Myr. The electron density is higher than typically expected for galaxies at similar redshift. This may indicate that the electron temperature is not particularly high, and that any additional pressure capable of driving thermal expansion does not appear to be reducing the electron density. Furthermore, the galaxy is relatively metal-rich for its stellar mass. The effective temperature of the ionizing stars is estimated to be T$_\mathrm{eff} > 40{,}000$~K, which is consistent with the gas-phase oxygen abundance of 12+log(O/H) = 8.017 $\pm$ 0.066.

Finally, thanks to the NIRSpec IFS observations, we uncovered a complex interstellar medium structure within this system. The primary star-forming clump hosts gas at very high densities, while the secondary clump exhibits lower excitation and lower gas densities. Moreover, the main emission features seen in the integrated spectrum arise from a compact region coincident with the primary star-forming clump. This resembles the compact, centrally concentrated, and bursty star formation observed in other low-mass galaxies and is consistent with an "outside-in" stellar mass assembly mode proposed at $z \sim 7$.

\begin{acknowledgements}
SC, BT, GV, and SZ acknowledge support from the European Union (ERC, WINGS, 101040227).

EB and GC acknowledge support of the INAF Large Grant 2022 ``The metal circle: a new sharp view of the baryon cycle up to Cosmic Dawn with the latest generation IFU facilities''.

MP, SA, BRP, and PPG acknowledge support from the research projects PID2021-127718NB-I00, PID2024-159902NA-I00, PID2024-158856NA-I00, and RYC2023-044853-I of the Spanish Ministry of Science and Innovation/State Agency of Research (MCIN/AEI/10.13039/501100011033) and FSE+.

AJB acknowledges funding from the ``FirstGalaxies'' Advanced Grant from the European Research Council (ERC) under the European Union’s Horizon 2020 research and innovation programme (Grant agreement No. 789056).

FDE and GCJ acknowledge support by the Science and Technology Facilities Council (STFC), by the ERC through Advanced Grant 695671 ``QUENCH'', and by the UKRI Frontier Research grant RISEandFALL.

IL acknowledges support from PRIN-MUR project “PROMETEUS”  financed by the European Union -  Next Generation EU, Mission 4 Component 1 CUP B53D23004750006.

H\"U acknowledges funding by the European Union (ERC APEX, 101164796). Views and opinions expressed are however those of the authors only and do not necessarily reflect those of the European Union or the European Research Council Executive Agency. Neither the European Union nor the granting authority can be held responsible for them.

\end{acknowledgements}

\bibliographystyle{aa} 
\bibliography{Article}

\appendix
\section{Cleaning procedure}\label{app:clean}

We adopted the Fourier-based cleaning procedure described by \citet[][]{zamora2023fourier}, adapting the classical cross-correlation technique of \citet{TonryandDavis} to the very high spectral-resolution regime of our data. This approach is particularly well suited to identify and remove residual artifacts such as cosmic-ray spikes or high-frequency noise patterns that remain after standard pipeline processing. We performed this procedure without taking into account any spatial information, cleaning each spectrum separately pixel-by-pixel. However, the method is very fast because, operationally, it only requires computing the Fast Fourier Transform (FFT) of the entire data cube along the spectral dimension.

The method consists of two main steps. First, the spectra are rebinned onto a logarithmic wavelength scale, which provides a uniform velocity step across the full spectral range ($n(\lambda) = A\cdot ln(\lambda)+B$). Based in our wavelength range, we used $A = 3370.29$ and $B = -3553.73$. This rebinning facilitates the identification of spurious features in velocity space and ensures that the subsequent Fourier transform operates on a uniform velocity grid.

\begin{figure}[h]
\centering
\includegraphics[width=0.9\linewidth]{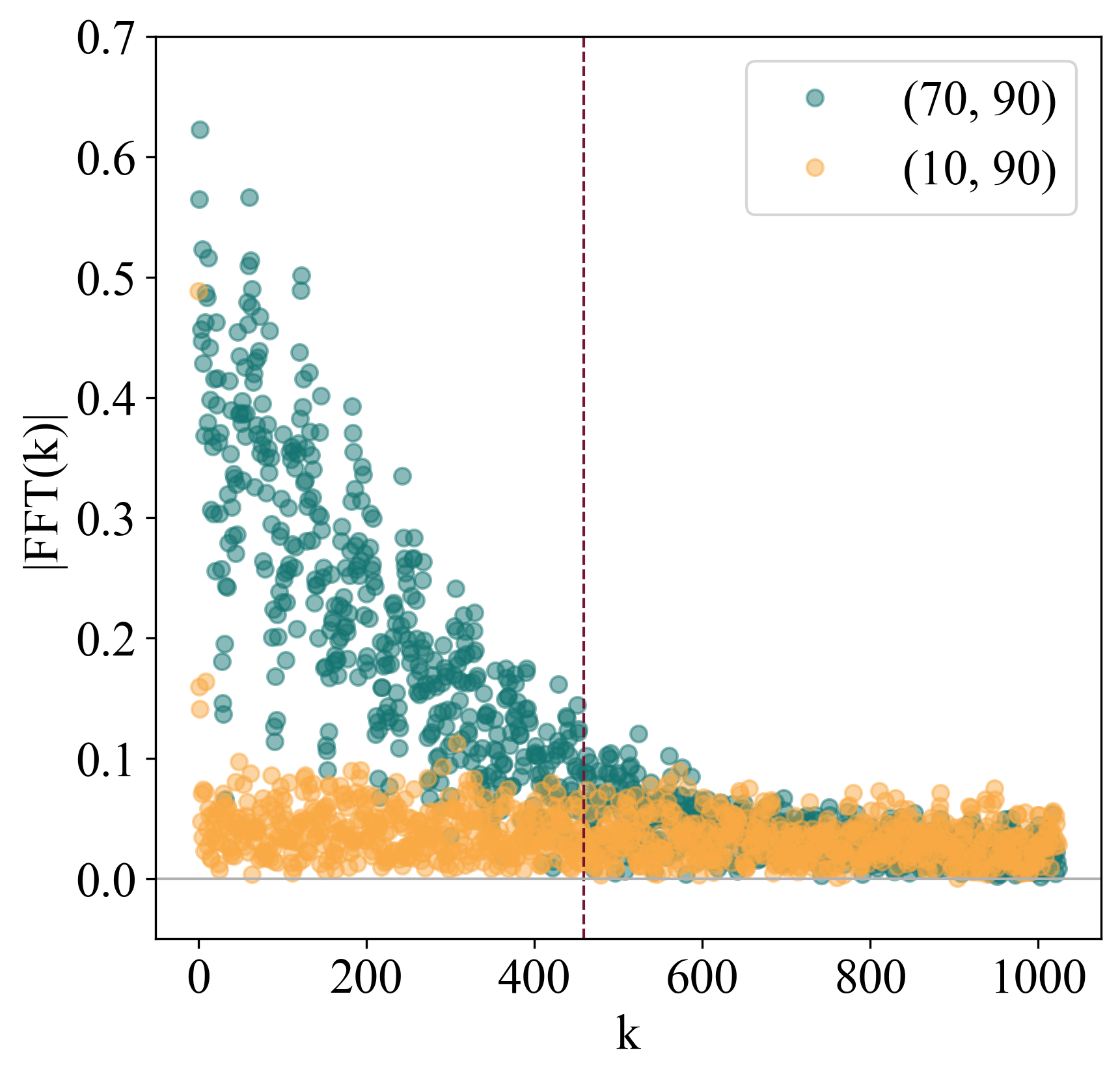}
\caption{Amplitude of the Fourier transform spectra of two different pixels, (70,90) and (10,90) with galaxy emission and without respectively. The parameter adopted for the cleaning procedure is showed with vertical lines.}\label{fig:fourier_}
\label{fig:fourier}
\end{figure}

Second, we suppressed high-frequency variations below the instrumental spectral resolution by applying a band-pass filter to the Fourier-transformed spectrum. For the G395H/F290LP configuration, we adopted the most conservative set of parameters appropriate for its spectral resolution \citep[$r \simeq 90~\mathrm{km~s^{-1}}$; $K_{\mathrm{min}} = 459$, see Equation 21 from][]{zamora2023fourier}. Figure \ref{fig:fourier_} shows the amplitude of the FFT spectra for two pixels in the G395H data cube: one located on the galaxy emission (at coordinates 70, 90) and another sampling the sky background (10, 90). The spectrum corresponding to the galaxy emission retains significant FFT amplitude up to the limit imposed by the filtering process, whereas the sky spectrum does not. Moreover, it is noteworthy that the sky spectrum shows amplitudes only at low frequencies, which correspond to the sky emission. These low frequencies can also be filtered to remove the sky emission, as shown in the lower panel of Fig. 5 in \citet[][]{zamora2023fourier}. However, in this work we used a more conventional approach to remove the sky contribution, using different apertures in regions free of galaxy emission.

\begin{figure}[h]
\centering
\includegraphics[width=\linewidth]{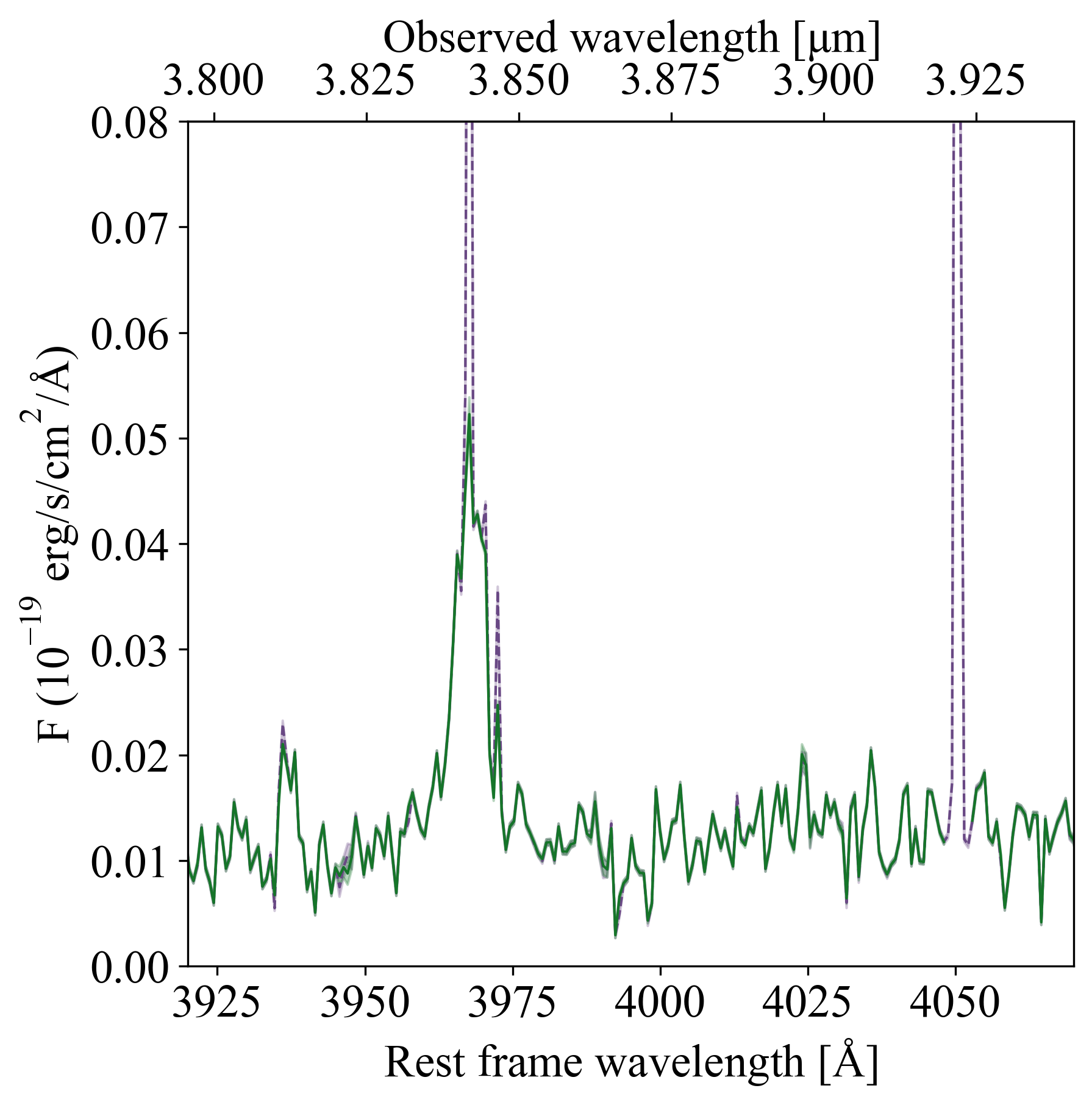}
\includegraphics[width=\linewidth]{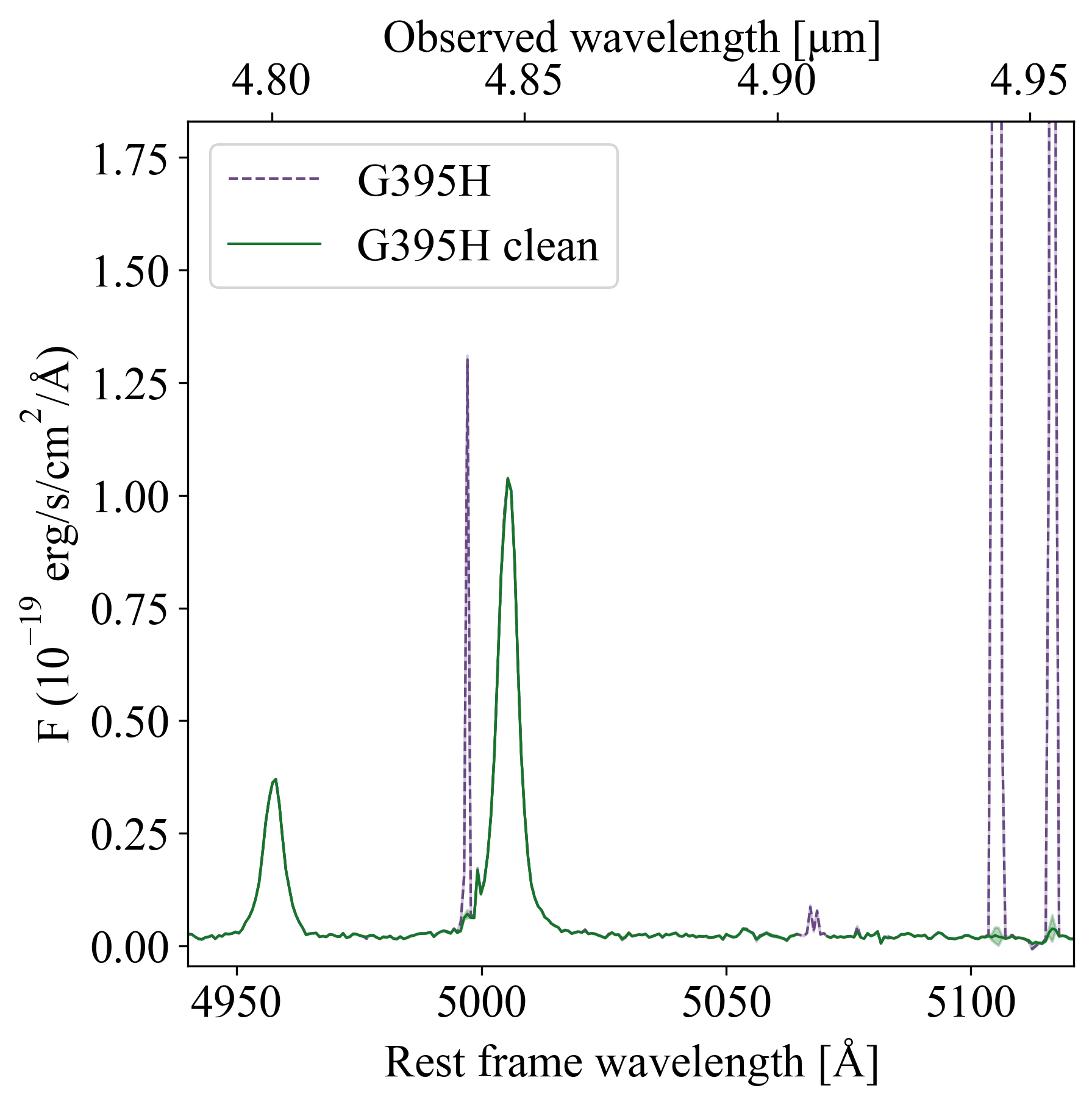}
\caption{Example of different emission lines in the G395H integrated spectrum of the galaxy, [NeIII]$\lambda$ 3968 \AA\ (upper panel) and [OIII]$\lambda\lambda$ 4959,5007 \AA\ (lower panel), with and without the cleaning procedure applied to the data cube.}
\label{fig:fourier}
\end{figure}

The result of this process is a significant reduction of narrow spikes and high-frequency artefacts in the reconstructed data cube, without degrading the intrinsic line profiles. The effectiveness of the procedure is illustrated in Figure \ref{fig:fourier}, where the cleaned spectra show markedly improved compared to the unfiltered data.

\newpage
\onecolumn
\section{Proposed scenarios for the broad H$\beta$ emission}
\begin{figure*}[h]
\centering
\includegraphics[width=1\textwidth]{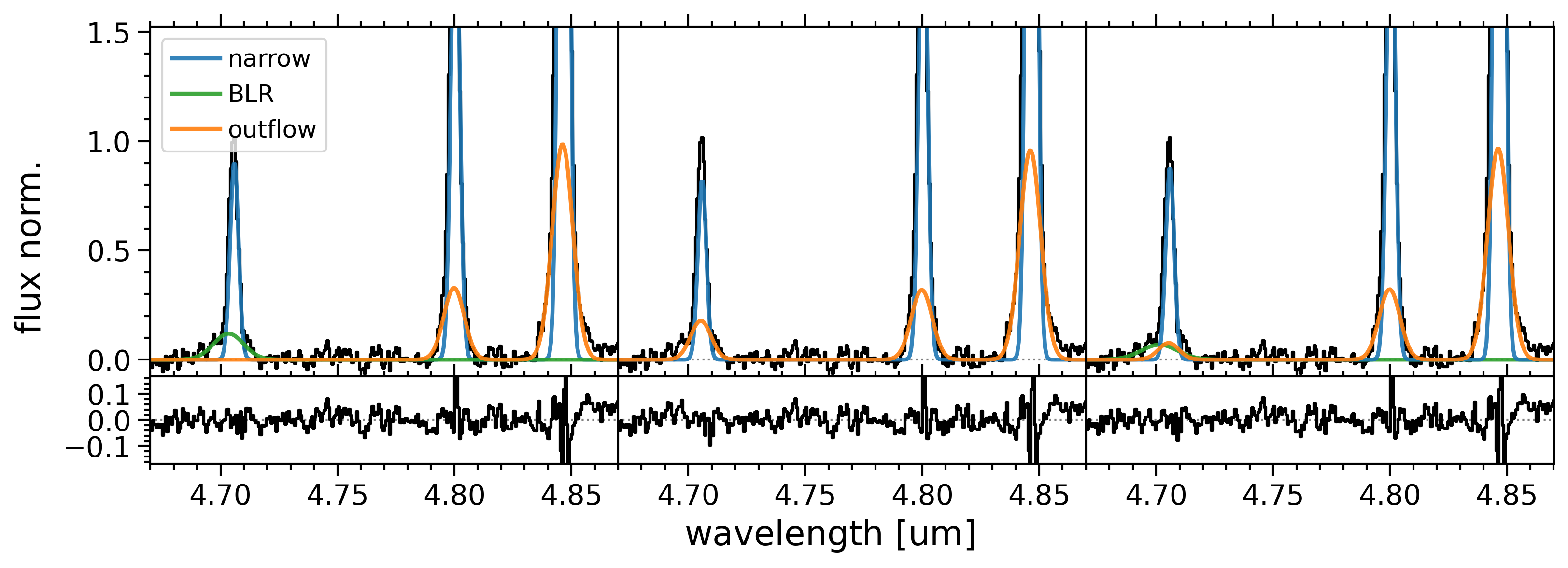}
\caption{Multi-Gaussian models fit to \hb\ and \oiii (top panels) and residual (bottom panels). Two components (narrow + outflows) are used to reproduce the \oiii\ profile in all panels. The narrow+BLR Gaussian model for \hb\ is shown in the left panel, narrow+outflow  model is reported in the middle, and narrow+BLR+outflows is shown in the right panel.}
\label{fig:bzoomin}
\end{figure*}
\begin{table}[h!]
\caption{Best-fits results for each model used for \hb\ broad component.}\label{tab:MOS_res}
\centering
\begin{tabular}{lccc}
\hline
    & BLR &  outflow & BLR   \\
    &     &          & + \\
    &     &          & outflow \\
\hline
$F_{\rm [OIII]}^{\rm outflow~\dagger}$ & $109\pm12$ & $108\pm12$ & $109\pm11$\\
$FWHM_{\rm [OIII]}^{\rm outflow ~\ddagger}$ &  $640\pm20$ & $650\pm20$ & $650\pm20$\\
$F_{\rm H\beta}^{\rm outflow ~\dagger}$  & - & $19\pm3$ & $8\pm7$ \\
$F_{\rm H\beta}^{\rm BLR ~\dagger}$  &$19\pm5$ & - &  $13\pm10$ \\
$FWHM_{\rm H\beta}^{\rm BLR ~\ddagger}$   & $960\pm110$ & - & $1160\pm200$\\
\hline
\end{tabular}
\\Note. $\dagger~{\rm 10^{-19}~erg/s/cm^{2}}$; $\ddagger~{\rm km/s} $
\end{table}

\newpage
\section{Emission lines measurements and maps}
\begin{figure*}[h]
\includegraphics[width=0.99\textwidth]{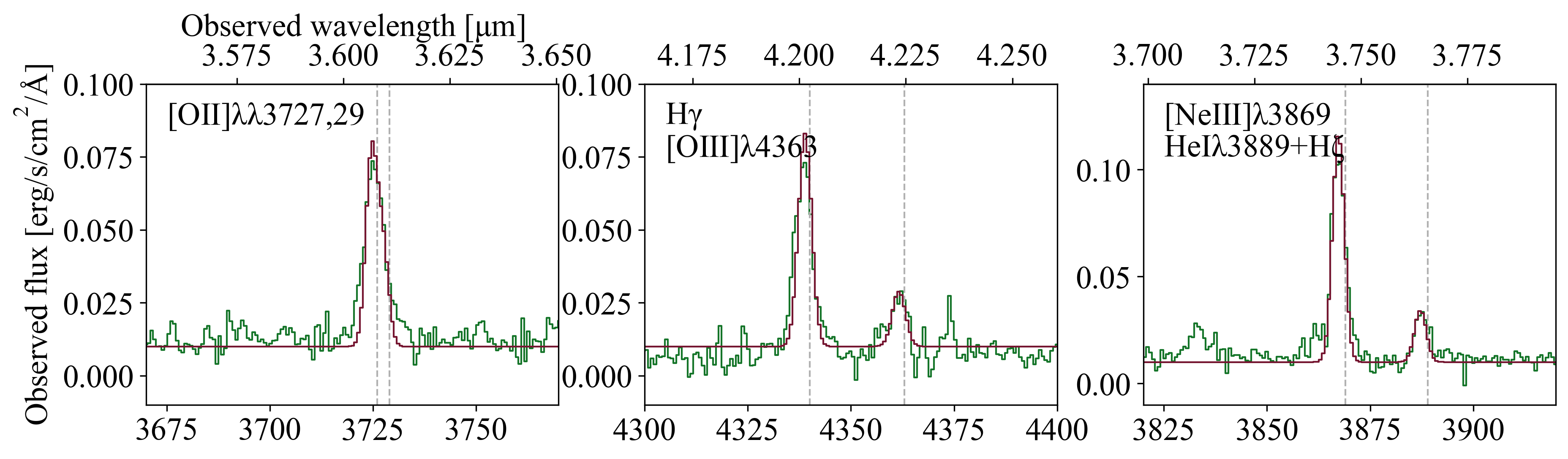}
\includegraphics[width=1\textwidth]{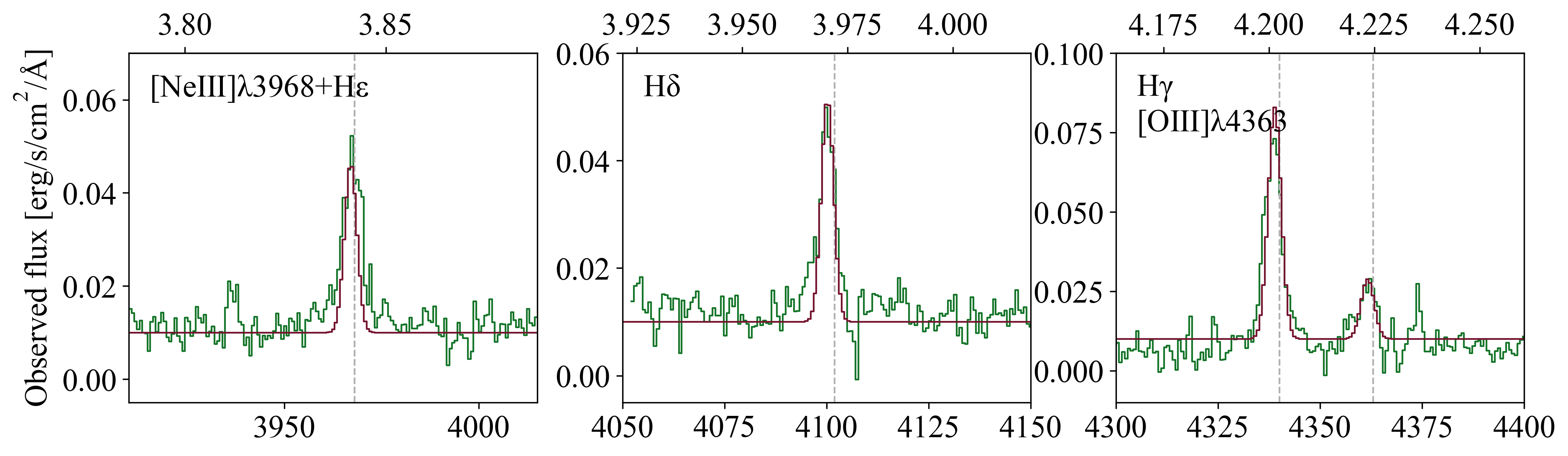}
\includegraphics[width=0.66\textwidth]{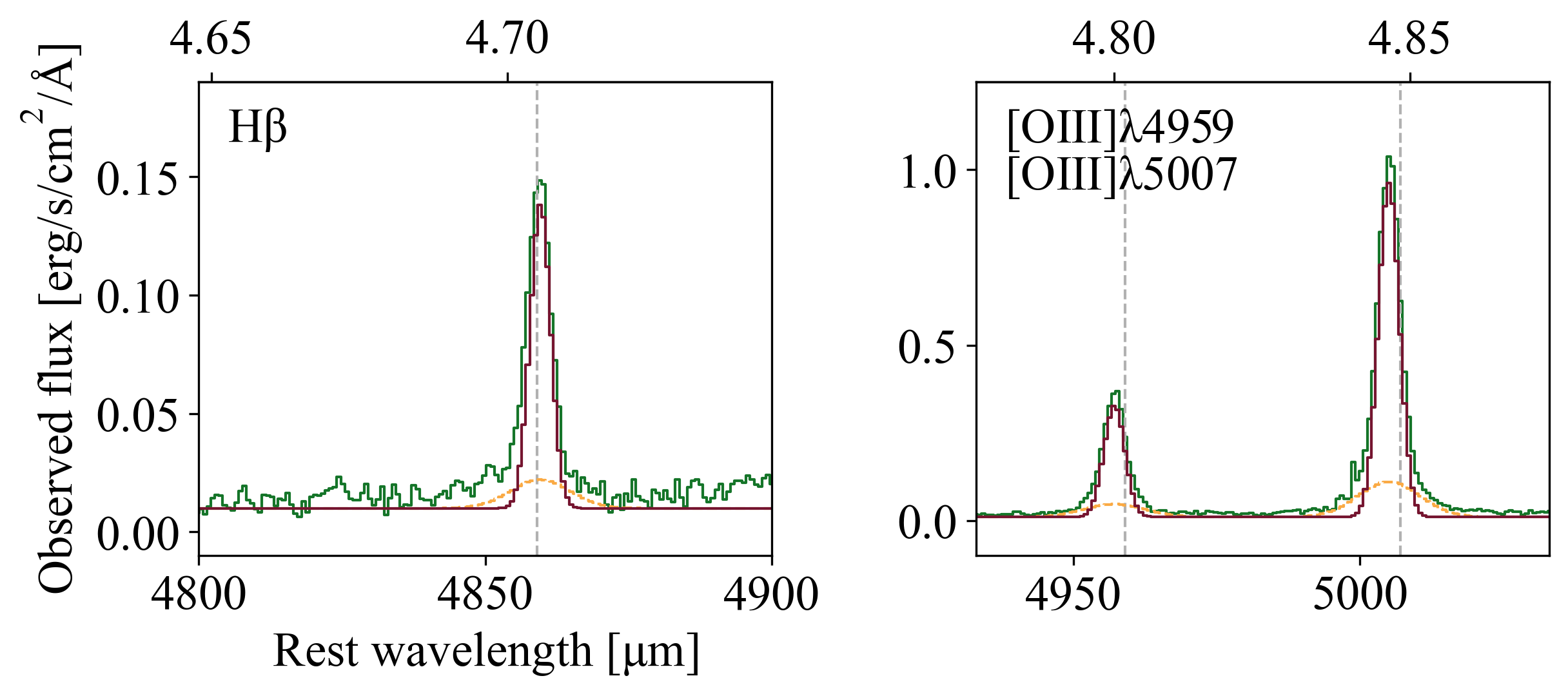}
\caption{JWST/NIRSpec G395H/F290LP emmision lines present in the spectrum of the galaxy. The fitted narrow and broad components are represented by the solid red and dashed yellow lines, respectively.}
\label{fig:lines-high}
\end{figure*}

\begin{figure*}[h]
\centering
\includegraphics[width=0.3\textwidth]{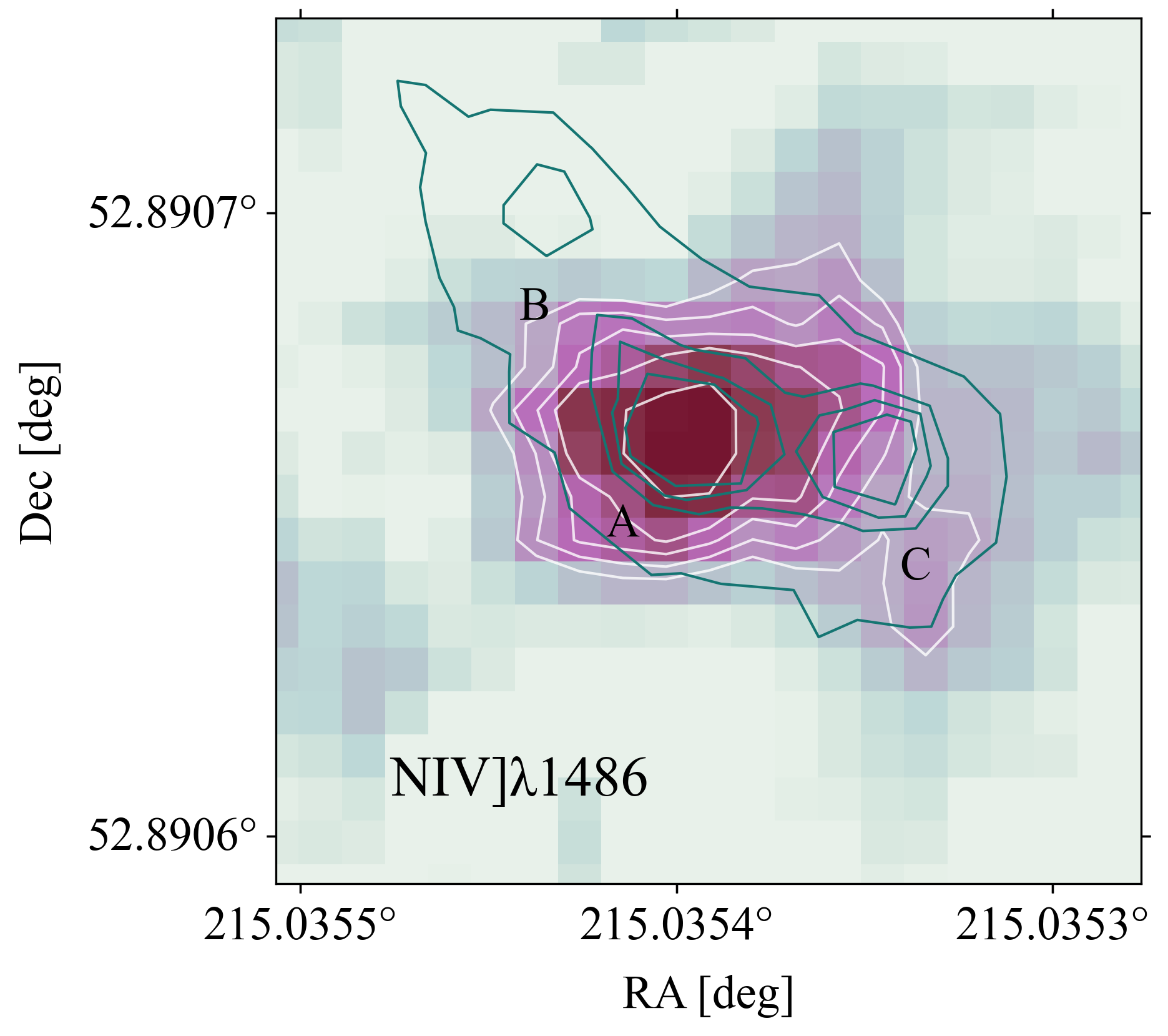}
\includegraphics[width=0.3\textwidth]{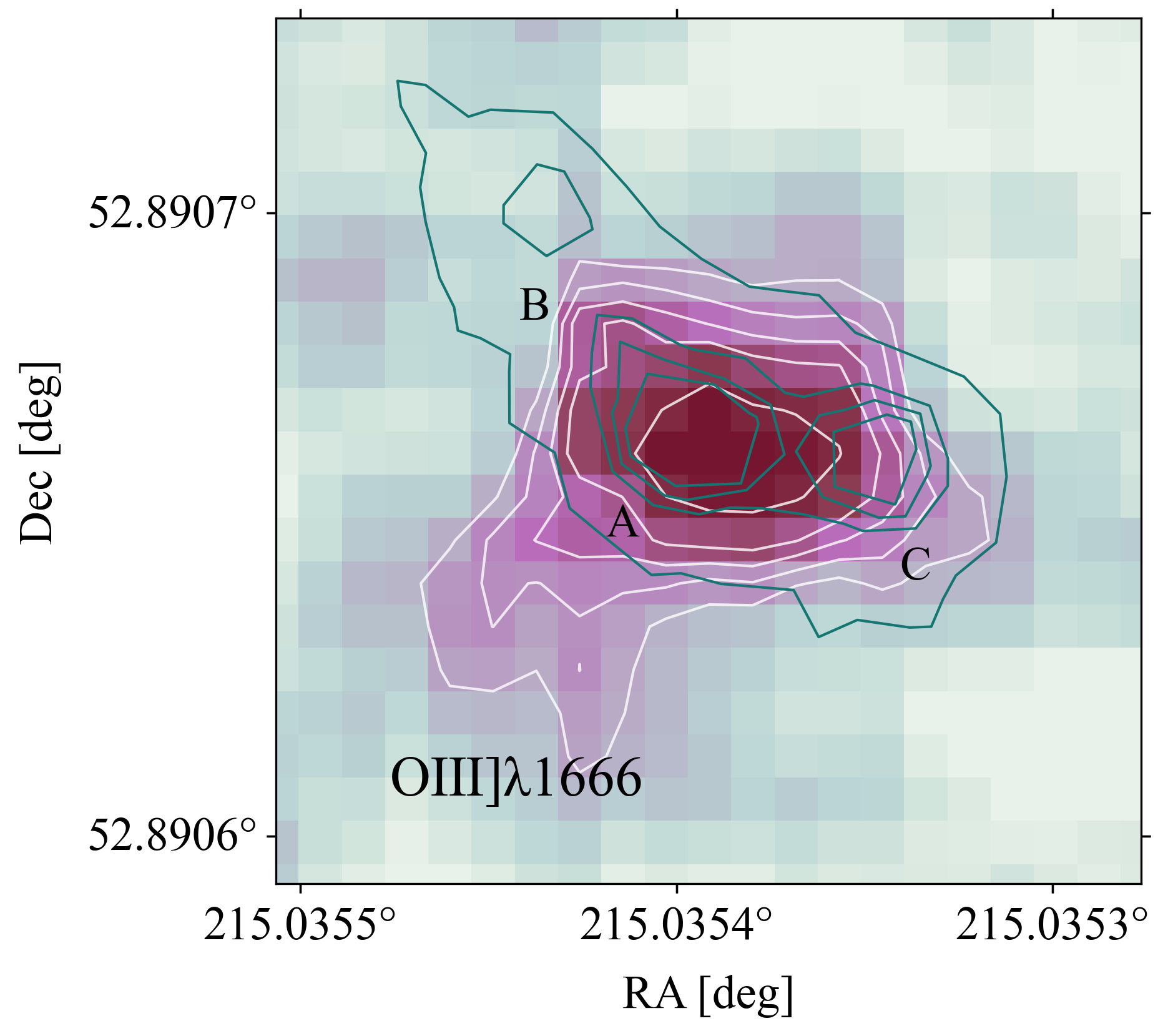}
\includegraphics[width=0.3\textwidth]{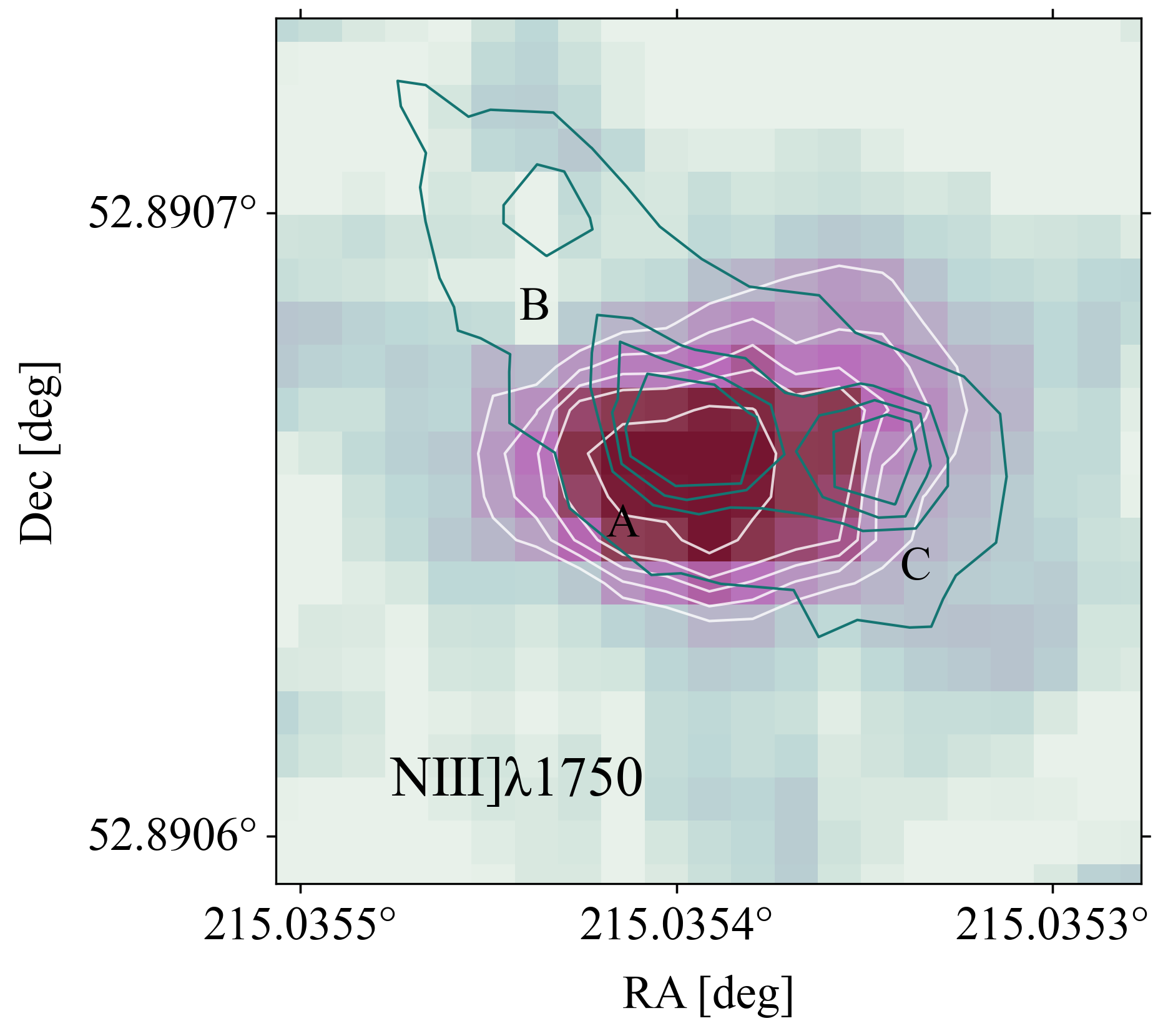}
\includegraphics[width=0.3\textwidth]{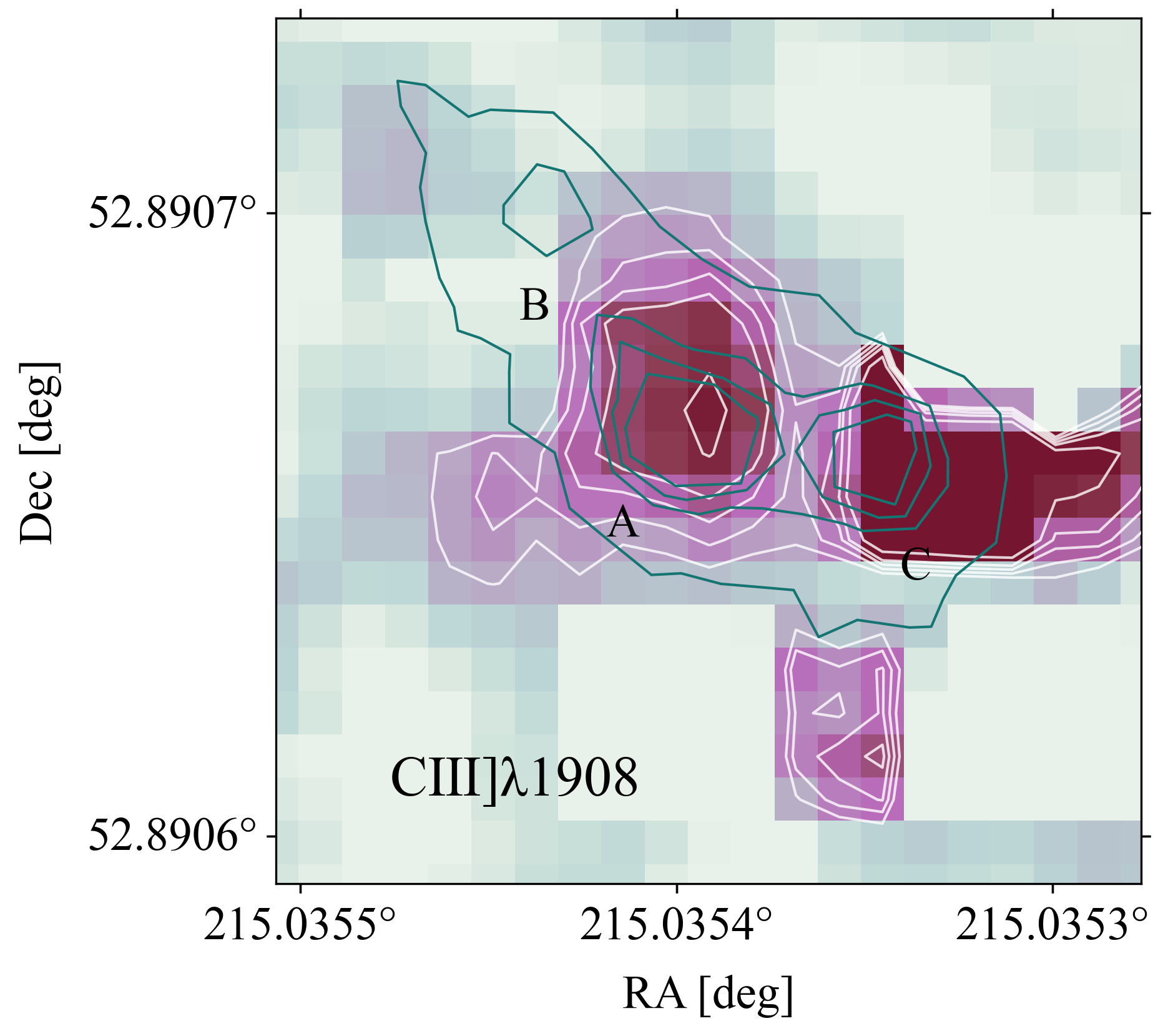}
\includegraphics[width=0.3\textwidth]{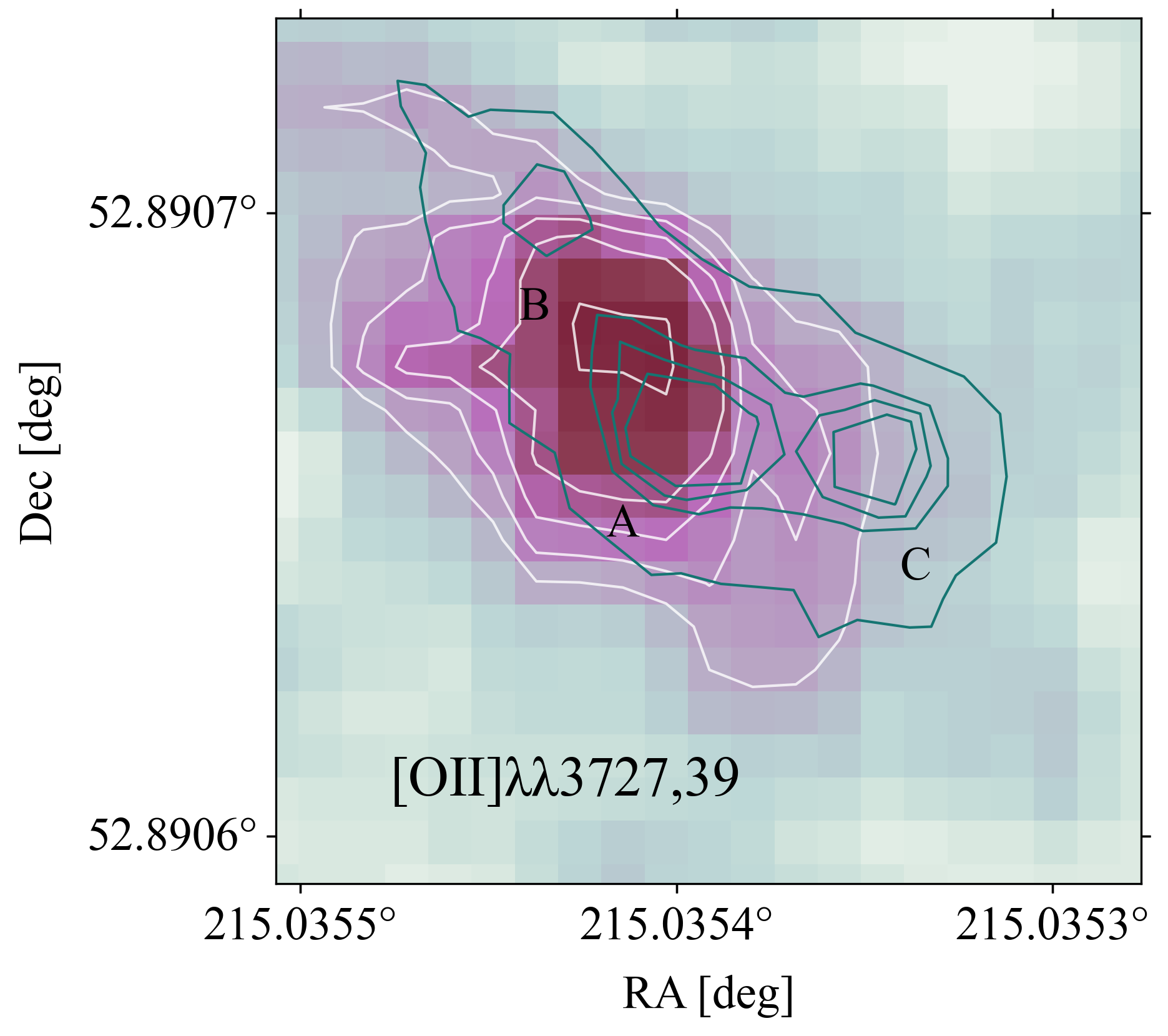}
\includegraphics[width=0.3\textwidth]{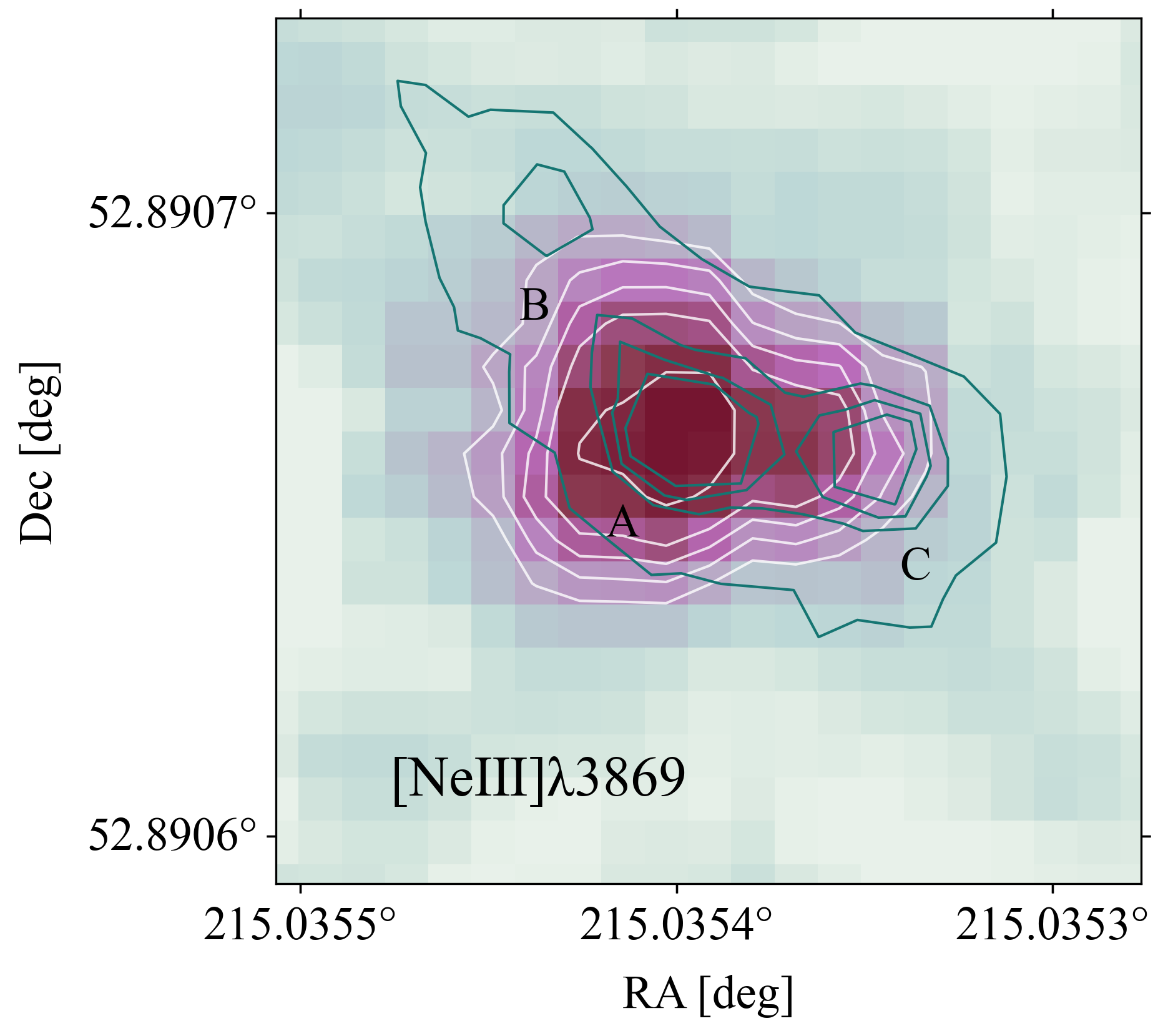}
\includegraphics[width=0.3\textwidth]{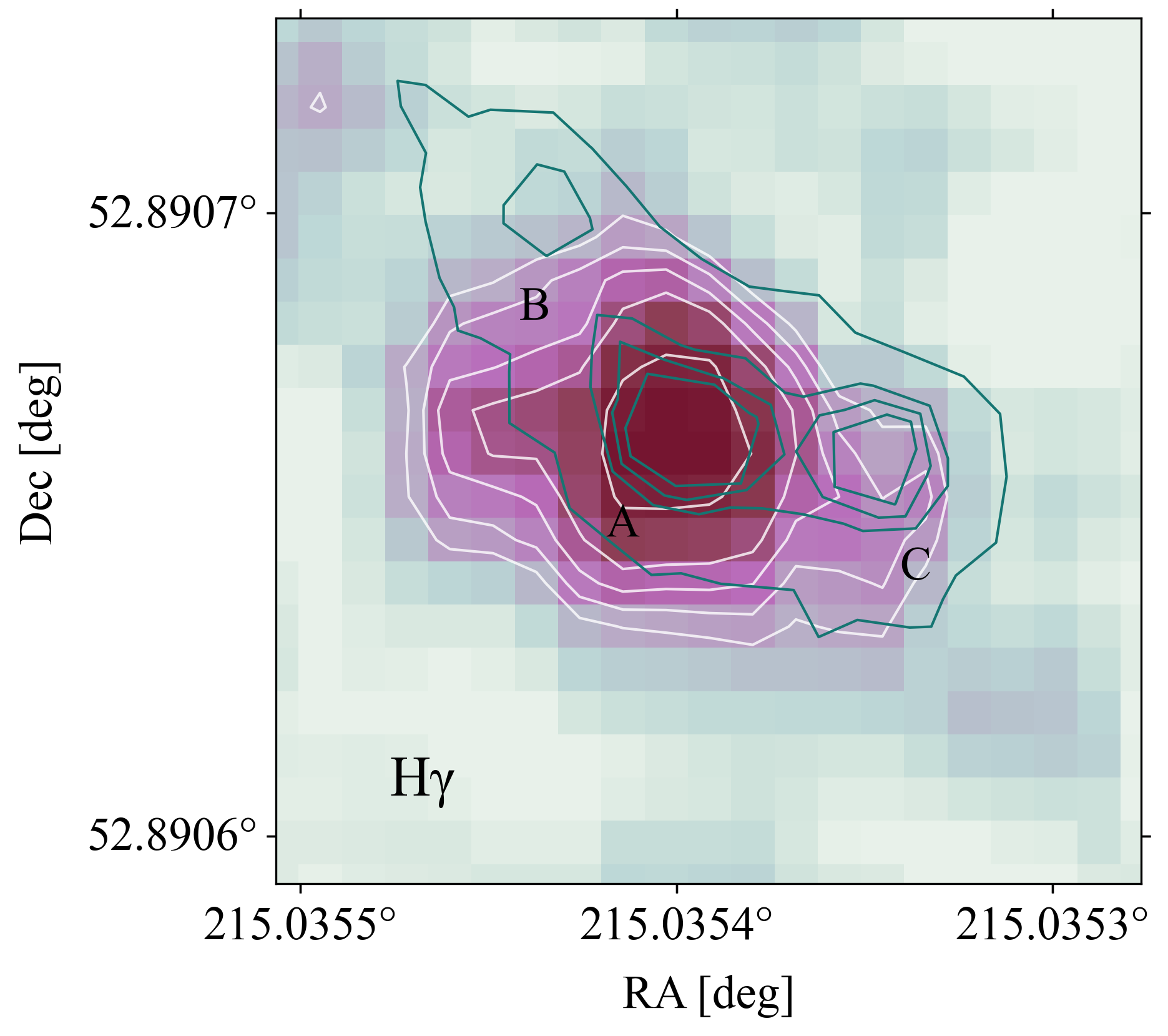}
\includegraphics[width=0.3\textwidth]{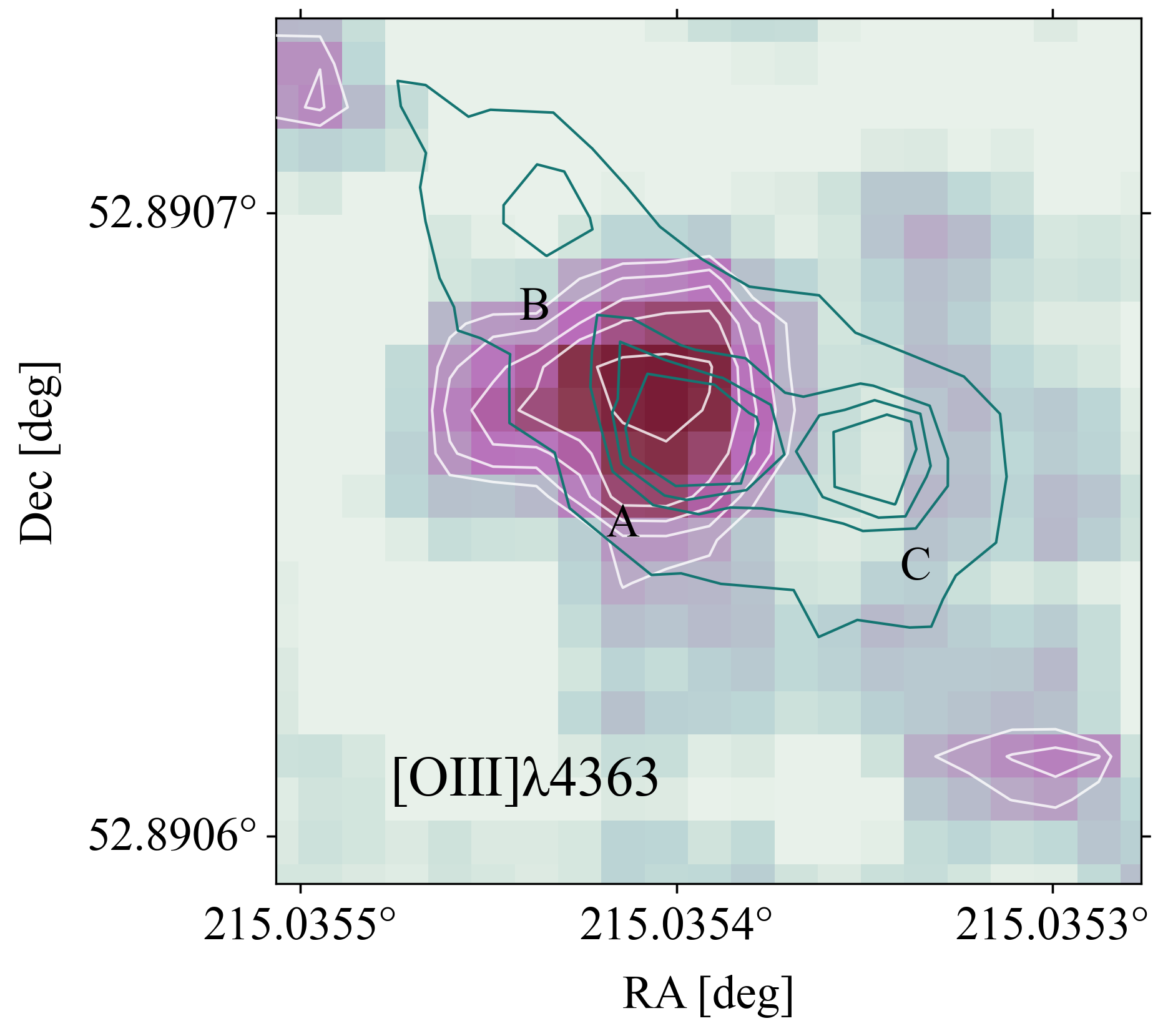}
\includegraphics[width=0.3\textwidth]{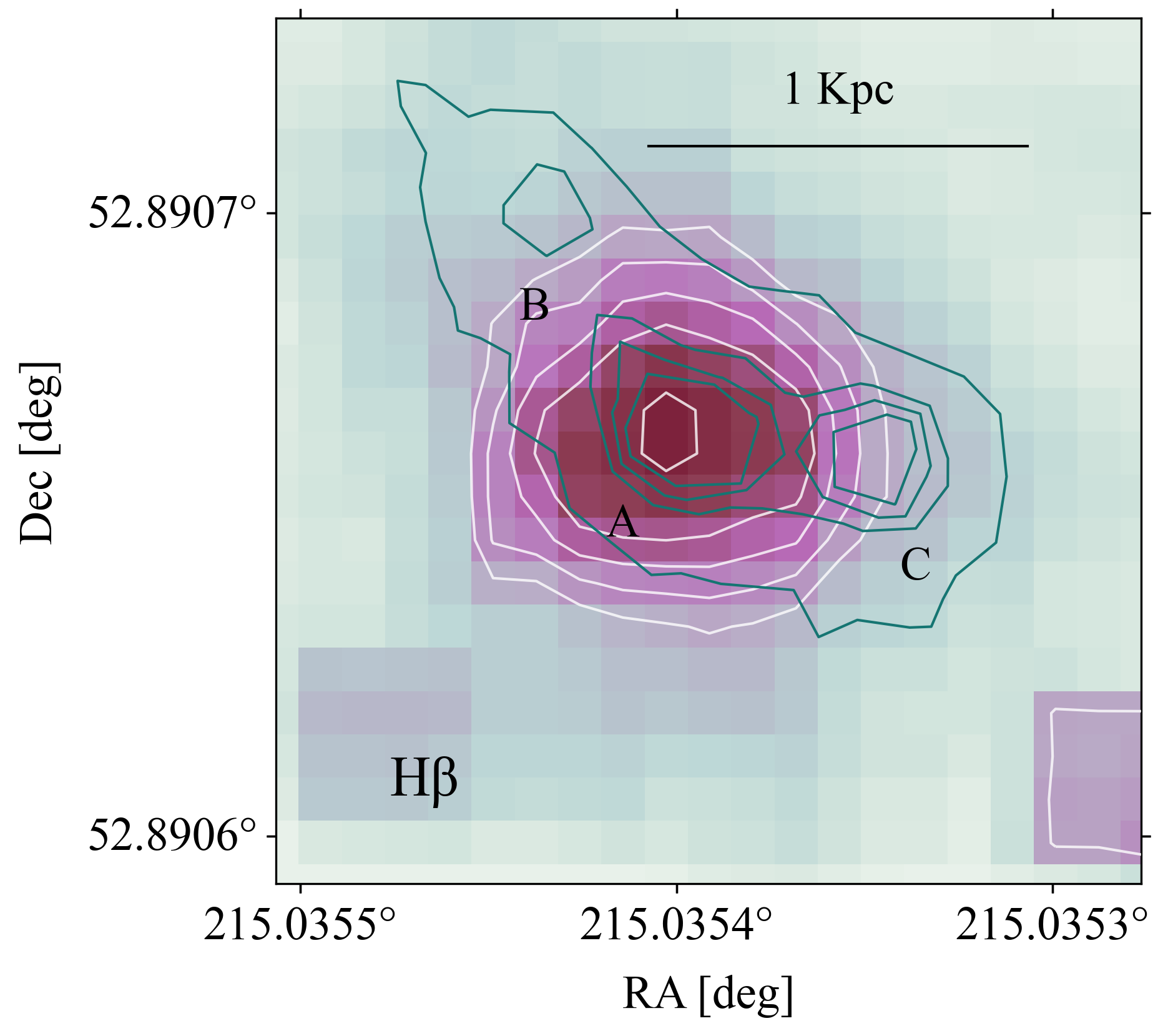}
\includegraphics[width=0.3\textwidth]{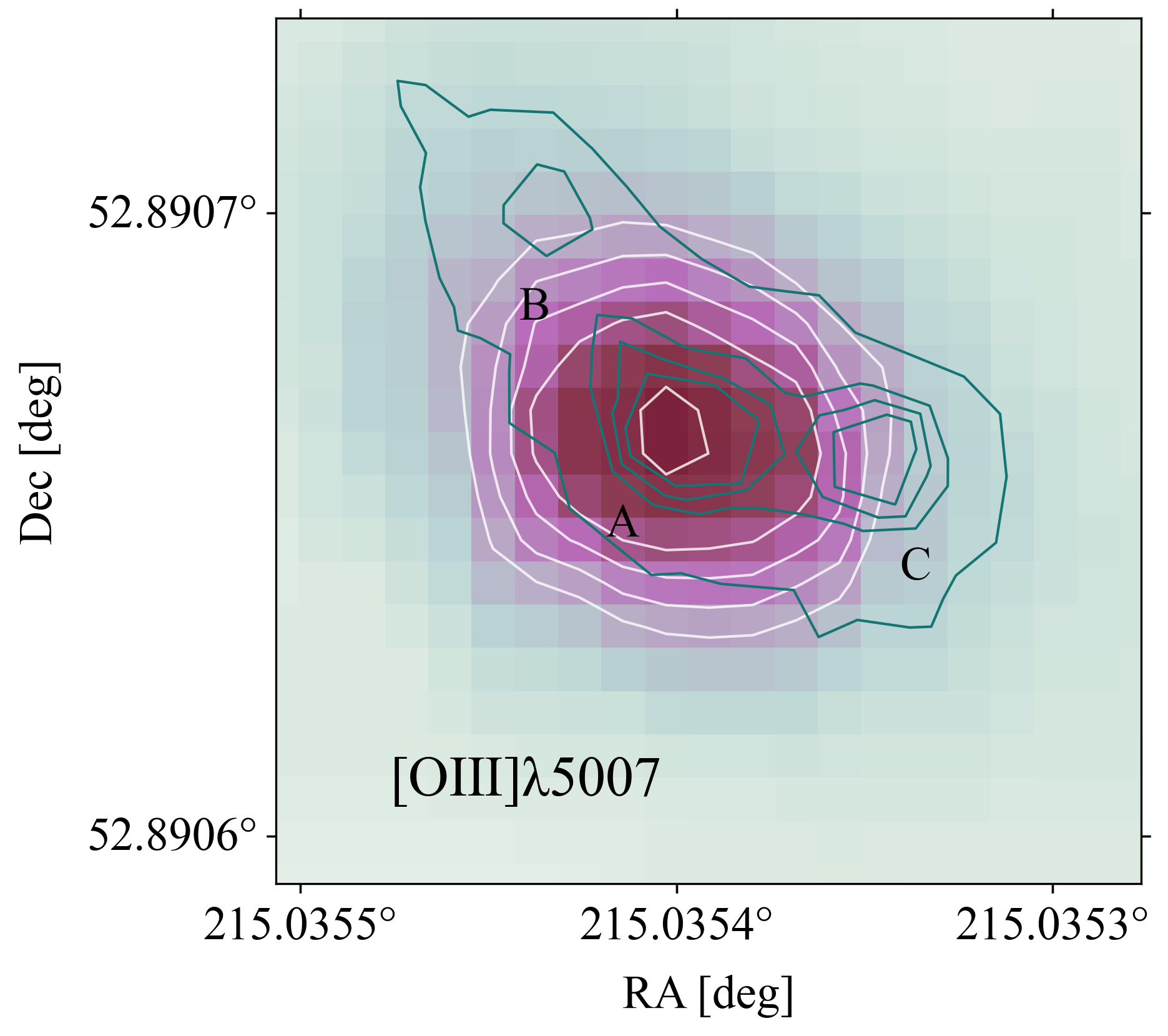}
\caption{JWST/NIRSpec PRISM maps of the different emission lines present in the spectra. The JWST/NIRCam F150W image is shown with blue contours, while the white contours correspond to the emission map presented. The strong CIII]$\lambda$ 1908 \AA\ emission at the position of clump C is the residual of a spike that was not removed during the data reduction process. The CIII]$\lambda$ 1908 emission detected in clump C is caused by a bad pixel.}\label{fig:line_maps}
\end{figure*}

\newpage

\onecolumn
\section{Spectro-astrometry analysis} \label{ap:spectrophot}
\begin{figure*}[h]
\centering
\includegraphics[width=0.3\textwidth]{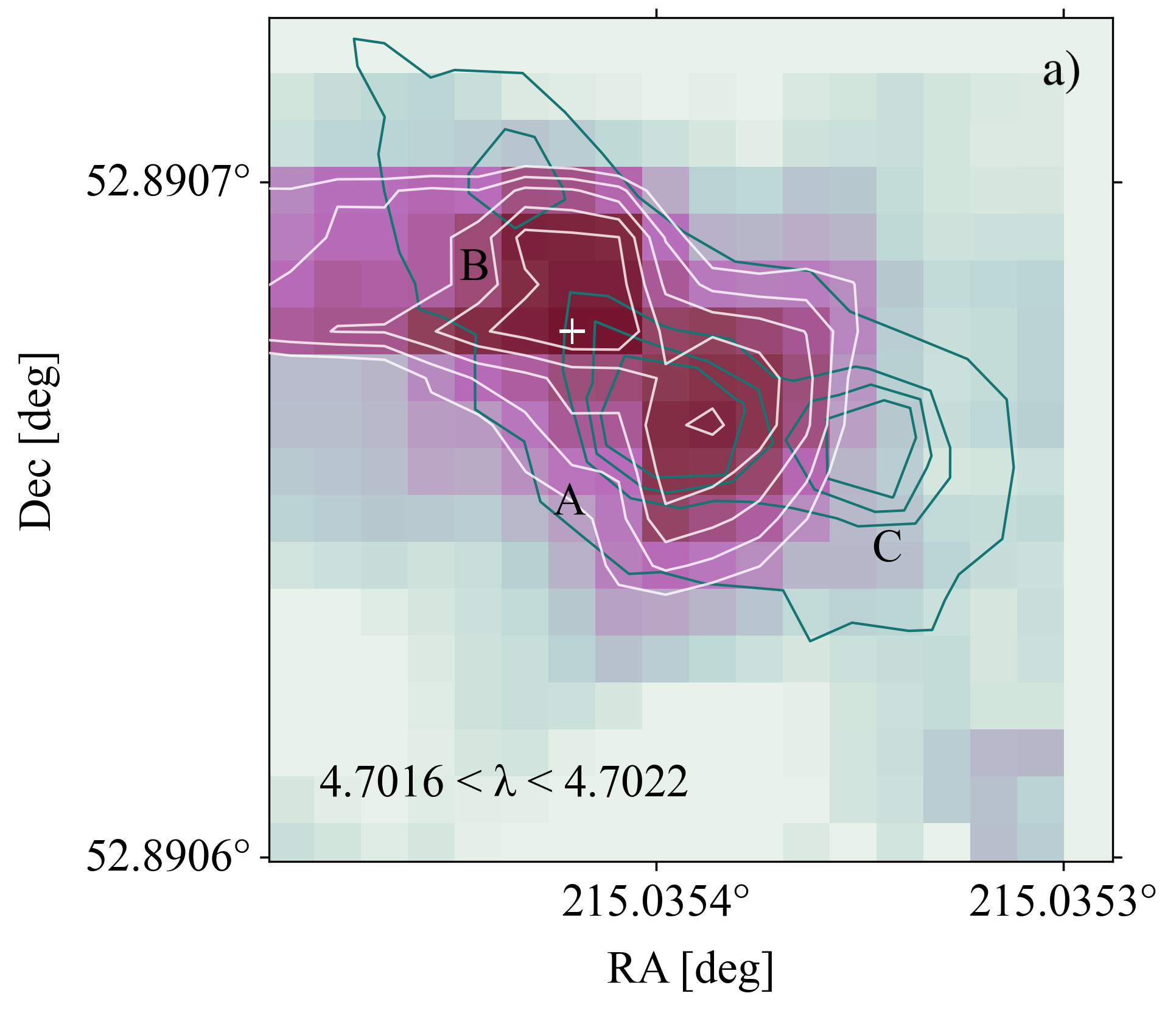}
\includegraphics[width=0.3\textwidth]{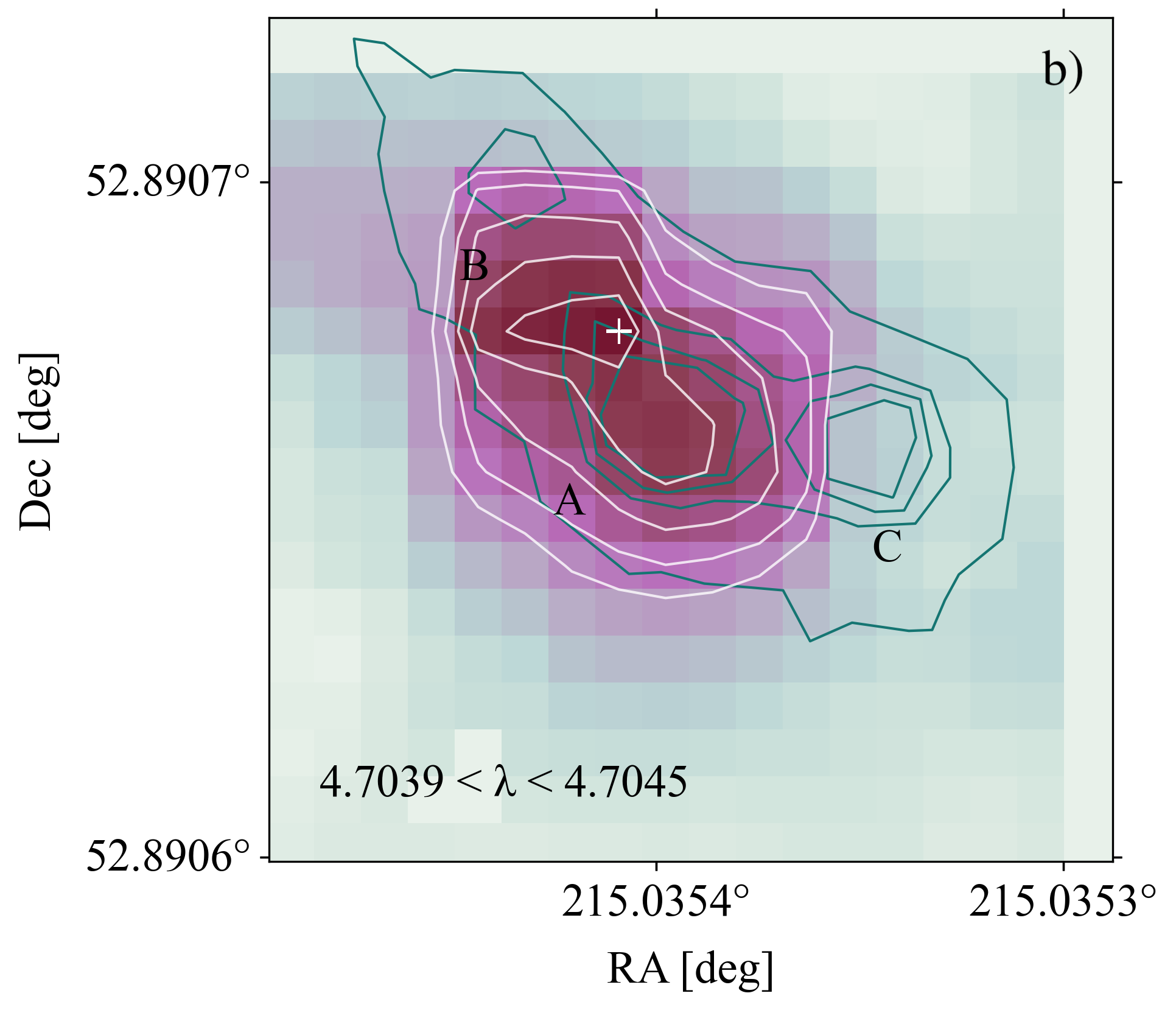}
\includegraphics[width=0.3\textwidth]{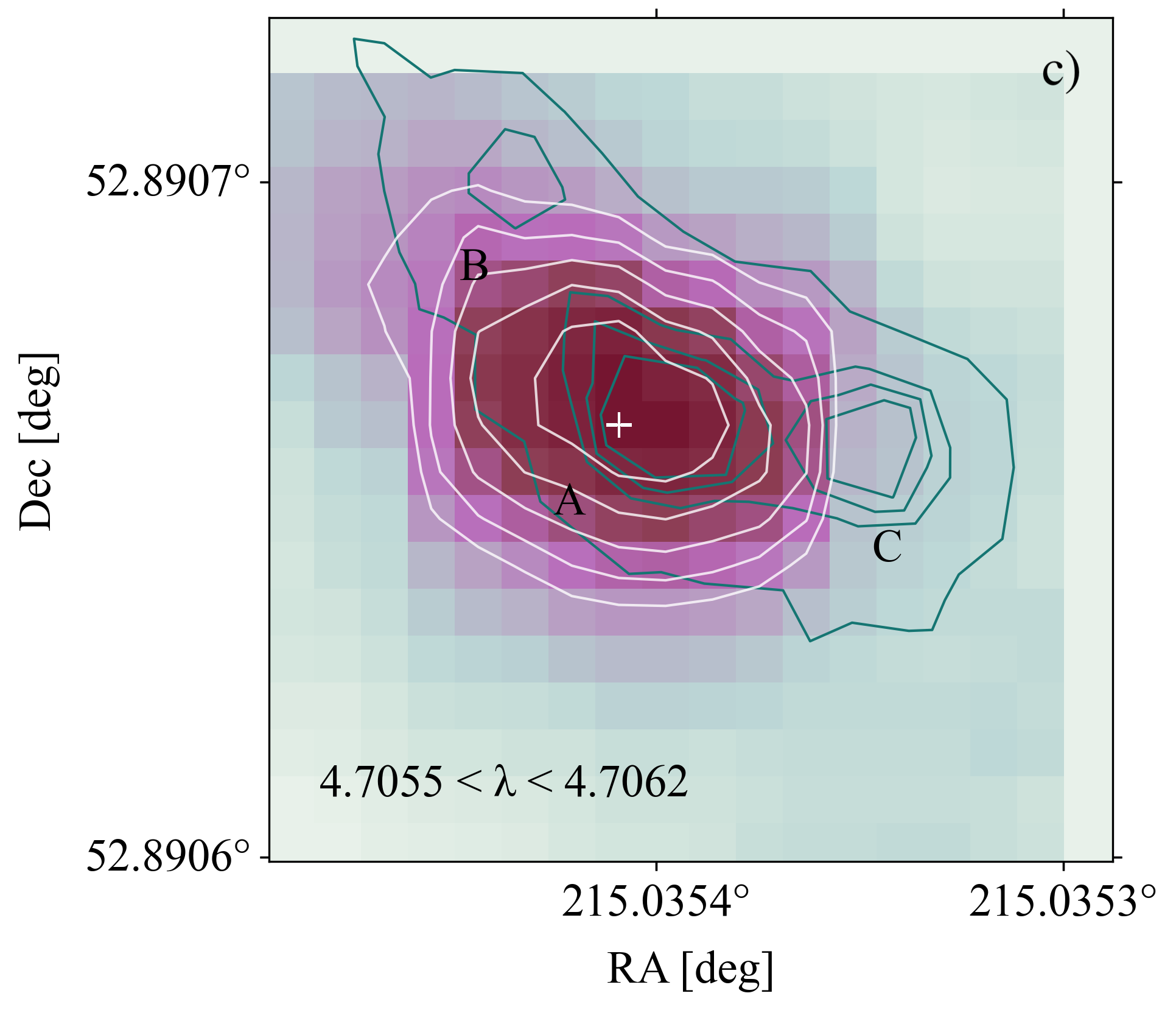}
\includegraphics[width=0.3\textwidth]{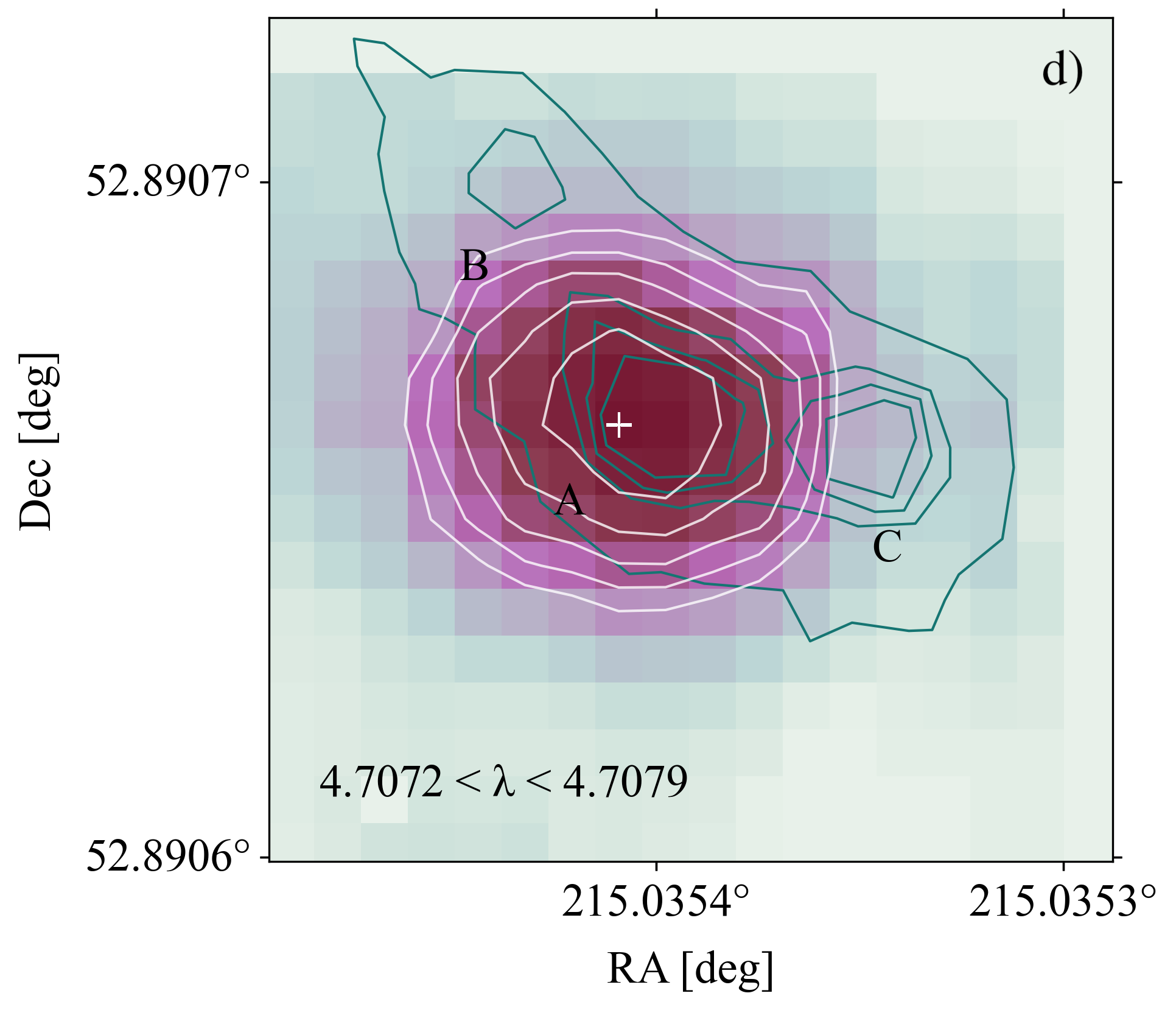}
\includegraphics[width=0.3\textwidth]{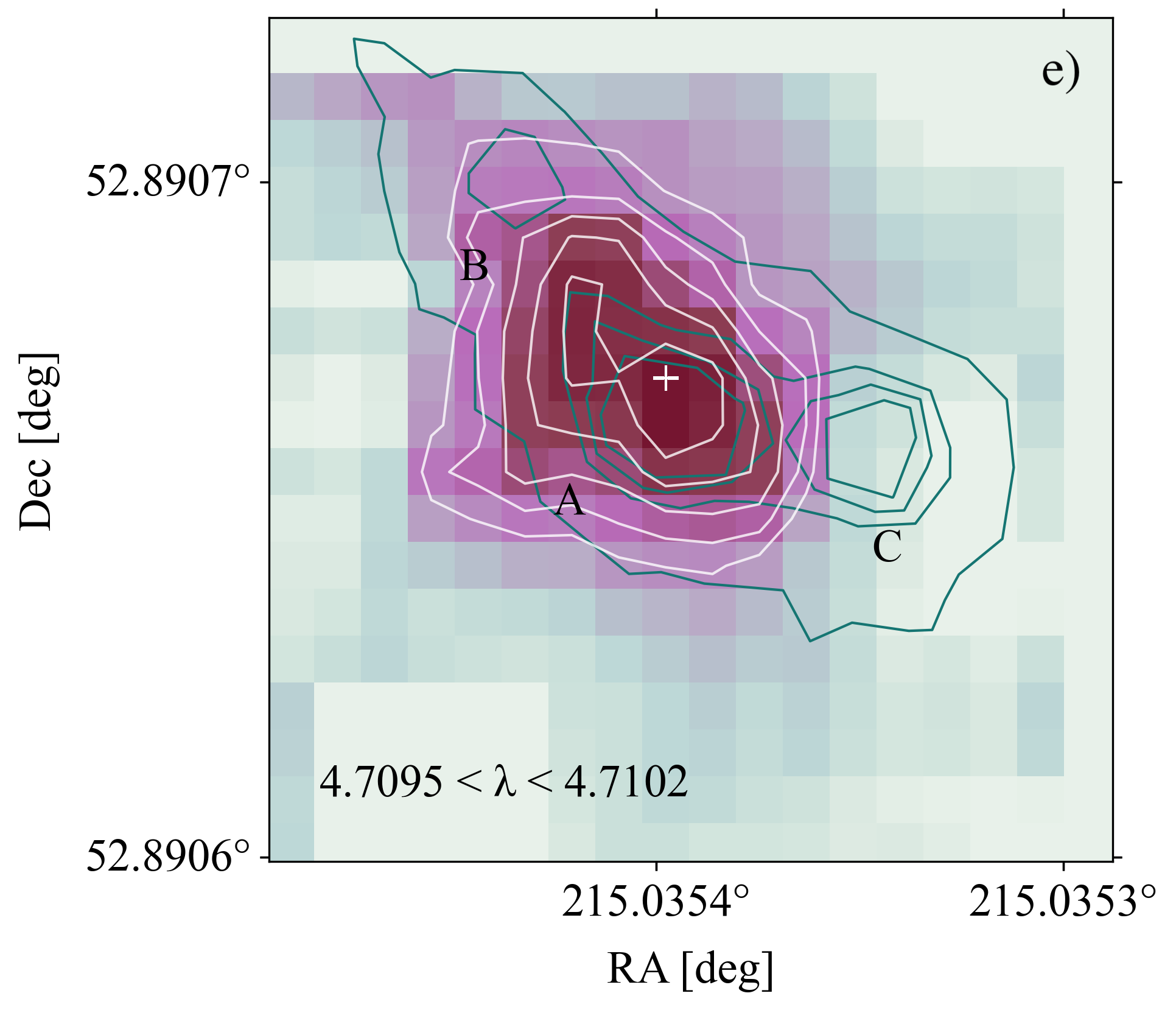}
\includegraphics[width=0.28\textwidth]{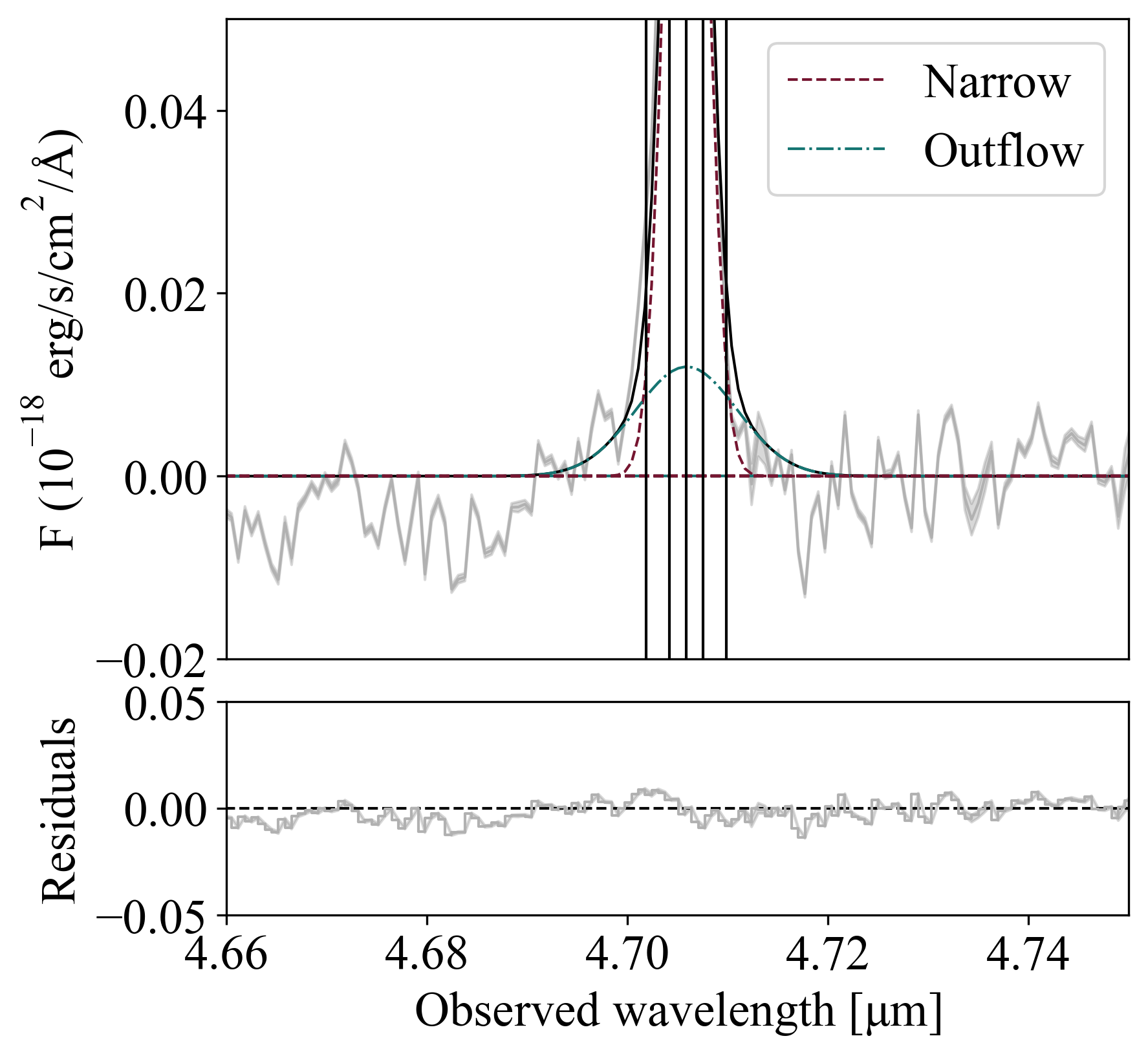}
\caption{JWST/NIRSpec G395H/F290LP maps of the H$\beta$ emission line, collapsed into different spectral channels centered at the wavelengths marked over the spectral fitting in the lower right panel of the figure and with a width of  6.65 $\times$ 10$^{-4}$ $\mu$m (2 spectral pixels). The wavelength ranges are indicated in the bottom left part of each panel. The JWST/NIRCam F150W image is shown with blue contours, while the white contours correspond to the emission map presented.}\label{fig:channel}
\end{figure*}

We tried to disentangle the nature of the broad H$\beta$ emission component, detected in the studies by \citet{Larson2023} and \citet{Marques-Chaves2024} as well as in the IFS data presented in this work, taking full advantage of the spatial resolution provided by this new dataset. However, the G395H/F290LP data do not have sufficient signal-to-noise to perform a pixel-by-pixel analysis of the broad H$\beta$ emission component. A direct spectro-astrometric analysis is also not possible due to the presence of a poorly subtracted spike, visible in the first panel of the figure (panel a) as an artificial extension from clump B toward the left side. This artifact likely produces the asymmetric profile observed in the MOS extraction, particularly in the blue wing of H$\beta$ shown in Figure \ref{fig:bzoomin}.

Then, we performed channel maps of the H$\beta$ emission to investigate the spatial origin of the broad component (see Figure \ref{fig:channel}). In panels c and d, where the emission is dominated by the narrow component, the signal is primarily centered on clump A. In contrast, in panels b and e, where the narrow and broad components contribute comparably, the maps display a clear double-peaked structure: one peak associated with clump A and a second peak located at the position of the outflow component. This behavior supports the interpretation that the broad H$\beta$ emission arises from a spatially offset component, consistent with an origin in the outflow.

\newpage
\section{Tables}

\begin{table}[h]
\caption{Indirect electron temperatures in units of 10$^4$ K.}\label{tab:temperatures}
\centering
\begin{tabular}{lcc}
\hline 
& Value                                & Calibration       \\ \hline 
T$_e$(OII)                 & 1.26361  $\pm$  0.00095 & \citet{Izotov2006}        \\
                            & 1.3287  $\pm$  0.0012   & \citet{stasinska1990}     \\
                           & 1.29253  $\pm$  0.00096 & \citet{campbell1986}      \\
                           & 1.4470  $\pm$  0.0011   & \citet{pilyugin2010}      \\ 

T$_e$(SIII)                & 1.3469  $\pm$  0.0011   & \citet{garnett1992}       \\
                           & 1.56 $\pm$  0.24        & \citet{croxall2016}       \\ 

T$_e$(NII)                 & 1.2980  $\pm$  0.0013   & \citet{perez-montero2009} \\
                           & 1.27  $\pm$  0.24       & \citet{berg2020}       \\ \hline 
\end{tabular}
\end{table}

\begin{table*}[h!]
\centering
\caption{Extraction parameters for emission-line maps.}
\begin{tabular}{lcccc}
\hline
Emission line                     & $\Delta\lambda_{line}$$^\dagger$ & $\Delta\lambda_{cont}^{left}$$^\dagger$ & $\Delta\lambda_{cont}^{right}$$^\dagger$ & S/N peak \\ \hline
NIV{]}$\lambda $1486              & 1448 - 1508            & 1384 - 1405                   & 1590 - 1611                    & 1.9      \\
OIII{]}$\lambda $1666             & 1618 - 1678            & 1601 - 1618                   & 1797 - 1815                    & 2.1      \\
NIII{]}$\lambda $1750             & 1722 - 1774            & 1601 - 1618                   & 1797 - 1815                    & 2.5      \\
CIII{]}$\lambda $1908             & 1858 - 1915            & 1837 - 1849                   & 1952 - 1983                    & 2.1      \\
{[}OII{]}$\lambda \lambda$3727,39 & 3708 - 3754            & 3646 - 3666                   & 4214 - 4234                    & 7.2      \\
{[}NeIII{]}$\lambda $3869         & 3831 - 3909            & 3646 - 3666                   & 4214 - 4234                    & 8.5      \\
H$\gamma$                         & 4314 - 4358            & 3646 - 3666                   & 4214 - 4234                    & 6.7      \\
{[}OIII{]}$\lambda $4363          & 4358 - 4389            & 3646 - 3666                   & 4214 - 4234                    & 3.8      \\
H$\beta$                           & 4841 - 4885            & 4792 - 4813                   & 5112 - 5133                    & 23.1    \\
{[}OIII{]}$\lambda $5007             & 4986 - 5036             & 4792 - 4813	              & 5112 - 5133	                 & 131.4\\\hline

\end{tabular}
\\Note. $\dagger~$All wavelengths are in rest frame and in \AA .
\label{tab:line_maps}
\end{table*}

\begin{table*}
\caption{Interstellar medium and stellar population properties}\label{tab:ism}
\centering
\begin{tabular}{lclcc}
\hline
Physical property        & Value                   & Method                                               & Value              & Reference         \\ \hline
t$_e$(OIII) [10$^{-4}$ K]             & 1.4179  $\pm$  0.0014   & {[}OIII{]}$\lambda \lambda $4959,5007/{[}OIII{]}$\lambda$4463 & 78.10  $\pm$  0.17 & \citet{hagele2008}        \\
12+log(O$^{+}$/H$^{+}$)$^1$  & 6.901  $\pm$  0.075     &                                                                     &                    &                   \\
12+log(O$^{2+}$/H$^{+}$) & 7.983  $\pm$  0.071     &                                                                     &                    &                   \\
12+log(O/H)              & 8.017  $\pm$  0.066     &                                                                     &                    &                   \\ 

log(u)                   & -1.980  $\pm$  0.066    & log({[}OIII{]}$\lambda $5007/{[}OIII{]}$\lambda $3727)              & 1.300 $\pm$ 0.033     & \citet{diaz2000}          \\
log(u)                   &       -2.216  $\pm$ 0.035                  & log({[}NeIII{]}$\lambda $3869/{[}OII{]}$\lambda$3727) & 0.234  $\pm$  0.034     & \citet{levesque2014}$^2$ \\ 

$n_{\rm e}$ {[}cm$^{-3}${]}    & 3.56 $^{-0.31}_{-0.21}$ & {[}OII{]}$\lambda$3729/{[}OII{]}$\lambda$3727                       & 0.542 $\pm$  0.092 & \citet{pyneb}  \\ \\

SFR$_{H\alpha}$ [M$_\odot$/yr] & 49 $\pm$ 3 & L(H$\alpha$) [erg/s] & (1.52 $\pm$ 0.11) $\times$ 10$^{43}$ & \citet{reddy2018} \\
SFR$_{UV}$ [M$_\odot$/yr] & 4.2 $\pm$ 0.7 & L(1500-2800\AA ) [erg/s] & (3.35 - 1.32) $\times$ 10$^{40}$ & \citet{kennicutt1998} \\

\hline
\end{tabular}
\\Note. $^1$ By assuming the [OII] temperature calculated. 
\end{table*}

\end{document}